\documentclass[twocolumn,epjc3]{svjour3}  
\RequirePackage{mathptmx}      
\RequirePackage{graphics}
\RequirePackage{amsmath}
\RequirePackage{amssymb}
\RequirePackage{listings}
\RequirePackage{multirow}
\RequirePackage[colorlinks=true,linkcolor=blue,citecolor=blue]{hyperref}
\renewcommand{\arraystretch}{1.18}

\def\slash#1{
\setbox0=\hbox{$#1$}                    
\dimen0=\wd0                            
\setbox1=\hbox{/} \dimen1=\wd1          
\ifdim\dimen0>\dimen1                   
\rlap{\hbox to \dimen0{\hfil/\hfil}}    
#1                                      
\else                                   
\rlap{\hbox to \dimen1{\hfil$#1$\hfil}} 
/                                       
\fi}
\newcommand{\GeV}{\mbox{$\,\mathrm{GeV}$}}
\newcommand{\TeV}{\mbox{$\,\mathrm{TeV}$}}
\newcommand{\Wp}{\mbox{$W^{+}$}}
\newcommand{\Wm}{\mbox{$W^{-}$}}

\newcommand{\Wmp}{\mbox{$W^{\mp}$}}
\newcommand{\WpWm}{\mbox{$\Wp\Wm$}}
\newcommand{\WpWp}{\mbox{$\Wp\Wp$}}
\newcommand{\WpZ}{\mbox{$\Wp Z$}}

\newcommand{\taup}{\mbox{$\tau^{+}$}}
\newcommand{\taum}{\mbox{$\tau^{-}$}}
\newcommand{\taupm}{\mbox{$\tau^{\pm}$}}
\newcommand{\tauptaum}{\mbox{$\taup\taum$}}
\newcommand{\lp}{\mbox{$\ell^{+}$}}
\newcommand{\lm}{\mbox{$\ell^{-}$}}
\newcommand{\lpm}{\mbox{$\ell^{\pm}$}}
\newcommand{\lmp}{\mbox{$\ell^{\mp}$}}
\newcommand{\lplm}{\mbox{$\lp\lm$}}
\newcommand{\uquark}{\mbox{$u$}}
\newcommand{\bquark}{\mbox{$b$}}
\newcommand{\tquark}{\mbox{$t$}}
\newcommand{\nul}{\mbox{$\nu_{\ell}$}}
\newcommand{\ubar}{\mbox{$\overline{\uquark}$}}
\newcommand{\bbar}{\mbox{$\overline{\bquark}$}}
\newcommand{\tbar}{\mbox{$\overline{\tquark}$}}
\newcommand{\nulbar}{\mbox{$\overline{\nul}$}}
\newcommand{\ttbar}{\mbox{$\tquark\tbar$}}
\newcommand{\mW}{\mbox{$m_{W}$}}
\newcommand{\mZ}{\mbox{$m_{Z}$}}
\newcommand{\mt}{\mbox{$m_{t}$}}
\newcommand{\mtt}{\mbox{$m_{t\overline{t}}$}}
\newcommand{\mH}{\mbox{$m_{H}$}}
\newcommand{\mHW}{\mbox{$m_{HW}$}}
\newcommand{\mHZ}{\mbox{$m_{HZ}$}}
\newcommand{\mHtt}{\mbox{$m_{Ht\overline{t}}$}}
\newcommand{\MSbar}{\mbox{$\overline{\mbox{\scriptsize MS}}$}}
\newcommand{\HWp}{\mbox{$HW^{+}$}}

\newcommand{\Wpdecay}{\mbox{$\Wp\!\!\rightarrow\!\lp\nul$}}
\newcommand{\Wpmdecay}{\mbox{$\Wp(\Wm)\!\rightarrow\!\lp\nul\,(\lm\nulbar)$}}
\newcommand{\Zdecay}{\mbox{$Z\!\rightarrow\!\lplm$}}
\newcommand{\tdecay}{\mbox{$t\!\rightarrow\!b\,\lp\nul$}}
\newcommand{\tbardecay}{\mbox{$\tbar\!\rightarrow\!\bbar\,\lm\nulbar$}}
\newcommand{\tmass}{\mbox{$\mt\!=\!175\GeV$}}
\newcommand{\tBr}{\mbox{$\Br(\tdecay)\!=\!0.216$}}
\newcommand{\Hdecay}{\mbox{$H\!\rightarrow\!\tauptaum$}}
\newcommand{\Hmass}{\mbox{$\mH\!=\!125\GeV$}}
\newcommand{\HBr}{\mbox{$\Br(\Hdecay)\!=\!0.0405$}}

\newcommand{\lemu}{\mbox{$(\ell\!=\!e,\mu)$}}
\newcommand{\lemutau}{\mbox{$(\ell\!=\!e,\mu,\tau)$}}

\newcommand{\pT}{\mbox{$p_{\mathrm{T}}$}}
\newcommand{\pTi}{\mbox{$p_{\mathrm{T}i}$}}
\newcommand{\pTcut}{\mbox{$p_{\mathrm{T}}^{\scriptstyle\mathrm{cut}}$}}
\newcommand{\pTjetcut}{\mbox{$p_{\mathrm{T,jet}}^{\scriptstyle\mathrm{cut}}$}}
\newcommand{\etai}{\mbox{$\eta^{\scriptstyle\mathrm{cut}}$}}
\newcommand{\etacut}{\mbox{$\eta^{\scriptstyle\mathrm{cut}}$}}

\newcommand{\ET}{\mbox{$E_\mathrm{T}$}}

\newcommand{\alphas}{\mbox{${\alpha}_\mathrm{s}$}}
\newcommand{\alphaszero}{\mbox{${\alpha}_\mathrm{s}^{0}$}}
\newcommand{\alphasone}{\mbox{${\alpha}_\mathrm{s}^{1}$}}
\newcommand{\alphastwo}{\mbox{${\alpha}_\mathrm{s}^{2}$}}
\newcommand{\alphasthree}{\mbox{${\alpha}_\mathrm{s}^{3}$}}
\newcommand{\alphasfour}{\mbox{${\alpha}_\mathrm{s}^{4}$}}
\newcommand{\gw}{\mbox{$g_\mathrm{\scriptscriptstyle W}$}}
\newcommand{\gwsq}{\mbox{$g_\mathrm{\scriptscriptstyle W}^{2}$}}
\newcommand{\gwwz}{\mbox{$g_\mathrm{\scriptscriptstyle WWZ}$}}
\newcommand{\gwwa}{\mbox{$g_\mathrm{\scriptscriptstyle WWA}$}}
\newcommand{\thetaw}{\mbox{$\theta_\mathrm{\scriptscriptstyle W}$}}
\newcommand{\sqrts}{\mbox{$\sqrt{s}$}}
\newcommand{\LHCenergy}{\mbox{$\sqrt{s}\!=\!14\TeV$}}
\newcommand{\Br}{\mbox{$\mathrm{Br}\,$}}

\journalname{Eur. Phys. J. C}

\begin{document}

\title{Fast computation of MadGraph amplitudes on graphics processing unit (GPU)}

\titlerunning{Fast computation of MadGraph amplitudes on GPU}

\author{K.~Hagiwara\thanksref{inst:KEK-theory}
   \and
   J.~Kanzaki\thanksref{e1,inst:KEK}
   \and
   Q.~Li\thanksref{e2,inst:Beijing}
   \and
   N.~Okamura\thanksref{e3,inst:Okamura}
   \and
   T.~Stelzer\thanksref{e4,inst:Illinois}
   }

\thankstext{e1}{junichi.kanzaki@kek.jp}
\thankstext{e2}{qliphy@gmail.com}
\thankstext{e3}{nokamura@iuhw.ac.jp}
\thankstext{e4}{tstelzer@illinois.edu}

\institute{KEK Theory Center and Sokendai, Tsukuba 305-0801, Japan\label{inst:KEK-theory}
    \and
    KEK and Sokendai, Tsukuba 305-0801, Japan \label{inst:KEK}
    \and
    Department of Physics and State Key, Laboratory of Nuclear Physics and Technology, Peking University, Beijing, 100871, China \label{inst:Beijing}
    \and
    Dep. of Radiological Sciences, International University of 
    Health and Welfare
    2600-1 Kitakenamaru, Ohtawara, Tochigi, Japan\label{inst:Okamura}
    \and 
    Dept. of Physics, University of Illinois, Urbana, IL 61801, USA \label{inst:Illinois}
   }
	    
\date{Received: date / Revised version: \today}

\maketitle

\begin{abstract}
Continuing our previous studies on QED and QCD processes, we use the graphics processing unit (GPU) for fast calculations of helicity amplitudes for general Standard Model (SM) processes. 
Additional HEGET codes to handle all SM interactions are introduced, as well as the program MG2CUDA that converts arbitrary MadGraph generated HELAS amplitudes (FORTRAN) into HEGET codes in CUDA.
We test all the codes by comparing amplitudes and cross sections for multi-jet processes at the LHC associated with production of single and double weak bosons, a top-quark pair, Higgs boson plus a weak boson or a top-quark pair, and multiple Higgs bosons via weak-boson fusion, where all the heavy particles are allowed to decay into light quarks and leptons with full spin correlations.
All the helicity amplitudes computed by HEGET are found to agree with those computed by HELAS within the expected numerical accuracy, and the cross sections obtained by gBASES, a GPU version of the Monte Carlo integration program, agree with those obtained by BASES (FORTRAN), as well as those obtained by MadGraph.
The performance of GPU was over a factor of 10 faster than CPU for all processes except those with the highest number of jets.
\end{abstract} 

 \section{Introduction}
 \label{sec:introduction}

 The start-up of the CERN Large Hadron Collider (LHC) opens a new discovery era of high energy particle physics. 
 With proton beams colliding at unprecedented energy, it provides us with great opportunities to discover the Higgs boson and new physics beyond the Standard Model (SM), with typical signals involving multiple high $p_{T}$ $\gamma$'s, jets, $W$'s and $Z$'s.
 Reliable searches for these signatures require a good understanding of all SM background processes, which is usually done by simulations performed by Monte Carlo (MC) event generators, such as MadGraph~\cite{madgraph5,madgraph4,madevent}. 
 However, the complicated event topology expected for some new physics signals makes their background simulation time consuming, and it is important to increase the computational speed for simulations in the LHC data analysis.

 In  previous studies~\cite{qed-paper,qcd-paper}, the GPU (Graphics Processing Unit) has been used to realize economical and powerful parallel computations of cross sections by introducing a C-language version of the HELAS~\cite{helas} codes, HEGET ({\bf H}ELAS {\bf E}valuation with {\bf G}PU {\bf E}nhanced {\bf T}echnology). \linebreak HEGET is based on the software development system \linebreak CUDA~\cite{cuda} introduced by NVIDIA~\cite{nvidia}. 
 For pure QED processes, $q\overline{q} \!\rightarrow\! n\gamma$, with $n\!=\!2$ to 8, the calculations ran  40-150 times faster on the GPU  than on the CPU~\cite{qed-paper}. 
 For pure QCD processes, $gg\!\rightarrow\! ng$ with $n$ up to 4, $q\overline{q}\!\rightarrow\! ng$ and $qq\!\rightarrow\! qq\!+\!(n-2)g$ with $n$ up to 5, 60-100 times better performance was achieved on the GPU~\cite{qcd-paper}. 
 In this paper, we extend these exploratory studies to cover general SM processes, opening the way to perform the complete matrix element computation of MadGraph on the GPU. 
 The  complexity of the calculations is increased due to new interaction types and complicated event topologies expected in background simulations for various new physics scenarios.
 We introduce additional HEGET functions to cover all of the SM particles and their interactions, and a phase space parameterization suited for GPU computations. 
 In order to test all of the new functions and the efficiency of the GPU computation in semi-realistic background simulations, we systematically study multi-jet processes associated with the production of SM heavy particles(s), followed by its (their) decay into final states including light quarks and leptons, including full spin correlations.
 In particular, we report numerical results on the following processes:
 \begin{subequations}
  \label{eq:process}
  \begin{align}
   & W/Z + n\mbox{-jets}       &(n\leq 4), \label{eq:process-V}  \\
   & WW/WZ/ZZ + n\mbox{-jets}  &(n\leq 3), \label{eq:process-VV}  \\
   & t\bar{t} + n\mbox{-jets}  &(n\leq 3), \label{eq:process-ttbar}  \\
   & HW/HZ + n\mbox{-jets}     &(n\leq 3),  \label{eq:process-HV}  \\ 
   & Ht\bar{t} + n\mbox{-jets} &(n\leq 2),  \label{eq:process-Httbar}  \\
   & H^k + (n\!-\!k)\mbox{-jets}\,\,{\rm via\, WBF} & (k\leq 3,n\leq 5).  \label{eq:process-HWBF}      
  \end{align}
 \end{subequations}
 For the processes (\ref{eq:process-V}) to (\ref{eq:process-Httbar}), we examine all of the major subprocesses at the LHC, while for the multiple Higgs production (\ref{eq:process-HWBF}), we study only the weak-boson fusion (WBF) subprocesses to test the Higgs self interactions.

 We present numerical results for the cross sections of processes in Eqs.~(\ref{eq:process-V})-(\ref{eq:process-HWBF}) computed by using the GPU version of the Monte Carlo integration program, gBASES~\cite{mcinteg}, with the new HEGET functions in the amplitude calculations, and compare the results with those obtained running two different programs on the CPU, the FORTRAN version BASES~\cite{bases} programs with HELAS subroutines and the latest version of MadGraph (ver.5)~\cite{madgraph5}.
 We also compare the performance of two versions of the BASES program, one on the GPU and the other on the CPU.

 The paper is structured as follows. 
 In Sect.~\ref{sec:physics_processes}, we present the cross section formulae for general SM production processes at the LHC, and list all of the subprocesses we study in this report.
 In Sect.~\ref{sec:algorithm}, we briefly describe a new phase space parameterization for efficient GPU computation. 
 In Sect.~\ref{sec:heget}, we introduce new HEGET functions for all SM particles and their interactions.
 In Sect.~\ref{sec:mg2cuda}, we introduce the software used to generate CUDA codes with HEGET functions from FORTRAN amplitude programs with HELAS subroutines obtained by MadGraph.
 In Sect.~\ref{sec:computation}, we review the computing environment, basic parameters of the GPU and CPU machines used in this analysis.
 Section~\ref{sec:results} gives numerical results of computations of cross sections and comparisons of performance of GPU and CPU programs.
 Section~\ref{sec:summary} summarizes our findings. 
 \ref{sec:ps} explains in more detail our phase space parameterization introduced in Sect.~\ref{sec:algorithm}. 
 \ref{sec:RNG} explains our method for generating random numbers on GPU. 
 \ref{sec:new-heget-codes} lists all the new HEGET codes introduced in this paper. 

 \section{Physics processes}
 \label{sec:physics_processes}
 In order to test not only the validity and efficiency of our GPU computation but also its robustness, we examine a series of multi-jet production processes in association with the SM heavy particle(s) ($W$, $Z$, $t$ and $H$), followed by their decays into light quarks and leptons, that can be backgrounds for discoveries in many new physics scenarios. 
 In this section, we list all the subprocesses we study in this paper and give the definition of multi-jet cross sections that are calculated both on the CPU and on the GPU at later sections.

 At the LHC with a collision energy of $\sqrt{s}$, the cross section for general production processes in the SM can be expressed as
 \begin{eqnarray}
  {\rm d\sigma} &=
   &\sum_{\{a,b\}}\!\iint \mathrm{d}x_a \mathrm{d}x_b
   \nonumber\\
  &\times&  D_{a\!/\!p}\left(x_{a},\!Q\right)
   D_{b\!/\!p}\left(x_{b},\!Q\right){\rm d\hat{\sigma}}(\hat{s}=sx_{a}x_{b}),
 \end{eqnarray}
 where $D_{a\!/\!p}$ and $D_{b\!/\!p}$ are the parton distribution functions (PDF's), $Q$ is the factorization scale, $x_{a}$ and $x_{b}$ are the momentum fractions of the partons $a$ and $b$, respectively, in the right- and left-moving protons, $\sqrt{\hat{s}}$ is the subprocess center of mass energy, and ${\rm d\hat{\sigma}}(\hat{s})$ gives the differential cross section for the $2\to n$ subprocess
 \begin{equation}
  {\renewcommand\arraystretch{1.5}
   \begin{array}{rcl}
    \lefteqn{\mathbf{a}(p_{a},\lambda_{a},c_{a}) +
     \mathbf{b}(p_{b},\lambda_{b},c_{b})} \\
    & \rightarrow &
     \mathbf{1}(p_{1},\lambda_{1},c_{1}) + \cdots
     + \mathbf{n} (p_{n},\lambda_{n},c_{n}).
   \end{array}
   }     \label{eq:hprocess}
 \end{equation}

 The subprocesses cross section can be computed in the leading order as
 \begin{eqnarray}
  {\rm d\hat{\sigma}}(\hat{s})
   &=&\dfrac{1}{2\hat{s}}\frac{1}{2\!\cdot\! 2}
   \sum_{\lambda_{i}}
   \dfrac{1}{n_{a}n_{b}}
   \sum_{c_{i}}
   \left|
    {\mathcal{M}}_{\lambda_{i}}
    ^{c_{i}}\right|^{2}
   {\rm d\Phi_{n}}\,,
   \label{eq:phase-space0}
 \end{eqnarray}
 where
 \begin{eqnarray}
  {\rm d\Phi_{n}}
   &=&
   \left(2\pi\right)^4
   \delta^4\left(\!p_{a}\!+\!p_{b}\!-\!\sum_{i=1}^{n}p_i\!\right)
   \prod_{i=1}^{n}
   \dfrac{\mathrm{d}^3 p_i}{(2\pi)^3\,2E_i}\,,
   \label{eq:phase-space}
 \end{eqnarray}
 is the invariant $n$-body phase space, $\lambda_{i}$ are the helicities of the initial and  final partons, $n_{a}$ and $n_{b}$ are the color degree of freedom of the initial partons, $a$ and $b$, respectively. 
 ${\mathcal{M}}_{\lambda_{i}}^{c_{i}}$ are the Helicity amplitudes for the process (\ref{eq:hprocess}), which can be generated automatically by MadGraph and expressed as
 \begin{equation}
  \mathcal{M}_{\lambda_{i}}^{c_{i}}
   = \sum_{l\in{\rm diagram}} \left( M_{\lambda_{i}}\right)_{l}^{c_{i}}
   \label{eq:matrix-element}
 \end{equation}
 where the summation is over all the Feynman diagrams.
 The subscripts $\lambda_{i}$ stand for a given combination of helicities ($0$ for Higgs bosons, $\pm 1/2$ for quarks and leptons, $\pm 1$ for photons and gluons), and the subscripts $c_{i}$ correspond to a set of color indices (none for colorless particles, 1, 2, 3 for flowing-IN quarks, $\overline{1}, \overline{2}, \overline{3}$ for flowing-OUT quarks, and 1 to 8 for gluons). 
 Details on color summation in MadGraph can be found in Ref.~\cite{qcd-paper} .

 In this paper, we are interested in general SM processes, typically involving the production of heavy resonances with decays, associated with multiple jet production, which often appear as major SM background for new physics signals.
The limitation in the number of extra jets, $n$ or $n\!-\!k$ in Eqs.~(\ref{eq:process-V})-(\ref{eq:process-HWBF}), is primarily due to limitations in the amount of memory available to the GPU as reported in previous studies~\cite{qed-paper,qcd-paper}.


Correlated decays of the heavy particles into the following channels are calculated:
 \begin{subequations}\label{eq:decay}
  \begin{align}
   & \Wmp\rightarrow \lmp\,\overset{\scriptscriptstyle (-\!\!-)}{\nul}     &(\ell=e, \mu),     \label{eq:decay-W}  \\
   & Z\rightarrow \lplm    &(\ell=e, \mu),  \label{eq:decay-Z}  \\
   & t\rightarrow b\,\lp\nul    &(\ell=e, \mu),  \label{eq:decay-t}  \\
   & \tbar\rightarrow \bbar\,\lm\nulbar &(\ell=e, \mu),  \label{eq:decay-tbar}  \\
   & H\rightarrow \tauptaum.   & \label{eq:decay-H}      
  \end{align}
 \end{subequations}
 We do not consider $\tau$ decays in this report for brevity.
 The same selection cuts are imposed for $\ell=e, \mu, \tau$, so that the Higgs boson production cross sections listed in this report can be used as a starting point for realistic simulations. 
 A Higgs boson mass of 125\GeV\ and its branching fraction into \tauptaum, 0.0405, are used throughout this report\footnote{It should be noted here that the polarized $\tau$ decay, based on the $\tau$ decay helicity amplitude is available in the framework of MadGraph5~\cite{taudecay}.}.

 For definiteness, we impose the following final state cuts at the parton level.
 For jets, we require the same conditions as in Refs.~\cite{qed-paper,qcd-paper}:
 \begin{subequations}
  \begin{eqnarray}
   |\eta_{i}| & < & \eta_{\mathrm{jet}}^{\scriptstyle\mathrm{cut}} = 5, 
   \label{eq:jet-cuts-eta}\\
   p_{\mathrm{T}i} & > & p_{\mathrm{T, jet}}^{\scriptstyle\mathrm{cut}} = 20\GeV, 
    \label{eq:jet-cuts-pt}\\
   p_{\mathrm{T}ij} & > & p_{\mathrm{T, jet}}^{\scriptstyle\mathrm{cut}} = 20\GeV. 
    \label{eq:jet-cuts-ptjj}
  \end{eqnarray}
  \label{eq:jet-cuts}
 \end{subequations}
 \negthinspace\negthinspace\negthinspace
 Where $\eta_{i}$ and $p_{\mathrm{T}i}$ are the pseudo-rapidity and the transverse momentum of the $i$-th jet, respectively, in the $pp$ collisions rest frame along the right-moving ($p_{\mathrm{z}}\!=\!|p|$) proton momentum direction, and $p_{\mathrm{T}ij}$ is the relative transverse momentum~\cite{durham-jet} between the jets $i$ and $j$ defined by
 \begin{subequations}
  \begin{eqnarray}
   p_{\mathrm{T}ij} & \equiv &
    \min(p_{\mathrm{T}i},p_{\mathrm{T}j})\,\Delta R_{ij}, \\
   \Delta R_{ij}  & = &
    \sqrt{(\Delta\eta_{ij})^{2}+(\Delta\phi_{ij})^{2}}.   
    \label{eq:isolation-cuts2}
  \end{eqnarray}
  \label{eq:isolation-cuts}
 \end{subequations} 
 \negthinspace\negthinspace\negthinspace
 Here $\Delta R_{ij}$ measures the boost-invariant angular separation between the $i$ and $j$ jets, and $\Delta\eta_{ij}$ and $\Delta\phi_{ij}$ are defined as differences of pseudorapidities and azimuthal angles of the two jets.

 For $b$ jets from $t$ decay, we require
 \begin{subequations}
  \begin{eqnarray}
   |\eta_{b}| & < & \eta_{b}^{\scriptstyle\mathrm{cut}} = 2.5, 
    \label{eq:bjet-cuts-eta}\\
   p_{\mathrm{T, }b} & > & p_{\mathrm{T, }b}^{\scriptstyle\mathrm{cut}} =
  20\GeV. \label{eq:bjet-cuts-pt}  
  \end{eqnarray}
  \label{eq:bjet-cuts}
 \end{subequations} 
 \negthinspace\negthinspace\negthinspace
 Note that we do not require them to be isolated from other jets via (\ref{eq:jet-cuts-ptjj}). 
 For charged leptons, we require
 \begin{subequations}
  \begin{eqnarray}
   |\eta_{\ell}| & < & \eta_{\ell}^{\scriptstyle\mathrm{cut}} = 2.5, \,\, \lemutau 
    \label{eq:lepton-cuts-eta}\\
   p_{\mathrm{T, }\ell} & > & p_{\mathrm{T, }\ell}^{\scriptstyle\mathrm{cut}} =
    20\GeV, \,\, \lemutau 
    \label{eq:lepton-cuts-pt}  
  \end{eqnarray}
  \label{eq:lepton-cuts}
 \end{subequations} 
 \negthinspace\negthinspace\negthinspace
 and for simplicity we treat $\tau$ the same as $e$ and $\mu$. 
 Like $b$ jets from $t$ decays, we do not impose isolation cuts for leptons, since performing realistic simulations is not the purpose of this study. 
 
 We use the set CTEQ6L1~\cite{cteq6} parton distribution functions (PDF) for all processes.

  \subsection{Single $W$ production}
  \label{sb:W}

  The following four types of $W$ production subprocesses are studied in this paper:
  \begin{subequations}
   \begin{align}
    u\bar{d} &\to \Wp + ng   &(n=0,1,2,3,4),      \label{eq1:W}  \\
    ug &\to \Wp + d + (n\!-\!1)g &(n=1,2,3,4),        \label{eq2:W}  \\
    uu &\to \Wp + ud + (n\!-\!2)g &(n=2,3,4),         \label{eq3:W}  \\
    gg &\to \Wp + d\bar{u} + (n\!-\!2)g &(n=2,3,4).   \label{eq4:W}  
   \end{align}
   \label{eq:W}
  \end{subequations}
  \negthinspace\negthinspace\negthinspace
  The subprocess (\ref{eq1:W}) starts with the leading order $\alphaszero$, the subprocess (\ref{eq2:W}) starts with the next order $\alphasone$, and those of (\ref{eq3:W}) and (\ref{eq4:W}) start with the $\alphastwo$ order.
  These subprocesses give the dominant contributions in $pp$ collisions.
  The corresponding \Wm\ production cross sections are smaller in $pp$ collisions since an incoming $u$-quark in the subprocesses (\ref{eq1:W}) to (\ref{eq3:W}) should be replaced by a $d$-quark, whose PDF is significantly softer than the up-quark PDF in the proton.

  \subsection{Single $Z$ production}
  \label{sb:Z}
  Similarly, the following subprocesses are studied for $Z$ production:
  \begin{subequations}
   \begin{align}
    u\bar{u} &\to Z + ng    &(n=0,1,2,3,4),        \label{eq1:Z} \\
    ug     &\to Z + u + (n\!-\!1)g  &(n=1,2,3,4),  \label{eq2:Z}        \\
    uu     &\to Z + uu + (n\!-\!2)g &(n=2,3,4),     \label{eq3:Z}      \\
    gg     &\to Z + u\bar{u} + (n\!-\!2)g &(n=2,3,4). \label{eq4:Z}
   \end{align}
   \label{eq:Z}
  \end{subequations}
  As in the case of $W^{+}\!$+$n$-jets, we examine all of the dominant contributions up to 4-jets.
  It should be noted that the down quark contribution to the $Z$+jets cross section is less suppressed than the $W^{-}\!$+jets case, since the $Z$-boson couples stronger to the down-quarks than the up-quarks (cf. $\Gamma(Z\rightarrow d\bar{d})/\Gamma(Z\rightarrow u\bar{u})\approx 1.3$).

  \subsection{$WW$ production}
  \label{sb:WW}
  For the $W$ boson pair production, we study the following subprocesses:
  \begin{subequations}
   \begin{align}
    u\bar{u} &\to \WpWm + ng    &(n=0,1,2,3),         \label{eq1:WW} \\
    ug     &\to \WpWm + u + (n\!-\!1)g  &(n=1,2,3),         \label{eq2:WW} \\
    uu     &\to \WpWm + uu + (n\!-\!2)g &(n=2,3),          \label{eq3:WW} \\
    uu     &\to \WpWp + dd + (n\!-\!2)g &(n=2,3),          \label{eq4:WW} \\
    gg     &\to \WpWm + u\bar{u} + (n\!-\!2)g &(n=2,3).    \label{eq5:WW}
   \end{align}
   \label{eq:WW}
  \end{subequations}
  \negthinspace\negthinspace\negthinspace
  The subprocess (\ref{eq1:WW}) starts with the leading order \alphaszero, the subprocess (\ref{eq2:WW}) starts with the next order \alphasone, and those of (\ref{eq3:WW}) to (\ref{eq5:WW}) start with the \alphastwo\ order. The subprocesses (\ref{eq4:WW}) are included as the dominant same sign $W$-pair production mechanism in $pp$ collisions.   

  \subsection{\WpZ\ production}
  \label{sb:WZ}
  Similarly, the following subprocesses are studied for \WpZ\ production: 
  \begin{subequations}
   \begin{align}
    u\bar{d} & \to \WpZ + ng     &(n=0,1,2,3),         \label{eq1:WZ}   \\
    ug       & \to \WpZ + d + (n\!-\!1)g   &(n=1,2,3), \label{eq2:WZ}         \\
    uu       & \to \WpZ + ud + (n\!-\!2)g  &(n=2,3),    \label{eq3:WZ}    \\
    gg       & \to \WpZ + d\bar{u} + (n\!-\!2)g &(n=2,3). \label{eq4:WZ}
   \end{align}
   \label{eq:WZ}
  \end{subequations}
  As in the $WW\!$+$n$-jets case, we consider all of the dominant $W^{+}Z$ production subprocesses up to 3 associated jets.
  Note again that the down quark contribution to the $W^{-}Z$+jets cross section can be significant because of the large $Z$ coupling to the $d$-quarks.

  \subsection{$ZZ$ production}
  \label{sb:ZZ}
  The following $ZZ$ production subprocesses are also studied:
  \begin{subequations}
   \begin{align}
    u\bar{u} &\to ZZ + ng    &(n=0,1,2,3), \label{eq1:ZZ} \\
    ug     &\to ZZ + u + (n\!-\!1)g  &(n=1,2,3), \label{eq2:ZZ} \\
    uu     &\to ZZ + uu + (n\!-\!2)g &(n=2,3), \label{eq3:ZZ}  \\
    gg     &\to ZZ + u\bar{u} + (n\!-\!2)g &(n=2,3) \label{eq4:ZZ} .
   \end{align}
   \label{eq:ZZ}
  \end{subequations}
  All of the dominant $ZZ$ production subprocesses up to 3 associated jets are studied.
  Note, however, that the down-quark contribution to the $qg$ collision subprocess (\ref{eq2:ZZ}) has the coupling enhancement factor of $(\Gamma(Z\rightarrow d\bar{d})/\Gamma(Z\rightarrow u\bar{u}))^{2}\approx 1.7$.
  Although we study only 4 lepton final states, $ZZ$+jets processes can be backgrounds for new physics signals with a $Z$ boson plus jets and large missing \ET.

  \subsection{$\ttbar$  production}
  \label{sb:ttbar}
  For $t\bar{t}$ production, we consider the following subprocesses: 
  \begin{subequations}
   \begin{align}
    u\bar{u} &\to t\bar{t} + ng      &(n=0,1,2,3),  \label{eq1:ttbar} \\
    ug       &\to t\bar{t} + u + (n\!-\!1)g  &(n=1,2,3),    \label{eq2:ttbar} \\
    uu       &\to t\bar{t} + uu + (n\!-\!2)g &(n=2,3),      \label{eq3:ttbar} \\
    gg       &\to t\bar{t} + ng      &(n=0,1,2,3).  \label{eq4:ttbar}
   \end{align}
   \label{eq:ttbar}
  \end{subequations}
  \negthinspace\negthinspace\negthinspace
  The subprocess (\ref{eq1:ttbar}) starts with the leading order \alphastwo, the subprocess (\ref{eq2:ttbar}) starts with the next order \alphasthree, and those of (\ref{eq3:ttbar}) and (\ref{eq4:ttbar}) start with the \alphasfour\ order.
  Again, only those subprocesses which give dominant contributions in $pp$ collisions are studied in each order. 

  \subsection{$W$ boson associated Higgs production}
  \label{sb:WH}
  As for the associate production of the Higgs boson and the $W$, we consider the following subprocesses:
  \begin{subequations}
   \begin{align}
    u\bar{d} &\to H\Wp + ng      &(n=0,1,2,3),   \label{eq1:WH} \\
    ug       &\to H\Wp + d + (n\!-\!1)g  &(n=1,2,3),     \label{eq2:WH} \\
    uu       &\to H\Wp + ud + (n\!-\!2)g &(n=2,3),       \label{eq3:WH} \\
    gg       &\to H\Wp + d\bar{u} + (n\!-\!2)g &(n=2,3). \label{eq4:WH}
   \end{align}
   \label{eq:WH}
  \end{subequations}
  \negthinspace\negthinspace\negthinspace
  All of the subprocesses (\ref{eq1:WH}) to (\ref{eq4:WH}) are obtained by replacing one gluon in the $W+$jets subprocesses (\ref{eq1:W}) to (\ref{eq4:W}) by $H$, respectively.  
  The $HW^{-}$+jets subprocesses corresponding to (\ref{eq1:WH}) to (\ref{eq3:WH}) are suppressed since an incoming $u$-quark would be replaced by a softer $d$-quark in the proton.

  \subsection{$Z$ boson associated Higgs production}
  \label{sb:ZH}
  Likewise, the following subprocesses are studied for $HZ$ production:
  \begin{subequations}
   \begin{align}
    u\bar{u} &\to HZ + ng  &(n=0,1,2,3), \label{eq1:HZ} \\
    ug     &\to HZ + u + (n\!-\!1)g &(n=1,2,3), \label{eq2:HZ} \\
    uu     &\to HZ + uu + (n\!-\!2)g &(n=2,3), \label{eq3:HZ} \\
    gg     &\to HZ + u\bar{u} + (n\!-\!2)g &(n=2,3). \label{eq4:HZ} 
   \end{align}
   \label{eq:ZH}
  \end{subequations}
  All of the four subprocesses are obtained from the corresponding $Z$+jets subprocesses (\ref{eq1:Z}) to (\ref{eq4:Z}), by replacing one final gluon by $H$.
  Again, the down quark contributions to the subprocesses (\ref{eq1:HZ}) to (\ref{eq3:HZ}) are less suppressed than those to the $HW^{+}$ production processes (\ref{eq1:HW}) to (\ref{eq3:HW}) due to the large $Z$ coupling to the $d$-quark.

  \subsection{Top quark associated Higgs production}
  \label{sb:tH}
  For the $Ht\bar{t}$ production process, the following subprocesses are examined in this paper:
  \begin{subequations}
  \begin{align}
   u\bar{u} &\to Ht\bar{t} + ng     &(n=0,1,2), \label{eq1:tH} \\
   ug       &\to Ht\bar{t} + u + (n\!-\!1)g &(n=1,2),   \label{eq2:tH} \\
   uu       &\to Ht\bar{t} + uu,                \label{eq3:tH} \\
   gg       &\to Ht\bar{t} + ng    &(n=0,1,2). \label{eq4:tH}
  \end{align}
  \label{eq:tH}
  \end{subequations}
  \negthinspace\negthinspace\negthinspace
  All of the subprocesses (\ref{eq1:tH}) to (\ref{eq4:tH}) are obtained, respectively, from the $t\bar{t}+$jets subprocesses (\ref{eq1:ttbar}) to (\ref{eq4:ttbar}) by replacing one gluon in the final state by a Higgs boson.

  \subsection{Higgs boson production via weak-boson fusion}
  \label{sb:HW}
  In all of the above subprocesses we consider only QCD interactions for production of jets and top quarks, while the weak interactions contribute only in $W$, $Z$, $H$ productions and in decays.
  For Higgs$+$jets processes, we study weak-boson fusion (WBF) subprocesses which can be identified at the LHC for various decay modes~\cite{awbf,tauwbf,wwbf}. 
  $WW$ fusion contributes to the subprocesses (\ref{eq1:HW}), $ZZ$ fusion contributes to the subprocesses (\ref{eq2:HW}), both contribute to the subprocesses (\ref{eq3:HW}) and (\ref{eq4:HW}):
  \begin{subequations}
 \begin{align}
  ud &\to H + ud + (n\!-\!2)g &(n=2,3,4),   \label{eq1:HW} \\
  uu &\to H + uu + (n\!-\!2)g &(n=2,3,4),   \label{eq2:HW}     \\
  ug &\to H + ud+ \bar{d} + (n\!-\!3)g &(n=3,4),  \label{eq3:HW}\\
  gg &\to H + u\bar{u} + d\bar{d}.        \label{eq4:HW}
 \end{align}
 \label{eq:HW}
  \end{subequations}
  \negthinspace\negthinspace\negthinspace
  The subprocesses (\ref{eq1:HW}) and (\ref{eq2:HW}) start with the leading order \alphaszero, those in (\ref{eq3:HW}) start with the next order \alphasone, the subprocess (\ref{eq4:HW}) occurs at the \alphastwo\ order. 
  We do not consider $gg$ fusion process, because its simulation requires non-renormalizable vertices~\cite{madgraph5} which are absent in the minimal set of HELAS codes~\cite{helas}.

  \subsection{Multiple Higgs bosons production via weak-boson fusion}
  \label{sb:HH}
  The following multiple Higgs boson production processes are studied in this paper as a test of cubic and quartic vertex functions among scalar and vector bosons:
  \begin{subequations}
   \begin{align}
    ud &\to HH + ud + (n\!-\!2)g &(n=2,3),    \label{eq1:HH}  \\
    uu &\to HH + uu + (n\!-\!2)g &(n=2,3),    \label{eq2:HH}  \\
    ud &\to HHH + ud, &               \label{eq3:HH}  \\
    uu &\to HHH + uu. &               \label{eq4:HH}  
   \end{align}
   \label{eq:HH}
  \end{subequations}
  \negthinspace\negthinspace\negthinspace
  The quartic scalar boson vertex appears only in the subprocesses (\ref{eq3:HH}) and (\ref{eq4:HH}). 

 \section{Algorithm for phase space generation}
 \label{sec:algorithm}
 In this section, we briefly introduce our phase space parame-trization, designed for efficient GPU computing. Details are given in \ref{sec:ps}.

Optimizing code to run efficiently on the GPU requires several special considerations.
In addition to the careful use of memory mentioned earlier, one needs to consider that each ``batch'' of calculations processed on the GPU must undergo identical operations.
That has particular consequences when one considers generating momentum in phase space that satisfy the appropriate cuts.
The efficiency of generating momenta that pass cuts is not particularly important on the CPU, since one can simply repeatedly generate momenta until a set is found that pass the cuts.
The structure of the GPU however does not allow such flexibility.
If the generation program is running on multiple processors, each processor has one opportunity to generate a valid phase space point before moving forward to calculate the amplitude.
So if the efficiency for generating a point in phase space that passes cuts is only 10\% then you loose a factor of 10 in computing speed.

 In previous studies~\cite{qed-paper,qcd-paper} for pure QED and QCD processes at hadron colliders:
 \begin{align}
  a+b\rightarrow 1+\cdots+n, \label{process}
 \end{align}
 the following phase space parameterization has been used including the integration over the initial parton momentum fractions (see \ref{sec:ps} for details):
 \begin{align}
  d\Sigma_n&\equiv dx_adx_bd\Phi_n. \nonumber \\
  &=\frac{\Theta(1-x_a)\Theta(1-x_b)}{2s(4\pi)^{2n-3}}
  dy_n\prod_{i=1}^{n-1}\bigg(dy_idp^2_{Ti}\frac{d\phi_i}{2\pi}\bigg).
  \label{ps02}
 \end{align}
 With this parameterization, when all the final particles are massless partons, we can generate phase space points that satisfy the rapidity constraints, $|\etai|\!<\!\etacut$, for all partons ($i\!=\!1$ to $n$) and the \pT\ constraints, $\pTi\!>\!\pTcut$, for $n\!-\!1$ partons ($i\!=\!1$ to $n\!-\!1$). 
 Only those phase space points which violate the conditions, $x_a\!<\!1$, $x_b\!<\!1$ or $p_{\mathrm{T}n}\!>\!\pTcut$, will be rejected.
 Studies in Refs.~\cite{qed-paper,qcd-paper} find that, for example, 78\%,  58\%, 42\% and 29\% of generated phase space points satisfy the final state cuts\footnote{Note this can easily be improved e.g. by introducing ordering in $p_{\mathrm{T}}$, $\pTcut<p_{\mathrm{T}1}<p_{\mathrm{T}2}<\cdots<p_{\mathrm{T}n}$, for $n$-photons or $n$-gluons. It is then only $p_{\mathrm{T}n}$ which can go up to $\sqrts/2$.}, for 2-5 photon productions at the LHC, when we choose $y_i$ ($i\!=\!1$ to $n$), $\ln p^2_{Ti}$ and $\phi_i$ ($i\!=\!1$ to $n\!-\!1$) as integration variables.

 We adopt this parameterization of the generalized phase space in this report, which is extended to account for the production and decay of several Breit-Wigner resonances. In particular, for process with $n$-jet plus $m$-resonance production
 \begin{align}
  pp\rightarrow (n-m)j+\sum_{k=1}^m R_k (\to f_k) + \mbox{\rm anything},
  \label{ps1-resonance}
 \end{align}
 when the resonance $R_k$ decays into a final state $f_k$, we parameterize the generalized phase space as follows:
 \begin{align}
  d\Sigma=
  d\Sigma_n\prod_{k=1}^m \frac{ds_k}{2\pi}d\Phi(R_k\to f_k \,\, {\rm at}\,\, p^2_{f_k}=s_k)\,.
  \label{ps2-resonance}
 \end{align}
 Here $d\Phi(R_k\to f_k)$ is the invariant phase space for the $R_k\to f_k$ decay when the invariant mass of the final state $f_k$ is $\sqrt{s_k}$, and $d\Sigma_n$ is the $n$-body generalized phase space for  $n-m$ massless particles and $m$ massive particles of masses $\sqrt{s_k}$ ($k$ = 1 to m).
 Integration over the invariant mass squared $s_k$ is made efficient by using the standard Breit-Wigner formalism
 \begin{align}
  ds_k = \frac{(s_k-m_k^2)^2+(m_k\Gamma_k)^2}{m_k\Gamma_k}\:
  d\tan^{-1}\bigg(\frac{s_k-m_k^2}{m_k\Gamma_k}\bigg)\,,
  \label{ps4-resonance}
 \end{align}
 and the transverse momenta are generated as
 \begin{align}
  dp^2_{Tk}=(p^2_{Tk}+s_k)\, d\ln(p^2_{Tk}+s_k)\,.
  \label{ps5-resonance}
 \end{align}

 Finally, we note that s-channel splitting of massless partons can be accommodated by using the same parameterization Eq.~(\ref{ps2-resonance}), where $R_k$ in Eq.~(\ref{ps2-resonance}) is a virtual parton of mass $\sqrt{s_k}$ and $f_k$ is a set of partons.
 Instead of the Breit-Wigner parameterization ~(\ref{ps4-resonance}), we simply use $\ln s_k$ as integration variables.


\section{HEGET functions}
\label{sec:heget}

 In this section, we explain the new HEGET functions for computing the helicity amplitudes for arbitrary SM processes.
 All of the HEGET functions that appear in this report are listed in \ref{sec:new-heget-codes} as List~\ref{list:ixxxxx} to List~\ref{list:hsssxx}.

  \subsection{Wave functions}
  \label{sec:WaveFunc}
  In the previous works~\cite{qed-paper,qcd-paper}, the wave functions for massless particles, which are named \texttt{ixxxx}$k$, \texttt{oxxxx}$k$ and \texttt{vxxxx}$k$ ($k=0,1,2$), have been introduced for fermions and vector bosons. 
  In this report, we introduce wave functions for massive spin 0, 1/2 and 1 particles, which can also be used for massless particles by setting the mass parameters to zero.
  The naming scheme for HEGET functions follows that of HELAS subroutines: the HEGET (HELAS) function names start with \texttt{i(I)} and \texttt{o(O)} for flow-IN and flow-OUT fermions, respectively, \texttt{v(V)} for vector boson wave functions, and \texttt{s(S)} for scalar boson wave functions.
  All of the HEGET functions for external lines are summarized in Table~\ref{tab:WAVEFC}.
  \begin{table}[tbh]
   \centering
    \caption{HEGET functions for external lines}
    \label{tab:WAVEFC}
    \begin{tabular}{ccc}  \hline
     External Line       & HEGET  & HELAS  \\ \hline
     Flowing-In  Fermion & \texttt{ixxxxx} & \texttt{IXXXXX} \\ 
     Flowing-Out Fermion & \texttt{oxxxxx} & \texttt{OXXXXX} \\ 
     Vector Boson        & \texttt{vxxxxx} & \texttt{VXXXXX} \\ 
     Scalar Boson        & \texttt{sxxxxx} & \texttt{SXXXXX} \\ \hline
    \end{tabular}
  \end{table}

  The spin 1/2 fermion wave functions with flowing-IN fermion number, \texttt{ixxxxx}, and flowing-OUT fermion number, \texttt{oxxxxx}, are listed in Appendix~\ref{sec:ioxxxxx}.
  The first 4 components of the output complex array, \texttt{fi[6]} and \texttt{fo[6]} of these functions are 4-spinors
  \begin{subequations}
   \begin{eqnarray}
    \texttt{|fi>} & = & u(\texttt{p},\texttt{nHEL}/2)\mbox{ for
     }\texttt{nSF} = +1 \\
    & =  & v(\texttt{p},\texttt{nHEL}/2)\mbox{ for }\texttt{nSF} = -1
   \end{eqnarray}
  \end{subequations}
  and
  \begin{subequations}
   \begin{eqnarray}
    \texttt{<fo|} & = & \overline{u}(\texttt{p},\texttt{nHEL}/2)\mbox{
     for }\texttt{nSF} = +1 \\
    & = &\overline{v}(\texttt{p},\texttt{nHEL}/2)\mbox{ for
     }\texttt{nSF} = -1
   \end{eqnarray}
  \end{subequations}
  respectively, just the same as in HELAS subroutines, where $\mathtt{nSF}\!=\!+1$ for particles, and $\mathtt{nSF}\!=\!-1$ for anti-particles.
  The helicities are given by $\mathtt{nHEL}/2\!=\!\pm 1/2$.
  The last 2 components\footnote{%
  In C language, a 6-dimensional array {\sf a[6]}
has 6 components {\sf a[0]} to {\sf a[5]}. Hence the first
four components are {\sf a[0]} to {\sf a[3]}, and the last two components
are {\sf a[4]} and {\sf a[5]}.} of the \texttt{fi[6]} and \texttt{fo[6]} give
  the 4-momentum of the fermions along the fermion-number flow,
  \begin{eqnarray}
   \texttt{fi[4]}=\texttt{fo[4]}&=&
    \texttt{nSF}\left(p_0 + i p_3\right)\nonumber\\
   \texttt{fi[5]}=\texttt{fo[5]}&=&
    \texttt{nSF}\left(p_1 + i p_2\right)\,.
  \end{eqnarray}

  Like the fermion wave functions, the first four components of the output complex array \texttt{vc[6]} in the HEGET function \texttt{vxxxxx} gives the wave function
  \begin{subequations}
   \begin{eqnarray}
    \texttt{(vc)}
     & = & \epsilon^\mu(\texttt{p},
      \texttt{nHEL})^{*}\mbox{ for }\texttt{nSV} = +1\,,  \\
     & = & \epsilon^\mu(\texttt{p},
      \texttt{nHEL})    \, \mbox{ for }\texttt{nSV} = -1\,,
   \end{eqnarray}
  \end{subequations}
  where $\texttt{nSV}\!=\! +1$ for final states, and $\texttt{nSV}\!=\! -1$ for initial states.  
  The helicity \texttt{nHEL} takes $\pm 1$ and 0 for vector bosons, while $\mathtt{nHEL}\!=\!\pm 1$ for massless vector bosons.
  The last 2 components of \texttt{vc[6]} give the flowing-out 4-momentum,
  \begin{eqnarray}
   \left(\texttt{vc[4]}, \texttt{vc[5]}\right) = \texttt{nSV}
    \left( p_0 + i p_3, p_1 + i p_2  \right)\,.
  \end{eqnarray}
  The vector boson wave function \texttt{vxxxxx} is listed in Appendix~\ref{sec:vxxxxx}.

  The function \texttt{sxxxxx} computes the wave function of a scalar boson, and is listed in Appendix \ref{sec:sxxxxx}.
  Since the scalar boson does not have any Lorentz indices, the first component of the output \texttt{sc[3]} is simply unity, $1+i0$.
  The last 2 components give the flowing-out 4-momentum:
  \begin{eqnarray}
   \left(\texttt{sc[1]}, \texttt{sc[2]}\right) = \texttt{nSS}
    \left( p_0 + i p_3, p_1 + i p_2  \right)\,,
  \end{eqnarray}
  where $\texttt{nSS}=+1(-1)$ when the scalar boson is in the final (initial) state.

  \subsection{Vertex functions}
  There are 8 types of renormalizable vertices among spin 0, 1/2 and 1 particles in the SM, as listed in Table~\ref{tab:HEGETFUC}.
  Here the capital letters \texttt{F}, \texttt{V} and \texttt{S} stand for spin 1/2, 1 and 0 particles respectively.

  As in the HELAS subroutines, we introduce two types of HEGET functions for each vertex, those which give an amplitude (a complex number) as output and those which give an off-shell wave function as output. 
  HEGET function names for off-shell wave functions of a fermion start with \texttt{f}, those for a vector boson start with \texttt{j}, and those for a scalar boson start with \texttt{h}.  
  The corresponding HELAS subroutine names are also shown in Table~\ref{tab:HEGETFUC}.

  \begin{table}[tbh]
   \centering
   \caption{HEGET functions for vertices}
   \label{tab:HEGETFUC}
   \begin{tabular}{ccccc} \hline
    Vertex & Inputs & Output & HEGET & HELAS \\
    \hline
    FFV
    & FFV & amplitude & \texttt{iovxxx} & \texttt{IOVXXX} \\
    & FF  & V         & \texttt{jioxxx} & \texttt{JIOXXX} \\
    & FV  & F         & \texttt{fvixxx} & \texttt{FVIXXX} \\
    &     &           & \texttt{fvoxxx} & \texttt{FVOXXX} \\
    \\
    FFS
    & FFS & amplitude & \texttt{iosxxx} & \texttt{IOSXXX} \\
    & FF  & S         & \texttt{hioxxx} & \texttt{HIOXXX} \\
    & FS  & F         & \texttt{fsixxx} & \texttt{FSIXXX} \\
    &     & F         & \texttt{fsoxxx} & \texttt{FSOXXX} \\ 
    \\
    VVV
    & VVV  & amplitude & \texttt{vvvxxx} & \texttt{VVVXXX} \\
    & VV   & V         & \texttt{jvvxxx} & \texttt{JVVXXX} \\ 
    \\
    VVVV
    & VVVV & amplitude & \texttt{wwwwxx} & \texttt{WWWWXX} \\
    &      &           &                 & \texttt{W3W3XX} \\
    &      &           & \texttt{ggggxx} & \\
    & VVV  & V         & \texttt{jwwwxx} & \texttt{JWWWXX}  \\
    &      &           &                 & \texttt{JW3WXX}  \\
    &      &           & \texttt{jgggxx} & \\ 
    \\\
    VVS
    & VVS & amplitude & \texttt{vvsxxx} & \texttt{VVSXXX} \\
    & VS  & V         & \texttt{jvsxxx} & \texttt{JVSXXX} \\
    & VV  & S         & \texttt{hvvxxx} & \texttt{HVVXXX} \\  
    \\
    SSS
    & SSS & amplitude & \texttt{sssxxx} & \texttt{SSSXXX} \\
    & SS  & S         & \texttt{hssxxx} & \texttt{HSSXXX} \\ 
    \\
    VVSS
    & VVSS & amplitude & \texttt{vvssxx} & \texttt{VVSSXX} \\
    & VSS  & V         & \texttt{jvssxx} & \texttt{JVSSXX} \\
    & VVS  & S         & \texttt{hvvsxx} & \texttt{HVVSXX} \\  
    \\
    SSSS
    & SSSS & amplitude & \texttt{ssssxx} & \texttt{SSSSXX} \\
    & SSS  & S         & \texttt{hsssxx} & \texttt{HSSSXX} \\ \hline
   \end{tabular}
  \end{table}
  
  \subsubsection{FFV: fermion-fermion-vector vertex}
  
  The HEGET functions for the \texttt{FFV} vertex are defined by the Lagrangian,
  \begin{equation}
   \mathcal{L}_{\mathrm{FFV}}
    \!= \overline{\psi}_{\mathrm{F}_{1}} \gamma^\mu
    \left[ \texttt{gc[0]} \frac{1\!-\!\gamma_{5}}{2}
    \!+\! \texttt{gc[1]} \frac{1\!+\!\gamma_{5}}{2} \right]
    \psi_{\mathrm{F}_{2}} V_{\mu}^{*}
  \label{HEGET-FFV}
  \end{equation}
  following the HELAS convention~\cite{helas}, where the boson names are defined by their flowing-out (final state) quantum number. 
  For instance, if ${\mathrm{F}_{1}}$ = up and ${\mathrm{F}_{2}}$ = down in Eq.~(\ref{HEGET-FFV}), then V should be \Wm\ ($V^{*}$ in Eq.~(\ref{HEGET-FFV}) is \Wp).  
  The amplitude function \texttt{iovxxx}, the off-shell fermion wave functions \texttt{fvixxx} and \texttt{fvoxxx}, and the off-shell vector current \texttt{jioxxx} obtained from the \texttt{FFV} Lagrangian~(\ref{HEGET-FFV}) are shown in Appendices~\ref{sec:iovxxx}, \ref{sec:fvixxx}, \ref{sec:fvoxxx} and \ref{sec:jioxxx}, respectively.

  For the $qqg$ vertex of QCD, we adopt the Lagrangian
  \begin{equation}
   \mathcal{L}_{qqg} =
    -g_s T^a_{i\bar{j}} A^{a}_{\mu}
    \bar{q}_{\bar i} \gamma^{\mu} q_j\,,
  \end{equation}
  where $g_s (=\!\sqrt{4\pi\alphas})$ is the strong coupling constant and $T^a_{i\bar{j}}$ ($a\!=\!1$ to 8) are the SU(3) generators in the fundamental representation, $i$ and $\bar{j}$ are the indices of the 3 and $\bar{3}$ representations, respectively.
  When calculating the amplitude of the $qqg$ vertex, we set
  \begin{equation}
   \texttt{gc[0]} = \texttt{gc[1]} = g_s
  \end{equation}
  for the couplings of Eq.~(\ref{HEGET-FFV}) in HEGET functions. 
  Correspondingly, the color amplitude should read
  \begin{equation}
   -T^a_{i \bar{j}} \times (\mathtt{vertex}),
  \end{equation}
  where $\mathtt{(vertex)}$ represents the output of \texttt{iovxxx} or that of any other off-shell wave functions for the \texttt{FFV} vertex.
  The color factor of $-T^a_{i\bar{j}}$ is then processed algebraically in integrand functions of the BASES program.

   \subsubsection{FFS: fermion-fermion-scalar vertex}

   The \texttt{FFS} vertex of the HEGET functions are defined by the Lagrangian
   \begin{equation}
    \mathcal{L}_{\mathrm{FFS}}
     \!= \overline{\psi}_{\mathrm{F}_{1}}
     \left[ \texttt{gc[0]} \frac{1\!-\!\gamma_{5}}{2}
      \!+\! \texttt{gc[1]} \frac{1\!+\!\gamma_{5}}{2} \right]
     \psi_{\mathrm{F}_{2}} S^{*}\,,
   \end{equation}
following the HELAS convention~\cite{helas}. The amplitude function \texttt{iosxxx}, the off-shell fermion wave functions \texttt{fsixxx} and \texttt{fsoxxx}, and the off-shell scalar current \texttt{hioxxx} are shown in Appendices \ref{sec:iosxxx}, \ref{sec:fsixxx}, \ref{sec:fsoxxx} and \ref{sec:hioxxx}, respectively.

   \subsubsection{VVV: three vector vertex}

   The HELAS subroutine for the \texttt{VVV} vertex functions  are defined by the following Yang-Mills Lagrangian
   \begin{eqnarray}
    \mathcal{L}_{\mathrm{VVV}}
     \!= -i\,\mathtt{gc}
     &&\left\{
      \left(\partial_\mu V^{\ast}_{1\nu}\right)
      \left(V^{\mu \ast}_{2}V^{\nu \ast}_{3}
      -V^{\nu \ast}_{2}V^{\mu \ast}_{3}\right) \nonumber \right.\\
   &&+
      \left(\partial_\mu V^{\ast}_{2\nu}\right)
      \left(V^{\mu \ast}_{3}V^{\nu \ast}_{1}
      -V^{\nu \ast}_{3}V^{\mu \ast}_{1}\right) \nonumber \\
   &&\left.\!
      +
      \left(\partial_\mu V^{\ast}_{3\nu}\right)
      \left(V^{\mu \ast}_{1}V^{\nu \ast}_{2}
      -V^{\nu \ast}_{1}V^{\mu \ast}_{2}\right)
     \right\}\,,
   \label{eq:HELAS-VVV}
   \end{eqnarray}
   with a real coupling $\texttt{gc}$, where the vector boson triple products are  anti-Hermitian.
  The amplitude of the \texttt{VVV} vertex is calculated by the HEGET function \texttt{vvvxxx} and the off-shell vector current is computed by \texttt{jvvxxx}, which are listed Appendices \ref{sec:vvvxxx} and \ref{sec:jvvxxx}, respectively. 

  For the electroweak gauge bosons, the coupling \texttt{gc} is chosen as
  \begin{eqnarray}
    \texttt{gc}
     &=&  \gwwz = \gw\cos\thetaw \quad\quad\, {\rm for \, WWZ},\nonumber\\
     &=&  \gwwa = \gw\sin\thetaw =e \,\,\, {\rm for \, WWA}.
  \end{eqnarray}
  Here the three vector bosons $(V_1,\, V_2, \, V_3)$ should be chosen as the cyclic permutations of $(\Wm,\, \Wp,\, Z/A)$, because of the HELAS convention which uses the flowing out quantum number for boson names: see Table~\ref{tab:vvvxxx} in Appendix \ref{sec:vvv-vertex} for details.
  Note $(V_{1\mu})^{*}=V_{2\mu}$ in the HELAS Lagrangian (\ref{eq:HELAS-VVV})

  The $ggg$ vertex in the QCD Lagrangian can be expressed as
  \begin{eqnarray}
   \mathcal{L}_{ggg} &=& g_s f^{abc}
    \left(\partial^\mu A^{a\nu}\right)
    \left(A_\mu^b A_\nu^c\right)\nonumber\\
   &=& (if^{abc})(-ig_s)\left(\partial^\mu A^{a\nu}\right)\left(A_\mu^b A_\nu^c\right)\,,
  \end{eqnarray}
  where now the vector boson triple products are Hermitian; $(A_{\mu}^{a})^{*}=A_{\mu}^{a}$.
  We can still use the same vertex functions \texttt{vvvxxx} and \texttt{jvvxxx} with the real coupling 
  \begin{equation}
   \mathtt{gc} = g_{s} \mbox{ for $ggg$}
  \end{equation}
  and by denoting the corresponding amplitude as
  \begin{equation}
   i f^{abc}\times\left( \mathtt{vertex}\right)\,,
  \end{equation}
  where (\texttt{vertex}) gives the output of the HEGET functions \texttt{vvvxxx} or \texttt{jvvxxx}. 
  The associated factor of $i f^{abc}$ is then treated as the color factor in MadGraph~\cite{madgraph4}.

  \subsubsection{VVVV: four vector vertex}

  There are two types of \texttt{VVVV} vertex in the SM Lagrangian, one is for SU(2) and the other for SU(3).
  
  As in the case for the \texttt{VVV} vertex, the SU(2) vector bosons are expressed in terms of 
  \begin{subequations}
   \begin{eqnarray}
    W_\mu^{\pm} & = & \frac{1}{\sqrt{2}} (W^{1}_{\mu} \mp i\, W^{2}_{\mu}) \\
    W^3_\mu & = & \cos\thetaw Z_\mu +\sin\thetaw A_\mu\,. 
  \end{eqnarray}
  \end{subequations}
  The contact four-point vector boson vertex for the SU(2) weak interaction is given by the Lagrangian
  \begin{eqnarray}
   \mathcal{L}_{\mathrm{W\!W\!W\!W}}
    \! &=& -\dfrac{\gwsq}{4}\epsilon_{ijm}\epsilon_{klm}
    W^i_\mu W^j_\nu W^{k\mu}W^{l\nu}
    \nonumber\\
   &=& -\,\frac{\gwsq}{2} \bigl\{(W^{-*}\!\cdot W^{+*})(W^{+*}\!\cdot W^{-*})
    \nonumber \\
   && \quad\ \ 
    -\ \>(W^{-*}\!\cdot W^{-*})(W^{+*}\!\cdot W^{+*})\bigr\} \nonumber \\ 
   && +\,\gwsq  \bigl\{(W^{-*}\!\cdot W^{3*})(W^{3*}\!\cdot W^{+*})
    \nonumber \\
   && \quad\ 
    -\ \>(W^{-*}\!\cdot W^{+*})(W^{3*}\!\cdot W^{3*})\bigr\}.
    \label{x4v}
  \end{eqnarray}
  Two distinct subroutines \texttt{wwwwxx} and \texttt{w3w3xx} (see Table~\ref{tab:HEGETFUC}) were introduced in HELAS, because there was an attempt to improve their numerical accuracy by combining the weak boson exchange contribution to the four-point vertices~\cite{helas}. 
  The attempt has not been successful and we have no motivation to keep the original HELAS strategy. 

  Instead, we introduce only one set of HEGET functions for the interactions among four distinct vector bosons $(V_{1},$ $V_{2},$ $\,V_{3},$ $V_{4})$;
  \begin{eqnarray}
    \mathcal{L}_{\mathrm{V\!V\!V\!V}}
     &=& \mathtt{gg}\,\bigl\{(V_{1}^{*}\!\cdot V_{4}^{*})
                             (V_{2}^{*}\!\cdot V_{3}^{*})  \nonumber \\
     & & \quad
      -\,(V_{1}^{*}\!\cdot V_{3}^{*})(V_{2}^{*}\!\cdot V_{4}^{*})\bigr\}.
   \label{xab}
  \end{eqnarray}
  The corresponding HEGET functions are named as \texttt{ggggxx} for the amplitude and \texttt{jgggxx} for the off-shell currents.

  Because the SU(2) weak boson vertices (\ref{x4v}) always have two identical bosons (or two channels for the vertex of $\WpWm\gamma Z$) we further introduce the HEGET functions \texttt{wwwwxx} and \texttt{jwwwxx} which sum over the two contributions internally. 
  The functions are listed in Appendices \ref{sec:wwwwxx} and \ref{sec:jwwwxx}, respectively, and the input fields and corresponding couplings are given in Table~\ref{tab:input_jwwwxx}.

  As for the QCD quartic gluon coupling
  \begin{subequations}
  \begin{eqnarray}
   \mathcal{L}_{\mathrm{gggg}}
    & = & -\dfrac{g_{s}^{2}}{4} f^{abe}f^{cde} A^a_\mu A^b_\nu A^{c\mu} A^{d\nu},
    \\
   & = & \frac{g_{s}^{2}}{2}\,f^{abe}f^{cde}
    \bigl\{ (A^{a}\!\cdot A^{d})(A^{b}\!\cdot A^{c}) \nonumber \\
   && \qquad\qquad\ \  -\,(A^{a}\!\cdot A^{c})(A^{b}\!\cdot A^{d})\bigr\},
    \label{x4g}
  \end{eqnarray}
  \end{subequations}
  we can use the HEGET functions \texttt{ggggxx} and \texttt{jgggxx} for the Lagrangian (\ref{xab}) to obtain the amplitudes and the off-shell gluon currents respectively.
  For instance, by using the four-vector vertex amplitude of $\Gamma(g^{2};V_{1},\!V_{2},\!V_{3},\!V_{4})$ for the Lagrangian (\ref{xab}), we can express the amplitude of $g_{1}^{a}$, $g_{2}^{b}$, $g_{3}^{c}$, $g_{4}^{d}$ as follows:
  \begin{eqnarray}
   \Gamma(g_{1}^{a},g_{2}^{b},g_{3}^{c},g_{4}^{d}) & = &
    f^{abe}f^{cde}\Gamma(g_{s}^{2};g_{1}^{a},g_{2}^{b},g_{3}^{c},g_{4}^{d}) \nonumber \\
   & + & f^{ace}f^{dbe}\Gamma(g_{s}^{2};g_{1}^{a},g_{2}^{c},g_{3}^{d},g_{4}^{b}) \nonumber \\
   & + & f^{adc}f^{bce}\Gamma(g_{s}^{2};g_{1}^{a},g_{2}^{d},g_{3}^{b},g_{4}^{c}). 
  \end{eqnarray}
  The off-shell gluon currents are obtained similarly by calling the HEGET function three times as explained in \cite{qed-paper,qcd-paper} and repeated in Appendix \ref{sec:vvvv-vertex} for completeness.

  \subsubsection{VVS: vector-vector-scalar vertex}

  The only \texttt{VVS} interaction of the SM appears in the Higgs Lagrangian in the form
  \begin{equation}
   \mathcal{L}_{\mathrm{V\!V\!S}}
    \!= \texttt{gc} \left(V_1^{*}\!\cdot V_{2}^{*}\right)S^{*},
  \end{equation}
  where the coupling \texttt{gc} is real and proportional to the Higgs vacuum expectation value (v.e.v.).
  The HEGET functions for the amplitude \texttt{vvsxxx}, the off-shell vector current \texttt{jvsxxx}, and the off-shell scalar current \texttt{hvsxxx} are given for a general complex \texttt{gc} (in GeV units), distinct complex vector bosons ($V_{1}$ and $V_{2}$), and a complex scalar field ($S$), and are listed in Appendices \ref{sec:vvsxxx}, \ref{sec:jvsxxx} and \ref{sec:hvvxxx}, respectively. 
  In the SM, only \texttt{WWH} and \texttt{ZZH} couplings appear.
  \subsubsection{VVSS: vector-vector-scalar-scalar vertex}

  The HEGET functions for the \texttt{VVSS} vertex are obtained from the Lagrangian
  \begin{equation}
   \mathcal{L}_{\mathrm{V\!V\!SS}}
    \!= \texttt{gc} \left(V_1^{\ast} \!\cdot V_{2}^\ast\right)
    S^{\ast}_{3} S^{\ast}_{4}\,.
  \end{equation}
  The amplitude function \texttt{vvssxx}, the off-shell vector current \texttt{jvssxx}, and the off-shell scalar current \texttt{hvvsxx} are listed in Appendices \ref{sec:vvssxx}, \ref{sec:jvssxx} and \ref{sec:hvvsxx}, respectively, for a complex \texttt{gc}, distinct complex vector bosons ($V_{1}$ and $V_{2}$), and for distinct complex scalars ($S_{3}$ and $S_{4}$).
  In the SM, only \texttt{WWHH} and \texttt{ZZHH} couplings appear, and the coupling \texttt{gc} is real and proportional to the squares of the electroweak gauge couplings.
   \subsubsection{SSS: three scalar vertex}

  In Appendix \ref{sec:sss-vertex}, we show the HEGET functions for the \texttt{SSS} vertex, which are obtained from the Lagrangian
  \begin{equation}
   \mathcal{L}_{\mathrm{SSS}}
    \!= \mathtt{gc}\,S^{*}_{1} S^{*}_{2} S^{*}_{3}\,.
  \end{equation}
  The HEGET functions for the amplitude \texttt{sssxxx} and the off-shell scalar current \texttt{hssxxx} are given for a complex \texttt{gc} (in GeV units) and for distinct complex scalars ($S_{1}, S_{2}, S_{3}$), and are listed in Appendices \ref{sec:sssxxx} and \ref{sec:hssxxx} respectively.
  In the SM, the coupling \texttt{gc} is real and proportional to the Higgs v.e.v.\ and only the $H^{3}$ coupling appears.

  \subsubsection{SSSS: four scalar vertex}

  The Lagrangian
  \begin{equation}
   \mathcal{L}_{\mathrm{SSSS}}
    \!= \mathtt{gc}\,S_1^{*} S_2^{*} S^{*}_{3} S^{*}_{4}\,
  \end{equation}
  gives the HEGET functions for the \texttt{SSSS} vertex: \texttt{ssssxx} for the amplitude and \texttt{hsssxx} for the off-shell scalar current, which are listed in Appendices \ref{sec:ssssxx} and \ref{sec:hsssxx}, respectively.
  Here again the HEGET functions are given for a complex \texttt{gc}, and for the four distinct complex scalar bosons ($S_{1}$, $S_{2}$, $S_{3}$, $S_{4}$).
  In the SM, only the $H^{4}$ coupling exists whose coupling \texttt{gc} is real and proportional to the ratio of the Higgs boson mass and the v.e.v.

 \section{Generation of CUDA functions for Monte Carlo integration}
 \label{sec:mg2cuda}

For the Monte Carlo integration of cross sections of the physics processes on the GPU, all integrand functions have to be coded using CUDA~\cite{cuda}.
In order to prepare these amplitude functions efficiently we develop an automatic conversion program, MG2CUDA.
As input this program takes the FORTRAN amplitude subroutine, \texttt{matrix.f} generated by MadGraph (ver.4)~\cite{madgraph4}, analyzes the source code and generates all CUDA functions necessary for the Monte Carlo integration on GPU.
MG2CUDA also optimizes generated CUDA codes for execution on the GPU by reducing unnecessary variables and dividing long amplitude functions into a set of smaller functions as necessary.

 In the following subsections, the major functions of \linebreak MG2CUDA are briefly described.

  \subsection{Generation of HEGET function calls from HELAS subroutines}
  MG2CUDA converts calling sequences of HELAS subroutines in \texttt{matrix.f} to those of HEGET functions in the integrand function of gBASES.
  All HEGET functions for the GPU are designed to have a one-to-one correspondence to HELAS subroutines with the same name, and their arguments have the same order and the same variables types.
  Hence HELAS subroutine calls are directly converted to HEGET function calls.

   \subsection{Decoding of initial and final state information}
   MG2CUDA decodes the physics process information, species of initial and final particles, the number of graphs and the number of color bases, written into \texttt{matrix.f} by MadGraph and adopts an appropriate phase space program and prepare header files to store process information and some constants.

   \subsection{Division of a long amplitude program}
   As the number of external particles increases, the number of Feynman diagrams contributing to the subprocess grows factorially and the amplitude program generated by MadGraph becomes very long.
   Due to the current limitation of the CUDA compiler, a very long amplitude program cannot be compiled~\cite{qed-paper,qcd-paper} by the CUDA compiler.
   MG2CUDA divides such a long amplitude function program into smaller functions which are successively called in the integrand function of gBASES.

   Among the processes listed in Eq.~(\ref{eq:process}), several processes with the maximum number of jets require such decomposition into smaller pieces by MG2CUDA.
   Those processes are denoted explicitly in the tables and plots in Section~\ref{sec:results} by an asterisk.

   \subsection{Decomposition of a color matrix multiplication}
   Compared to CPU, the memory resources of GPU are quite limited.
   Hence, if calculations on GPU require a large amount of data, the data must reside on slower memory (global memory), and the access to the data becomes a cause of the degradation of performance of GPU programs.
   When the number of independent color bases of a physics process becomes large, the data size necessary for the multiplication of color factors also becomes large.
   For example, the number of color bases for $u\bar{u}\to t\bar{t}\!+\! ggg$~(\ref{eq1:ttbar}) is 144, and the data size and the total number of multiplications in the color matrix multiplication becomes the order of $(144)^2~\sim 20000$.
   In order to avoid degradation of performance of GPU, MG2CUDA decomposes arguments of the color matrix into a set of independent color factors and combines multiplications which have the same factors.
   This significantly reduces the number of color factors and the total number of multiplications~\cite{qcd-paper}.
   For the $u\bar{u}\to t\bar{t}\!+\! ggg$ case, the number of independent factors is only 51 and the reduced number of multiplications becomes $\sim 3600$.
   These color factors are stored in the read-only (e.g. constant) memory which GPU can access more quickly than the global memory\footnote{It should be introduced here that the different approach to the computation of color factors was tried also using GPU and good performance was obtained~\cite{color-gpu}.}.
   
   \subsection{Reduction of the number of temporary wave functions}
  During the computation of amplitudes, temporary variables of wave functions are necessary to keep intermediate particle information.
   MG2CUDA analyzes the use of these temporary variables and recycles variables which are not used any more in the latter part of the program.
   This greatly relaxes the memory resource requirement.
   Again for the $u\bar{u}\to t\bar{t}\!+\! ggg$ case, the number of variables used for wave functions in original \texttt{matrix.f} is 1607, and it becomes only 83 after recycling temporary variables.


 \section{Computing environment}
 \label{sec:computation}

 In this section, we introduce our computing environment used for all computations presented in this paper.

  \subsection{Computations on GPU}
  \label{sec:gpu-env}
  
  We used a Tesla C2075 GPU processor board produced by NVIDIA \cite{nvidia} to compute cross sections of the physics processes listed in Eqs.~(\ref{eq1:W})-(\ref{eq4:HH}).
  %
  The Tesla C2075 has 448 processors (CUDA cores) in one GPU chip, which delivers up to 515 GFLOPS of double-precision peak performance.
  Other parameters of the board are listed in Table~\ref{tab:gpu}.
  The Tesla C2075 is controlled by a Linux PC with Fedora 14 (64bit) operating system.
  CUDA codes executed on the GPU are developed on the host PC with the CUDA 4.2~\cite{cuda} software development kit.
   \begin{table}[tbh]
    \centering
    \caption{Parameters of Tesla C2075 and CUDA tools}
    \label{tab:gpu}       
    \smallskip
    \begin{tabular}{ll}     \hline
     Number of CUDA core & 448  \\  
     Total amount of & \multirow{2}{*}{5.4 GB} \\
     global memory   &                        \\ 
     Total amount of & \multirow{2}{*}{64 kB}  \\
     constant memory &                        \\ 
     Total amount of shared & \multirow{2}{*}{48 kB} \\
     memory per block       &                       \\ 
     Total number of registers & \multirow{2}{*}{32768} \\
     available per bloc        &                        \\ 
     Clock rate & 1.15 GHz \\ \\
     nvcc CUDA compiler & Rel. 4.2 (V0.2.1221)  \\ 
     CUDA Driver  & Ver.4.2 \\ 
     CUDA Runtime & Ver 4.2 \\ \hline
    \end{tabular}
   \end{table}

   For the computation of cross sections we use the Monte Carlo integration program, BASES~\cite{bases}.
   The GPU version of BASES, gBASES, has originally been developed in single precision~\cite{mcinteg}.
   In this paper, however, we use the newly developed double precision version of gBASES for all GPU computations throughout this report.
     
 \subsection{CPU environment}
  \label{sec:cpu-env}

  As references of cross section computations and also for purposes of comparisons of process time, we use the BASES program in FORTRAN on the CPU~\cite{bases}.
  The measurement of total execution time is performed on the Linux PC with Fedora 13 (64 bit) operating system.
  The hardware parameters and the version of the software used for the execution of the FORTRAN BASES programs are summarized in Table~\ref{tab:host-pc}.
   \begin{table}[tbh]
    \centering
    \caption{CPU environment}
    \label{tab:host-pc}       
    \begin{tabular}{ll} \hline
     CPU & Intel Core i7 2.67 GHz \\ 
     Cache size & 8192 KB  \\ 
     Memory & 6 GB  \\ \\
     OS & Fedora 13 (64 bit)  \\ 
     gcc & 4.4.5 (Red Hat 4.4.5-2) \\ \hline
    \end{tabular}
   \end{table}

   As another reference of cross sections we also use the latest version of MadGraph (ver.5)~\cite{madgraph5} which has been released in 2011.
   All numerical results appear in Sec.~\ref{sec:results} as ``MadGraph'' are obtained by this new version of MadGraph.
   During computations it shows good performance in the execution time and gives us stable results.


 \section{Results}
 \label{sec:results}

 In this section, we present numerical results of the computations of the total cross sections and a comparison of total process time of gBASES programs on the GPU for the SM processes listed in Sec.~\ref{sec:physics_processes}. 
 As references for the cross sections and process time, we also present results obtained by BASES and MadGraph on a CPU. 
 Recent improvements in double precision calculations on the GPU allows us to perform all computations in this paper at double precision accuracy.

In addition, in the previous papers~\cite{qed-paper,qcd-paper}, a simple program composed of a single event loop without any optimizations for the computation of cross sections was used both on the GPU and the CPU and their process time for a single event loop was compared.
In this paper, we compare the total execution time of BASES programs running on the GPU and on a CPU with the same integration parameters.
Since the total execution time of gBASES includes both process time on the GPU and the CPU, the comparison gives a more practical index of the gain in the total process time by using the GPU.

 For all three programs, gBASES, BASES and MadGraph, we use the same final state cuts, PDF's and the model parameters as explained in Sec.~\ref{sec:physics_processes}. 
 All results for the various SM processes are obtained for the LHC at \LHCenergy, and summarized in Tables~\ref{tab:result-W}-\ref{tab:result-HHqq}, and Figs.~\ref{fig:time-W}-\ref{fig:time-HHqq}. 
 Some general comments are in order here.

 First, we test all of the HEGET functions listed in \ref{sec:new-heget-codes} against the HELAS subroutines~\cite{helas} and the amplitude subroutines for MadGraph (ver.5) by comparing the helicity amplitudes of all subprocess listed in Sec.~\ref{sec:physics_processes}.
 We generally find agreement within the accuracy of double precision computation, except for multiple Higgs production processes via weak boson fusion, Eqs.~(\ref{eq3:HH}) and (\ref{eq4:HH}).
 For these processes, the MadGraph amplitude codes give significantly smaller amplitudes.
 We identify the cause of the discrepancy as subtle gauge theory cancellation among weak boson exchange amplitudes.
 After modifying both the HELAS and HEGET codes to respect the tree-level gauge invariance strictly, by replacing all $m_{V}^{2}$ in the weak boson propagators and in the vertices\footnote{In our calculations with MadGraph, subroutines for amplitude computation are slightly modified to replace all squared massive vector boson mass, $m_{V}^{2}$, with $m_{V}^{2}-i\,m_{V}\Gamma_{V}$, which is only partly realized in the original codes}, and by setting $m_{W}^{2}-i\,m_{W}\Gamma_{W} = (g_{W}^{2}/g_{Z}^{2}) (m_{Z}^{2}-i\,m_{Z}\Gamma_{Z})$ as a default, we find agreement for all the amplitudes.
 Except for the triple Higgs boson production processes of Eqs.~(\ref{eq3:HH}) and (\ref{eq4:HH}), these modifications on the code do not give significant difference in the amplitudes.

  In Tables~\ref{tab:result-W}-\ref{tab:result-HHqq}, the results of the process time ratio with the divided amplitude functions are denoted by an asterisk.
  In the plots of Figs.~\ref{fig:time-W}-\ref{fig:time-HHqq} they are indicated by open circles.
  By comparing the numbers with and without asterisks in the Tables, and also by comparing the heights of the blobs with and without circles in the Figures for the same number of jets, we can clearly observe the loss of efficiency in the GPU computation when the amplitude function is so long that its division into smaller pieces is needed.
  For example, among the $Z$+4-jets processes only the amplitude function of the process, $uu\!\to\! Z\!+\!uu\!+\!gg$~(\ref{eq3:Z}), has to be divided.
  It is clearly seen from Table~\ref{tab:result-Z} and Fig.~\ref{fig:time-Z} that the GPU gain over the CPU is significantly lower ($\sim\!5$) for this process, as compared to the other $Z$+4-jets processes.
  A similar trend is observed for all other processes with divided amplitude program.

  From Tables \ref{tab:result-W}-\ref{tab:result-HHqq}, we find that the results obtained by gBASES with HEGET functions agree with those by the BASES programs with HELAS within the statistics of generated number of events.
  On the other hand we observe some deviations of MadGraph results from BASES results as the number of jets in the final state, $n$, increases.
  It amounts to about 5-7\% level for processes with the maximum number of jets.
  These deviations may be attributed to the difference of the phase space generation part for multi-jet productions, and require further studies.
  The program gSPRING~\cite{gspring} which generates events on the GPU by making use of the grid information of variables optimized by gBASES~\cite{mcinteg} is being developed, and more detailed comparison between MadGraph and its GPU version will be reported elsewhere.
 
 In the following subsections, let us briefly summarize our findings for each subprocesses as listed in Sect.~\ref{sec:physics_processes}.

  \subsection{Single $W$ production}
   \label{sec:result-W}
   
\begin{figure}[htb]
     \centering
     \resizebox{0.48\textwidth}{!}{%
     \includegraphics{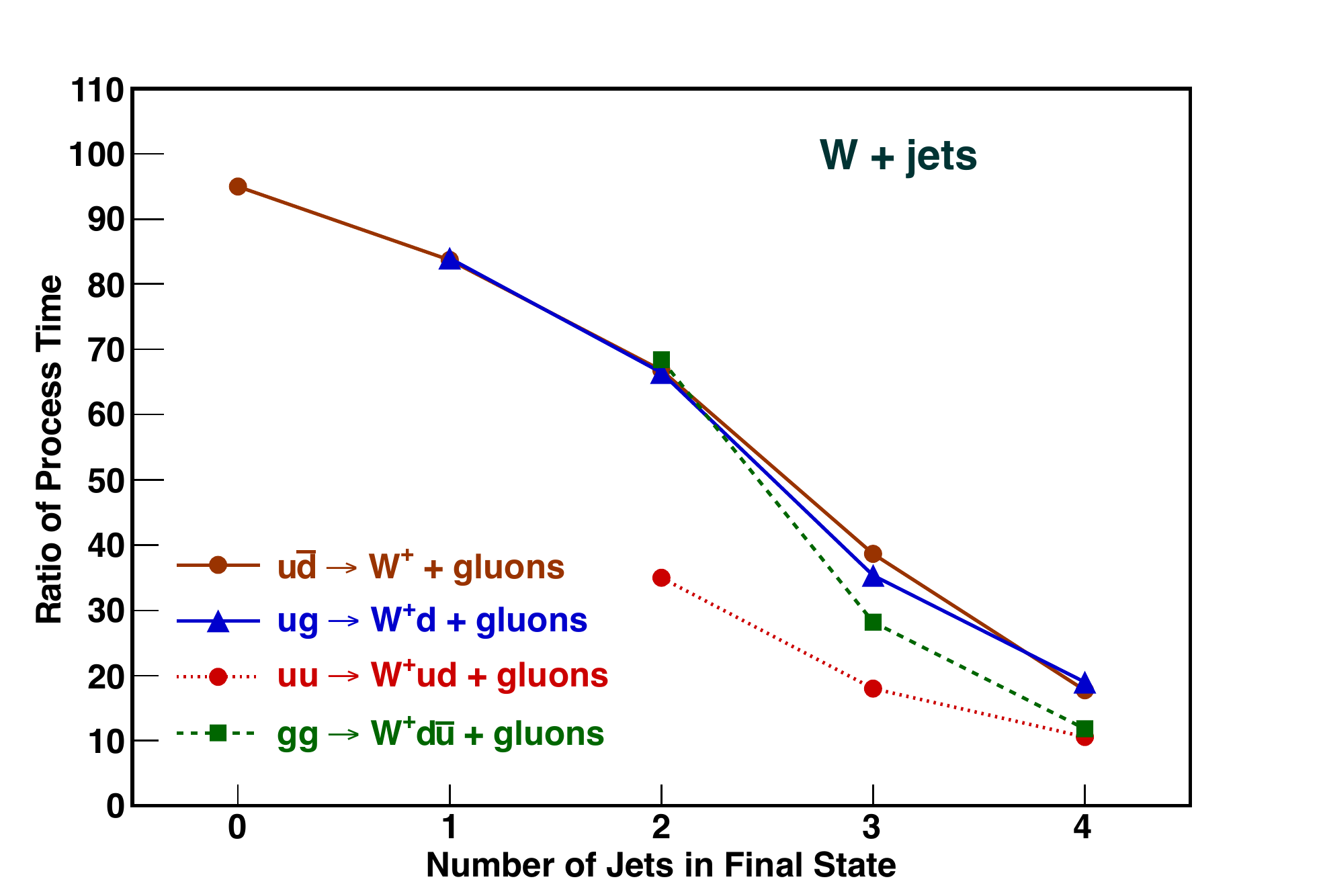}
     }
     \caption{Ratio of BASES process time (CPU/GPU) for $\Wp\!+\!n$-jet production with $\Wp \!\to\! \lp\nu_l\, \lemu$ in $pp$ collisions at \LHCenergy.
     Event selection cuts are given by Eqs.~(\ref{eq:jet-cuts-eta})-(\ref{eq:jet-cuts-ptjj}) and (\ref{eq:lepton-cuts-eta})-(\ref{eq:letpon-cuts-pt}), the parton distributions of CTEQ6L1~\cite{cteq6} at the factorization scale of $Q\!=\!\pTjetcut\!=\!20\GeV$ and the fixed QCD coupling at $\alphas(20\GeV)_{\mathrm{LO}}\!=\!0.171$ are used, except for $n\!=\!0$ for which the factorization scale is chosen as $Q\!=\!\mW$.
     }
     \label{fig:time-W}       
    \end{figure}

   The results of the total cross section and the process time for $\Wp\!+\! n$-jet production processes of Eqs.~(\ref{eq1:W})-(\ref{eq4:W}) followed by $\Wpdecay\,\lemu$ decays at the LHC with $\sqrt{s}=14\TeV$ are presented in Table \ref{tab:result-W} and the ratio of process time between GPU and CPU is shown in Fig.~\ref{fig:time-W}.
   All the $n$-jets are required to satisfy the conditions Eqs.~(\ref{eq:jet-cuts-eta})-(\ref{eq:jet-cuts-ptjj}) while leptons \lemu\ from $W$ decays satisfy the cuts Eqs.~(\ref{eq:lepton-cuts-eta})-(\ref{eq:lepton-cuts-pt}).
 The QCD coupling is fixed as $\alphas(20\GeV)_{\mathrm{LO}}\!=\!0.171$ and the CTEQ6L1 parton distribution functions~\cite{cteq6} are evaluated at the factorization scale of $Q\!=\!p_{\mathrm{T, jet}}^{\rm cut}\!=\!20$\,GeV, except for $n\!=\!0$ for which the factorization scale is chosen as $Q\!=\!\mW$.
   
   As for the integrated results for the $W^{+}\!+\!n$-jet processes presented in Table \ref{tab:result-W}, we find that the GPU results obtained by gBASES with the HEGET functions agree well with the corresponding CPU results from the other programs, especially with those obtained by BASES with HELAS, within the statistics of generated number of events. 
   For $n\!=\!3$ or $4$ cases, some discrepancies between BASES and MadGraph results are found. 
   They may be due to the difference in the phase space generation, as mentioned above.

   Their performance on GPU was significantly better than that on CPU, as clearly shown by the total process time ratios of BASES program on CPU over that on GPU with the same integration parameters listed in Table \ref{tab:result-W}.
   In Fig.~\ref{fig:time-W}, we show this ratio as a function of the number of jets in the final state, $n$.
   The ratio gradually decreases from $\sim 100$ at $n\!=\!0$ as $n$ grows, but it still exceeds 10 even for $n\!=\!4$.
   Compared with other processes, the gain for the process, $uu\!\rightarrow\! W^{+}ud\!+\!\mbox{gluons}$, is small.
   That is simply because these processes have more Feynman diagrams and a larger color bases than the other processes with the same number of jets.

    \begin{table*}[hbt]
     \centering
     \caption{Total cross sections and BASES process time ratios for $\Wp\!+\!n$-jet production with $\Wpdecay\,\lemu$ at the LHC (\LHCenergy).
     Event selection cuts are given by Eqs.~(\ref{eq:jet-cuts-eta})-(\ref{eq:jet-cuts-ptjj}) and (\ref{eq:lepton-cuts-eta})-(\ref{eq:lepton-cuts-pt}), the parton distributions of CTEQ6L1~\cite{cteq6} at the factorization scale of $Q\!=\!\pTjetcut\!=\!20\GeV$ and the fixed QCD coupling at $\alphas(20\GeV)_{\mathrm{LO}}\!=\!0.171$ are used, except for $n\!=\!0$ for which the factorization scale is chosen as $Q\!=\!\mW$.
     }
     \label{tab:result-W}       
     \begin{tabular}{c|c|cccl|c} \hline
      \multirow{2}{*}{$n$} & \multirow{2}{*}{Subprocess} & \multicolumn{4}{c|}{Cross section [fb]} &
      Process time ratio \\ \cline{3-7}
      &  & gBASES & BASES & MadGraph &
      & BASES/gBASES  \\ \hline
      
      0
      & $u\bar{d}\to W^{+}$
      & $7.653 \pm 0.008$ & $7.660 \pm 0.007$ & $7.662 \pm 0.007$
      & $\times 10^{6}$
      & 95.0 \\ \hline

      \multirow{2}{*}{1}
      & $u\bar{d}\to W^{+} + g$
      & $6.545 \pm 0.005$ & $6.542 \pm 0.005$ & $6.541 \pm 0.008$
      & $\times 10^{5}$
      & 83.7 \\ 
      
      & $ug\to  W^{+} + d$
      & $1.234 \pm 0.001$ & $1.235 \pm 0.001$ & $1.234 \pm 0.002$
      & $\times 10^{6}$
      & 83.9 \\ \hline

      \multirow{4}{*}{2}
      & $u\bar{d}\to W^{+} + gg$
      & $1.008 \pm 0.001$ & $1.010 \pm 0.001$ & $1.004 \pm 0.001$
      & $\times 10^{5}$
      & 66.8 \\ 
      
      & $ug \to W^{+} + dg$
      & $9.381 \pm 0.006$ & $9.384 \pm 0.006$ & $9.248 \pm 0.014$
      & $\times 10^{5}$
      & 66.5 \\ 
      
      & $uu \to W^{+} + ud$
      & $5.536 \pm 0.002$ & $5.539 \pm 0.002$ & $5.516 \pm 0.008$
      & $\times 10^{4}$
      & 35.0 \\ 
      
      & $gg \to W^{+}+ d\bar{u}$
      & $6.744 \pm 0.005$ & $6.747 \pm 0.006$ & $6.734 \pm 0.011$
      & $\times 10^{4}$
      & 68.4 \\ \hline

      \multirow{4}{*}{3}
      & $u\bar{d}\to W^{+} + ggg$
      & $2.267 \pm 0.002$ & $2.264 \pm 0.006$ & $2.230 \pm 0.002$
      & $\times 10^{4}$
      & 38.6 \\
      
      & $ug \to W^{+} + dgg$
      & $6.033 \pm 0.002$ & $6.032 \pm 0.002$ & $5.947 \pm 0.007$
      & $\times 10^{5}$
      & 35.3 \\ 
      
      & $uu \to W^{+} + udg$
      & $7.221 \pm 0.010$ & $7.215 \pm 0.006$ & $7.086 \pm 0.008$
      & $\times 10^{4}$
      & 18.0 \\ 
      
      & $gg \to W^{+}+ d\bar{u}g$
      & $5.960 \pm 0.003$ & $5.963 \pm 0.004$ & $5.883 \pm 0.008$
      & $\times 10^{4}$
      & 28.2 \\ \hline

      \multirow{4}{*}{4}
      & $u\bar{d}\to W^{+} + gggg$
      & $7.234 \pm 0.011$ & $7.246 \pm 0.004$ & $6.937 \pm 0.007$
      & $\times 10^{3}$
      & 17.7 \\
      
      & $ug \to W^{+} + dggg$
      & $3.918 \pm 0.002$ & $3.918 \pm 0.001$ & $3.718 \pm 0.004$
      & $\times 10^{5}$
      & 19.0 \\
      
      & $uu \to W^{+} + udgg$
      & $7.398 \pm 0.009$ & $7.389 \pm 0.006$ & $6.917 \pm 0.007$
      & $\times 10^{4}$
      & 10.6 \\ 
      
      & $gg \to W^{+}+ d\bar{u}gg$ 
      & $3.662 \pm 0.001$ & $3.664 \pm 0.001$ & $3.502 \pm 0.004$
      & $\times 10^{4}$
      & 11.9 \\ \hline
      
     \end{tabular}
    \end{table*}


   \subsection{Single $Z$ production}
   \label{sec:result-Z}

    \begin{figure}[htb]
     \centering
     \resizebox{0.48\textwidth}{!}{%
     \includegraphics{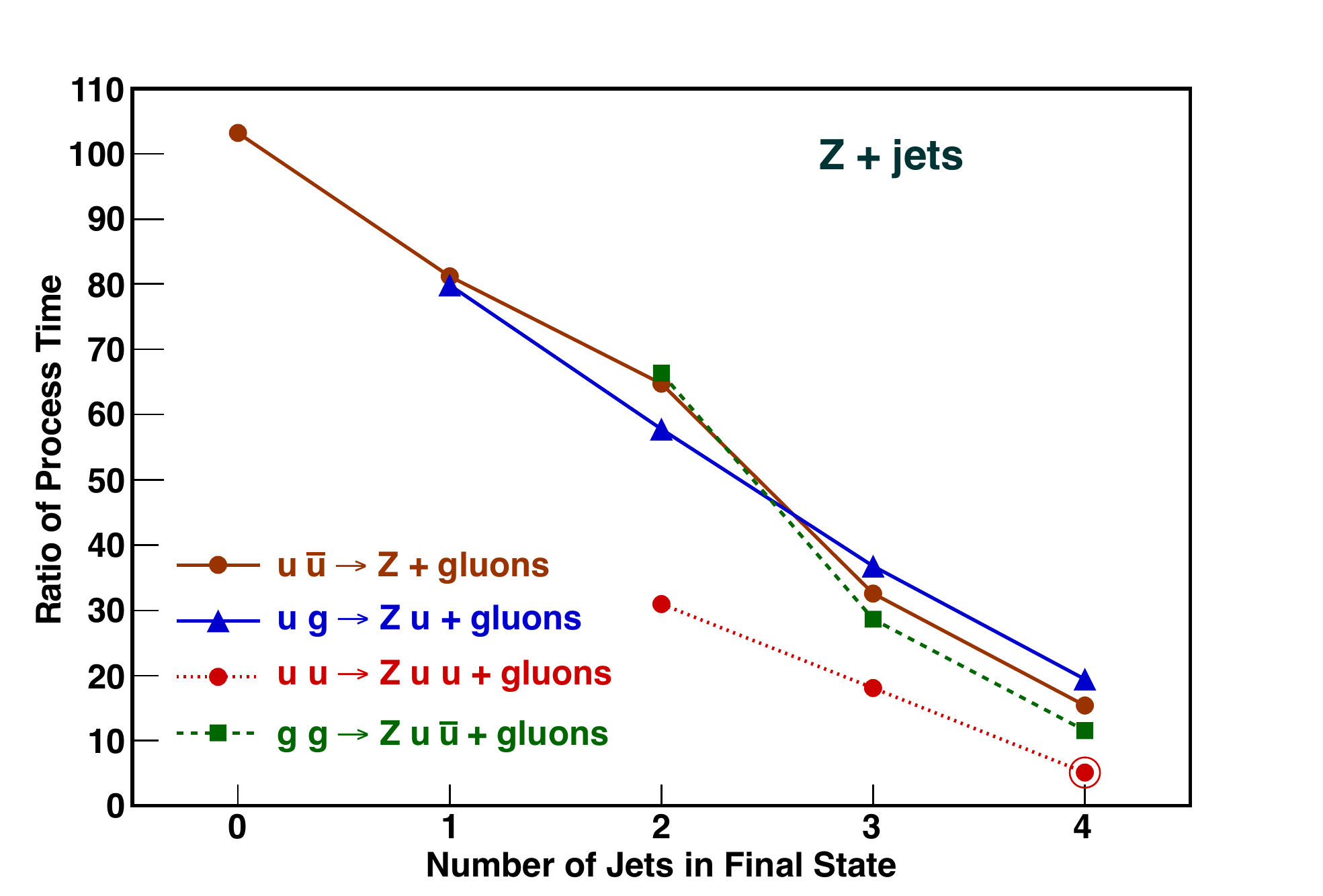}
     }
     \caption{Ratio of BASES process time (CPU/GPU) for $Z + n$-jet production with $\Zdecay\,\lemu$ in $pp$ collisions at \LHCenergy.
     Event selection cuts, PDF and \alphas\ are the same as in Fig.~\ref{fig:time-W}, except for $n\!=\!0$ for which the factorization scale is $Q\!=\!\mZ$.
     }
     \label{fig:time-Z}       
    \end{figure}

   \begin{table*}[htb]
    \centering
     \caption{Total cross sections and BASES process time ratios for $Z + n$-jet production
     with $\Zdecay\,\lemu$ at the LHC (\LHCenergy).
     Event selection cuts, PDF and \alphas\ are the same as in Table~\ref{tab:result-W}, except for $n\!=\!0$ for which the factorization scale is $Q\!=\!\mZ$.
     }
     \label{tab:result-Z}       
    \begin{tabular}{c|c|cccl|c} \hline
     \multirow{2}{*}{$n$} & \multirow{2}{*}{Subprocess} & \multicolumn{4}{c|}{Cross section [fb]} &
      Process time ratio \\ \cline{3-7}
      &  & gBASES & BASES & MadGraph &
      & BASES/gBASES  \\ \hline

      0
      & $u\bar{u} \to Z$
      & $4.333 \pm 0.004$ & $4.330 \pm 0.005$ & $4.334 \pm 0.004$
      & $\times 10^{5}$
      & 103.2 \\ \hline

     \multirow{2}{*}{1}
      & $u\bar{u} \to Z + g$
      & $4.143 \pm 0.004$ & $4.135 \pm 0.004$ & $4.136 \pm 0.005$
      & $\times 10^{4}$
      & 81.2 \\

      & $ug\to Z + u$
      & $7.161 \pm 0.007$ & $7.171 \pm 0.007$ & $7.162 \pm 0.009$
      & $\times 10^{4}$
      & 79.9 \\ \hline

     \multirow{4}{*}{2}
      &  $u\bar{u} \to Z + gg$
      & $7.283 \pm 0.007$ & $7.290 \pm 0.008$ & $7.258 \pm 0.010$
      & $\times 10^{3}$
      & 64.7 \\

      & $ug \to Z + ug$ 
      & $5.738 \pm 0.003$ & $5.743 \pm 0.003$ & $5.718 \pm 0.009$
      & $\times 10^{4}$
      & 57.8 \\

      & $uu \to Z +uu$
      & $3.503 \pm 0.003$ & $3.511 \pm 0.003$ & $3.475 \pm 0.004$
      & $\times 10^{3}$
      & 31.0 \\ 

      & $gg \to Z + u\bar{u}$
      & $5.301 \pm 0.007$ & $5.301 \pm 0.007$ & $5.292 \pm 0.007$
      & $\times 10^{3}$
      & 66.3 \\ \hline

     \multirow{4}{*}{3}
      & $u\bar{u} \to  Z+ ggg$
      & $1.758 \pm 0.004$ & $1.766 \pm 0.002$ & $1.724 \pm 0.002$
      & $\times 10^{3}$
      & 32.6 \\

      & $ug \to Z + ugg$
      & $3.918 \pm 0.002$ & $3.917 \pm 0.002$ & $3.838 \pm 0.005$
      & $\times 10^{4}$
      & 36.8 \\

      & $uu \to Z + uug$
      & $4.897 \pm 0.004$ & $4.898 \pm 0.005$ & $4.804 \pm 0.005$
      & $\times 10^{3}$
      & 18.1 \\

      & $gg \to Z + u\bar{u}g$
      & $4.832 \pm 0.002$ & $4.839 \pm 0.006$ & $4.764 \pm 0.006$
      & $\times 10^{3}$
      & 28.6 \\ \hline

     \multirow{4}{*}{4}
      & $u\bar{u} \to Z + gggg$
      & $5.738 \pm 0.011$ & $5.746 \pm 0.004$ & $5.514 \pm 0.006$
      & $\times 10^{2}$
      & 15.4 \\
     
      & $ug \to  Z + uggg$
      & $2.694 \pm 0.002$ & $2.694 \pm 0.001$ & $2.557 \pm 0.003$
      & $\times 10^{4}$
      & 19.5 \\
     
      & $uu \to Z + uugg$
      & $5.250 \pm 0.005$ & $5.259 \pm 0.004$ & $4.964 \pm 0.005$
      & $\times 10^{3}$
      &  5.1* \\

      & $gg \to  Z + u\bar{u}gg$
      & $3.038 \pm 0.002$ & $3.038 \pm 0.001$ & $2.901 \pm 0.003$
      & $\times 10^{3}$
      & 11.6 \\ \hline

     \end{tabular}
    \end{table*}
  
  Similarly, results for $Z\!+n$-jet production processes of Eqs.~(\ref{eq1:Z})-(\ref{eq4:Z}) with $\Zdecay\,\lemu$ are presented in Fig.~\ref{fig:time-Z} and Table \ref{tab:result-Z}. 
   All of the selection cuts and the SM parameters are the same as in the previous subsection.
   The factorization scale is chosen as $Q\!=\!\mZ$ for $n\!=\!0$, while $Q\!=\!\pTjetcut\!=\!20\GeV$ and $\alphas\!=\!\alphas(\pTjetcut)_{\mathrm{LO}}\!=\!0.171$ for $n\!\geq\!1$.

  As in the case of $\Wp\!+\!n$-jet processes, the integrated cross sections obtained by gBASES with HEGET and by BASES with HELAS as well as those from MadGraph are more or less consistent. Small discrepancies may be attributed to differences in phase space parameterization or in its optimization processes of integrations. 
  The ratios of total process time of BASES between CPU and GPU are given in Fig.~\ref{fig:time-Z} and show very similar behavior to those in Fig.~\ref{fig:time-W} for $\Wp\!+\!n$-jet processes.  
  A factor over 100 is obtained for $n\!=\!0$.
  It decreases as $n$, but still exceeds 10 for $n\!=\!4$, except for $uu\!\rightarrow\! Z\!+\!uu\!+\!gg$.
  The gains  for $uu\!\rightarrow\! Z\!+\!uu\!+\!\mbox{gluons}$~(\ref{eq3:Z}) are smaller than the other processes with the same number of jets, because of their larger number of graphs and color bases.
  The amplitude program for $uu\!\rightarrow\! Z\!+\!uu\!+\!gg$ cannot be compiled as a single GPU function call, and it has to be divided into smaller function calls of GPU programs.
  The gain for this process, which is denoted with an asterisk in Table~\ref{tab:result-Z} and with an open circle in Fig.~\ref{fig:time-Z}, is only a factor of $\sim\!5$.

   \subsection{$WW$ production}
   \label{sec:result-WW}
   
    \begin{figure}[htb]
     \centering
     \resizebox{0.48\textwidth}{!}{%
     \includegraphics{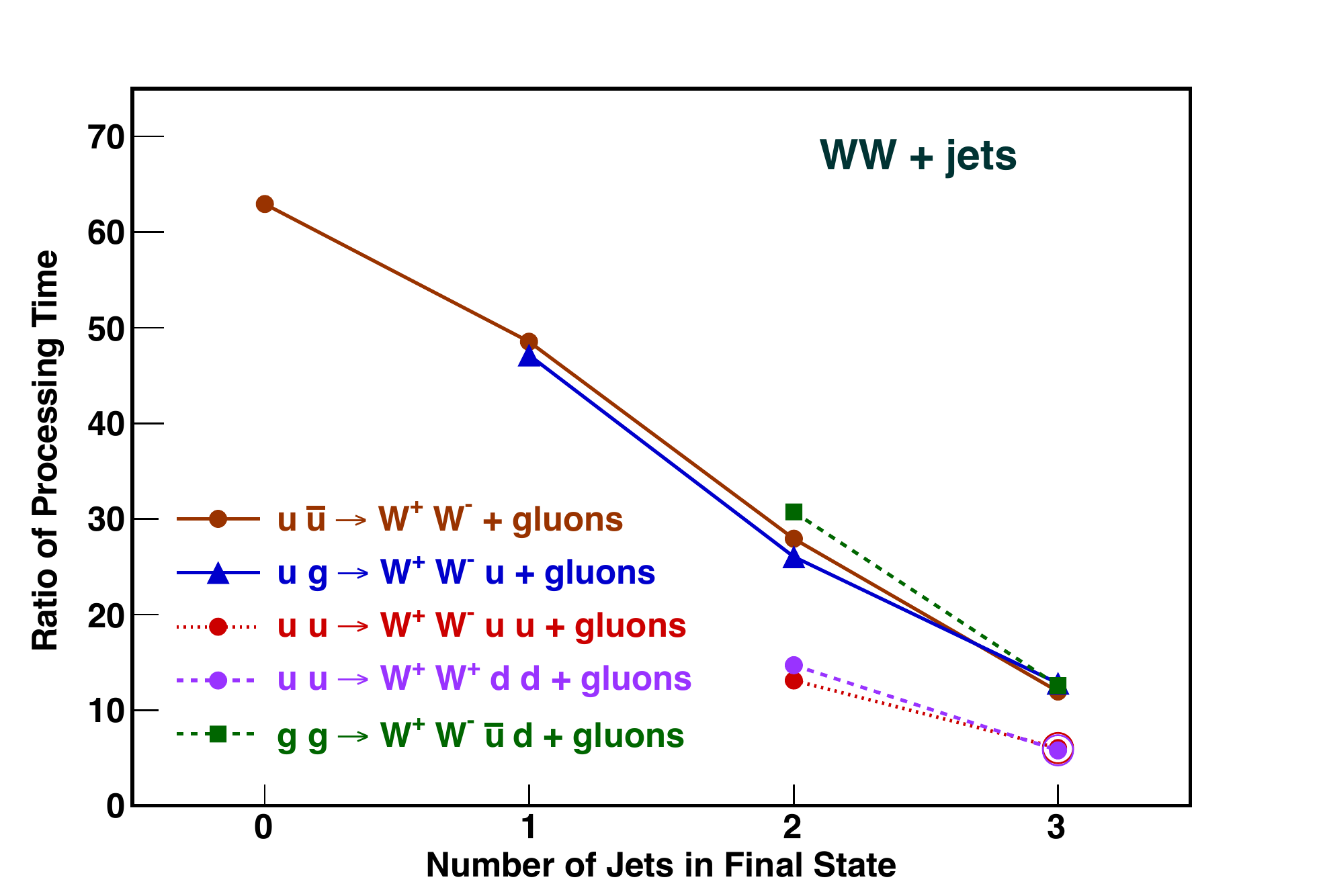}
     }
     \caption{Ratio of BASES process time (CPU/GPU) for $WW + n$-jet production
     with $\Wpmdecay\,\lemu$ in $pp$ collisions at \LHCenergy.
     Event selection cuts, PDF and \alphas\ are the same as in Fig.~\ref{fig:time-W}.
     }
     \label{fig:time-WW}       
    \end{figure}

    \begin{table*}[htb]
    \centering
     \caption{Total cross sections and the total process time ratios for $WW + n$-jet production with $\Wpmdecay$ with \lemu\ at the LHC (\LHCenergy).
     Event selection cuts, PDF and \alphas\ are the same as in Table~\ref{tab:result-W}.
    }
     \label{tab:result-WW}       
     \begin{tabular}{c|c|cccl|c} \hline
     \multirow{2}{*}{$n$} & \multirow{2}{*}{Subprocess} & \multicolumn{4}{c|}{Cross section [fb]} &
      Process time ratio \\ \cline{3-7}
      &  & gBASES & BASES & MadGraph &
      & BASES/gBASES  \\ \hline

      0
      & $u\bar{u} \to \WpWm$
      & $5.801 \pm 0.004$ & $5.800 \pm 0.004$ & $5.797 \pm 0.009$
      & $\times 10^{2}$
      & 63.0 \\ \hline

      \multirow{2}{*}{1}
      & $u\bar{u} \to \WpWm + g$
      & $2.275 \pm 0.001$ & $2.276 \pm 0.002$ & $2.258 \pm 0.003$
      & $\times 10^{2}$
      & 48.5 \\

      & $ug \to \WpWm + u$
      & $2.746 \pm 0.002$ & $2.752 \pm 0.002$ & $2.740 \pm 0.004$
      & $\times 10^{2}$
      & 47.2 \\ \hline

      \multirow{5}{*}{2}
      & $u\bar{u} \to \WpWm + gg$
      & $9.063 \pm 0.009$ & $9.037 \pm 0.009$ & $8.954 \pm 0.009$
      & $\times 10^{1}$
      & 27.9 \\

      & $ug \to \WpWm + ug$
      & $3.890 \pm 0.003$ & $3.889 \pm 0.005$ & $3.860 \pm 0.005$
      & $\times 10^{2}$
      & 26.0 \\

      & $uu \to \WpWm + uu$
      & $3.115 \pm 0.001$ & $3.114 \pm 0.001$ & $3.045 \pm 0.002$
      & $\times 10^{1}$
      & 13.1 \\

      & $uu \to \WpWp + dd$
      & $2.070 \pm 0.001$ & $2.070 \pm 0.001$ & $1.955 \pm 0.002$
      & $\times 10^{1}$
      & 14.7 \\

      & $gg \to\WpWm + u\bar{u}$
      & $1.451 \pm 0.001$ & $1.451 \pm 0.001$ & $1.442 \pm 0.002$
      & $\times 10^{1}$
      & 30.8 \\ \hline

      \multirow{5}{*}{3}
      & $u\bar{u} \to \WpWm + ggg$
      & $3.857 \pm 0.005$ & $3.839 \pm 0.006$ & $3.720 \pm 0.004$
      & $\times 10^{1}$
      & 12.0 \\

      & $ug \to \WpWm + ugg$
      & $3.899 \pm 0.005$ & $3.900 \pm 0.001$ & $3.773 \pm 0.004$
      & $\times 10^{2}$
      & 12.8 \\

      & $uu \to \WpWm + uug$
      & $6.066 \pm 0.007$ & $6.074 \pm 0.003$ & $5.691 \pm 0.006$
      & $\times 10^{1}$
      &  6.1* \\

      & $uu \to \WpWp + ddg$
      & $3.192 \pm 0.006$ & $3.192 \pm 0.003$ & $3.001 \pm 0.003$
      & $\times 10^{1}$
      &  5.8* \\

      & $gg \to \WpWm + u\bar{u}g$
      & $1.999 \pm 0.002$ & $1.996 \pm 0.001$ & $1.952 \pm 0.002$
      & $\times 10^{1}$
      & 12.6 \\ \hline

     \end{tabular}
    \end{table*}
   
   Results for $WW\! +\! n$-jet production processes of Eqs.~(\ref{eq1:WW})-(\ref{eq4:WW}) are presented in Fig.~\ref{fig:time-WW} and Table \ref{tab:result-WW}. 
   Here both $W$'s decay leptonically, with $\Wpmdecay$ with \lemu. 
   In addition to the $\WpWm$  production processes, we also give results for $\WpWp$ production processes of Eq.~(\ref{eq4:WW}).
   The factorization scale is chosen as $Q\!=\!\mW$ for $n\!=\!0$, while $Q\!=\!\pTjetcut\!=\!20\GeV$ and $\alphas\!=\!\alphas(\pTjetcut)_{\mathrm{LO}}\!=\!0.171$ for $n\!\geq\!1$.
   The GPU gain over the CPU computation is $\gtrsim\!50$ for $n\!=\!0$ and 1, where it becomes $\sim 12$ for $n\!=\!3$ except for $uu\!\rightarrow\!\WpWm uu\!+\!\mbox{gluons}$ and $uu\!\rightarrow\!\WpWp dd\!+\!\mbox{gluons}$.
   Smaller gains of $\sim 6$ are observed for these two types, due to the large size of the programs which have been divided into several smaller pieces, as indicated by the asterisk besides the number in the right most column of Table~\ref{tab:result-WW}.
   The trends are similar to those observed for $qq\rightarrow Vqqgg$ subprocesses for $V=W$ or $Z$, which has been discussed in subsections~\ref{sec:result-W} and \ref{sec:result-Z}, respectively.

   \subsection{$\WpZ$ production}

    \begin{figure}[htb]
     \centering
     \resizebox{0.48\textwidth}{!}{%
     \includegraphics{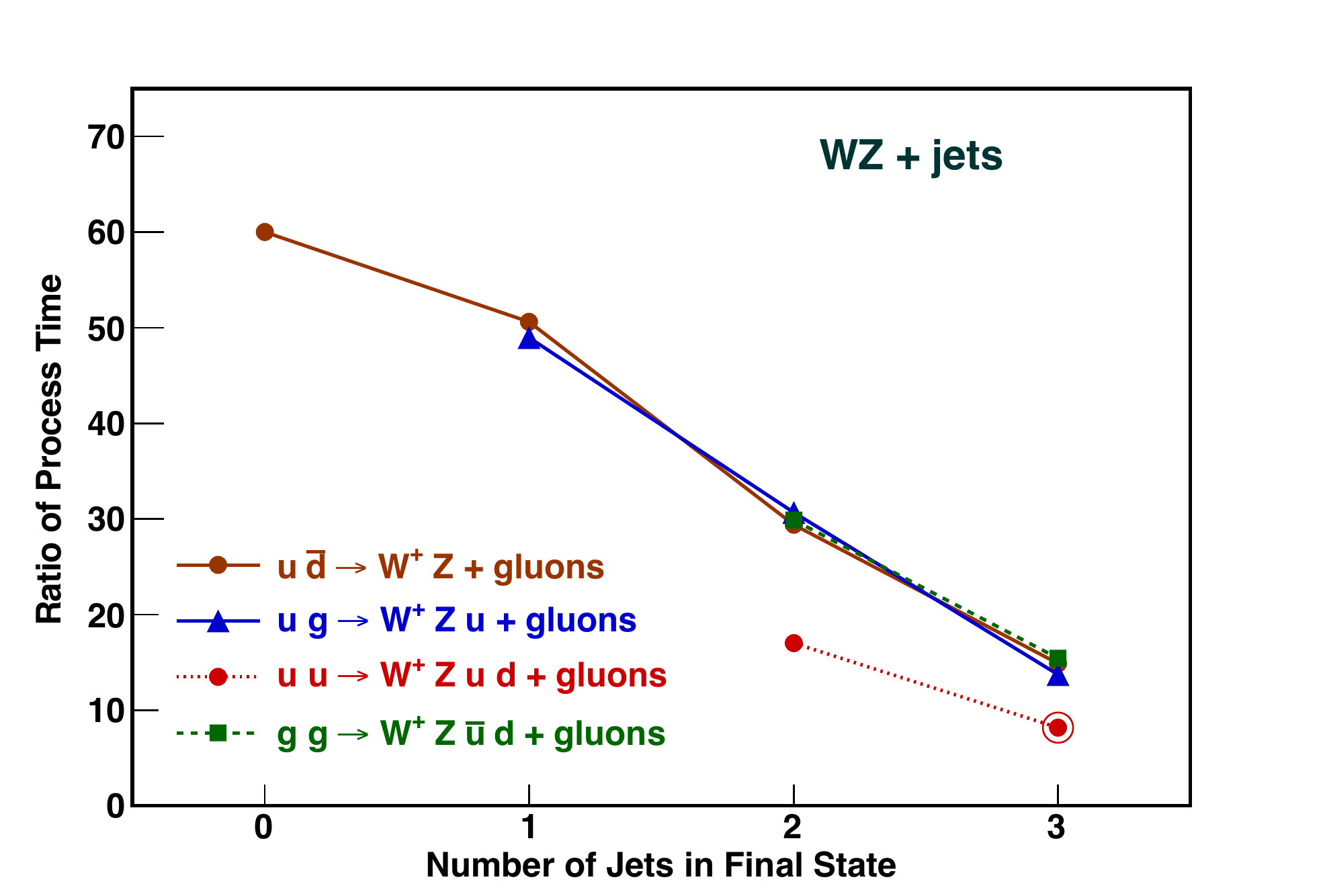}
     }
     \caption{Ratio of BASES process time (CPU/GPU) for $\WpZ + n$-jet production with $\Wpdecay$, $\Zdecay\,\lemu$ in $pp$ collisions at \LHCenergy.
     Event selection cuts, PDF and \alphas\ are the same as in Fig.~\ref{fig:time-W}.
     }
     \label{fig:time-WZ}       
    \end{figure}

    \begin{table*}[htb]
     \centering
     \caption{Total cross sections and BASES process time ratios for $\WpZ + n$-jet production with $\Wpdecay$, $\Zdecay\,\lemu$ at the LHC (\LHCenergy).
     Event selection cuts, PDF and \alphas\ are the same as in Table~\ref{tab:result-W}.
     }
     \label{tab:result-WZ}       
     \begin{tabular}{c|c|cccl|c} \hline
     \multirow{2}{*}{$n$} & \multirow{2}{*}{Subprocess} & \multicolumn{4}{c|}{Cross section [fb]} &
      Process time ratio \\ \cline{3-7}
      &  & gBASES & BASES & MadGraph &
      & BASES/gBASES  \\ \hline

      0
      & $u\bar{d} \to \WpZ$
      & $6.021 \pm 0.006$ & $6.008 \pm 0.006$ & $5.997 \pm 0.008$
      & $\times 10^{1}$
      & 60.0 \\ \hline

      \multirow{2}{*}{1}
      & $u\bar{d} \to \WpZ + g$
      & $2.949 \pm 0.001$ & $2.945 \pm 0.003$ & $2.928 \pm 0.004$
      & $\times 10^{1}$
      & 50.6 \\

      & $ug \to \WpZ + d$
      & $4.597 \pm 0.004$ & $4.591 \pm 0.004$ & $4.582 \pm 0.007$
      & $\times 10^{1}$
      & 49.0 \\ \hline

      \multirow{4}{*}{2}
      & $u\bar{d} \to \WpZ +gg$
      & $1.482 \pm 0.002$ & $1.487 \pm 0.002$ & $1.472 \pm 0.002$
      & $\times 10^{1}$
      & 29.4 \\

      & $ug \to \WpZ + dg$
      & $7.407 \pm 0.004$ & $7.410 \pm 0.004$ & $7.352 \pm 0.009$
      & $\times 10^{1}$
      & 30.7 \\

      & $uu \to \WpZ + ud$
      & $6.880 \pm 0.003$ & $6.881 \pm 0.003$ & $6.670 \pm 0.007$
      & $\times 10^{0}$
      & 17.0 \\

      & $gg\to \WpZ + \bar{u}d$
      & $2.741 \pm 0.002$ & $2.741 \pm 0.002$ & $2.719 \pm 0.004$
      & $\times 10^{0}$
      & 29.9 \\ \hline

      \multirow{4}{*}{3}
      & $u\bar{d} \to \WpZ + ggg$
      & $7.925 \pm 0.028$ & $7.933 \pm 0.008$ & $7.656 \pm 0.008$
      & $\times 10^{0}$
      & 14.9 \\ 

      & $ug \to \WpZ + dgg$
      & $8.075 \pm 0.007$ & $8.080 \pm 0.012$ & $7.830 \pm 0.009$
      & $\times 10^{1}$
      & 13.7 \\

      & $uu \to \WpZ +udg$
      & $1.447 \pm 0.004$ & $1.446 \pm 0.001$ & $1.338 \pm 0.002$
      & $\times 10^{1}$
      &  8.2* \\

      & $gg \to \WpZ +\bar{u}dg$
      & $4.092 \pm 0.003$ & $4.089 \pm 0.005$ & $4.004 \pm 0.005$
      & $\times 10^{0}$
      & 15.5 \\ \hline

     \end{tabular}
    \end{table*}
   
   Results for $\WpZ+n$-jet production processes of Eqs.~(\ref{eq1:WZ})-(\ref{eq4:WZ}), with $\Wpdecay$, $\Zdecay\,\lemu$, are presented in Fig.~\ref{fig:time-WZ} and Table \ref{tab:result-WZ}.
   The final state selection cuts for jets and leptons are given by Eqs.~(\ref{eq:jet-cuts-eta})-(\ref{eq:jet-cuts-ptjj}) and (\ref{eq:lepton-cuts-eta})-(\ref{eq:lepton-cuts-pt}).
   Here again the factorization scale is chosen as $Q\!=\!\mW$ for $n\!=\!0$, while $Q\!=\!\pTjetcut\!=\!20\GeV$ and $\alphas\!=\!\alphas(\pTjetcut)_{\mathrm{LO}}\!=\!0.171$ for $n\!\geq\!1$.

   All of the cross sections in Table~\ref{tab:result-WZ} are consistent between GPU and CPU within the statistical error of Monte Carlo integrations, and the performance improvement of GPU over CPU again depends on $n$.
   The improvement  for $uu\!\rightarrow\!W^{+}Zud\!+\!\mathrm{gluon}$ is also the smallest among those of $W^{+}Z\!+\!3$-jets processes, because the amplitude function has to be divided into smaller pieces as indicated by the asterisk besides the factor of 8.2.


   \subsection{$ZZ$ production}

    \begin{figure}[htb]
     \centering
     \resizebox{0.48\textwidth}{!}{%
     \includegraphics{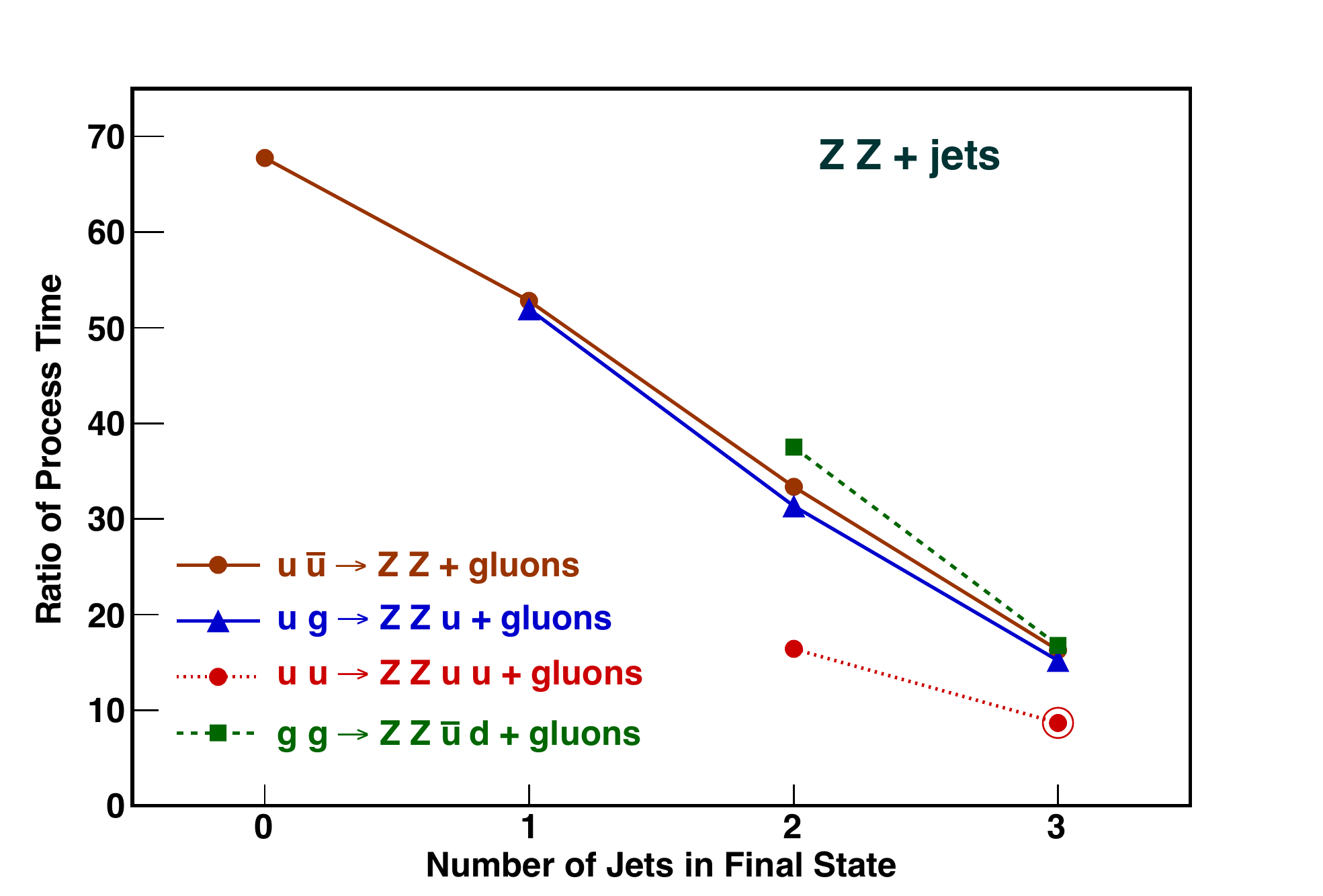}
     }
     \caption{Ratio of BASES process time (CPU/GPU) for $ZZ\!+\!n$-jet production
     with $\Zdecay\,\lemu$ in $pp$ collisions at \LHCenergy.
     Event selection cuts, PDF and \alphas\ are the same as in Fig.~\ref{fig:time-Z}.
     }
     \label{fig:time-ZZ}       
    \end{figure}

    \begin{table*}[htb]
     \centering
     \caption{Total cross sections and BASES process time ratios for $ZZ + n$-jet production with $\Zdecay\,\lemu$ at the LHC (\LHCenergy).
     Event selection cuts, PDF and \alphas\ are the same as in Table~\ref{tab:result-Z}.
     }
     \label{tab:result-ZZ}       
     \begin{tabular}{c|c|cccl|c} \hline
     \multirow{2}{*}{$n$} & \multirow{2}{*}{Subprocess} & \multicolumn{4}{c|}{Cross section [fb]} &
      Process time ratio \\ \cline{3-7}
      &  & gBASES & BASES & MadGraph &
      & BASES/gBASES  \\ \hline

      0
      & $u\bar{u} \to ZZ$
      & $8.214 \pm 0.009$ & $8.200 \pm 0.009$ & $8.192 \pm 0.011$
      & $\times 10^{0}$
      & 67.7 \\ \hline

      \multirow{2}{*}{1}
      & $u\bar{u} \to ZZ + g$
      & $3.508 \pm 0.003$ & $3.513 \pm 0.003$ & $3.494 \pm 0.004$
      & $\times 10^{0}$
      & 52.8 \\

      & $ug \to ZZ + u$
      & $2.504 \pm 0.002$ & $2.504 \pm 0.002$ & $2.503 \pm 0.004$
      & $\times 10^{0}$
      & 52.0 \\ \hline

      \multirow{4}{*}{2}
      & $u\bar{u} \to ZZ + gg$
      & $1.437 \pm 0.001$ & $1.439 \pm 0.002$ & $1.426 \pm 0.002$
      & $\times 10^{0}$
      & 33.3 \\

      & $ug \to ZZ + ug$
      & $3.263 \pm 0.003$ & $3.266 \pm 0.003$ & $3.241 \pm 0.004$
      & $\times 10^{0}$
      & 31.3 \\

      & $uu \to ZZ + uu$
      & $2.747 \pm 0.003$ & $2.745 \pm 0.004$ & $2.710 \pm 0.002$
      & $\times 10^{-1}$
      & 16.4 \\

      & $gg \to ZZ + u\bar{u}$
      & $1.466 \pm 0.001$ & $1.466 \pm 0.001$ & $1.462 \pm 0.002$
      & $\times 10^{-1}$
      & 37.5 \\ \hline

      \multirow{4}{*}{3}
      & $u\bar{u} \to ZZ + ggg$
      & $6.026 \pm 0.005$ & $6.026 \pm 0.002$ & $5.863 \pm 0.006$
      & $\times 10^{-1}$
      & 16.3 \\

      & $ug \to ZZ + ugg$
      & $3.108 \pm 0.003$ & $3.108 \pm 0.001$ & $3.036 \pm 0.003$
      & $\times 10^{0}$
      & 15.2 \\

      & $uu \to ZZ + uug$
      & $5.812 \pm 0.022$ & $5.810 \pm 0.003$ & $5.552 \pm 0.005$
      & $\times 10^{-1}$
      &  8.7* \\

      & $gg \to ZZ + u\bar{u}g$
      & $1.911 \pm 0.001$ & $1.912 \pm 0.001$ & $1.871 \pm 0.002$
      & $\times 10^{-1}$
      & 16.8 \\ \hline

     \end{tabular}
    \end{table*}
   
   Results for $ZZ\!+\!n$-jet production processes of Eqs.~(\ref{eq1:ZZ})-(\ref{eq4:ZZ}) where both $Z$ bosons decay as $\Zdecay\,\lemu$ are presented in Fig.~\ref{fig:time-ZZ} and Table \ref{tab:result-ZZ}.
   All of the selection cuts and the SM parameters are the same as in the previous subprocesses.
   The factorization scale is chosen as $Q\!=\!\mZ$ for $n\!=\!0$, while $Q\!=\!\pTjetcut\!=\!20\GeV$ and $\alphas\!=\!\alphas(\pTjetcut)_{\mathrm{LO}}\!=\!0.171$ for $n\!\geq\!1$.
   All of the cross sections in Table~\ref{tab:result-ZZ} are consistent between the GPU and the CPU computations, and the GPU gain over CPU for the total process time of BASES shown in Fig.~\ref{fig:time-ZZ} ranges from $\sim\! 70$ for $n\!=\!0$ to $\sim\! 16$ for $n\!=\!3$ except for $uu\!\rightarrow\!ZZuu\!+\!\mathrm{gluon}$ whose gain is 8.7 with asterisk in Table~\ref{tab:result-ZZ}, just as in the case of the other $qq\!\rightarrow\!VVqqg$ processes whose amplitude functions are too large to be executed as a single function.

   \subsection{$\ttbar$  production}

    \begin{figure}[htb]
     \centering
     \resizebox{0.48\textwidth}{!}{%
     \includegraphics{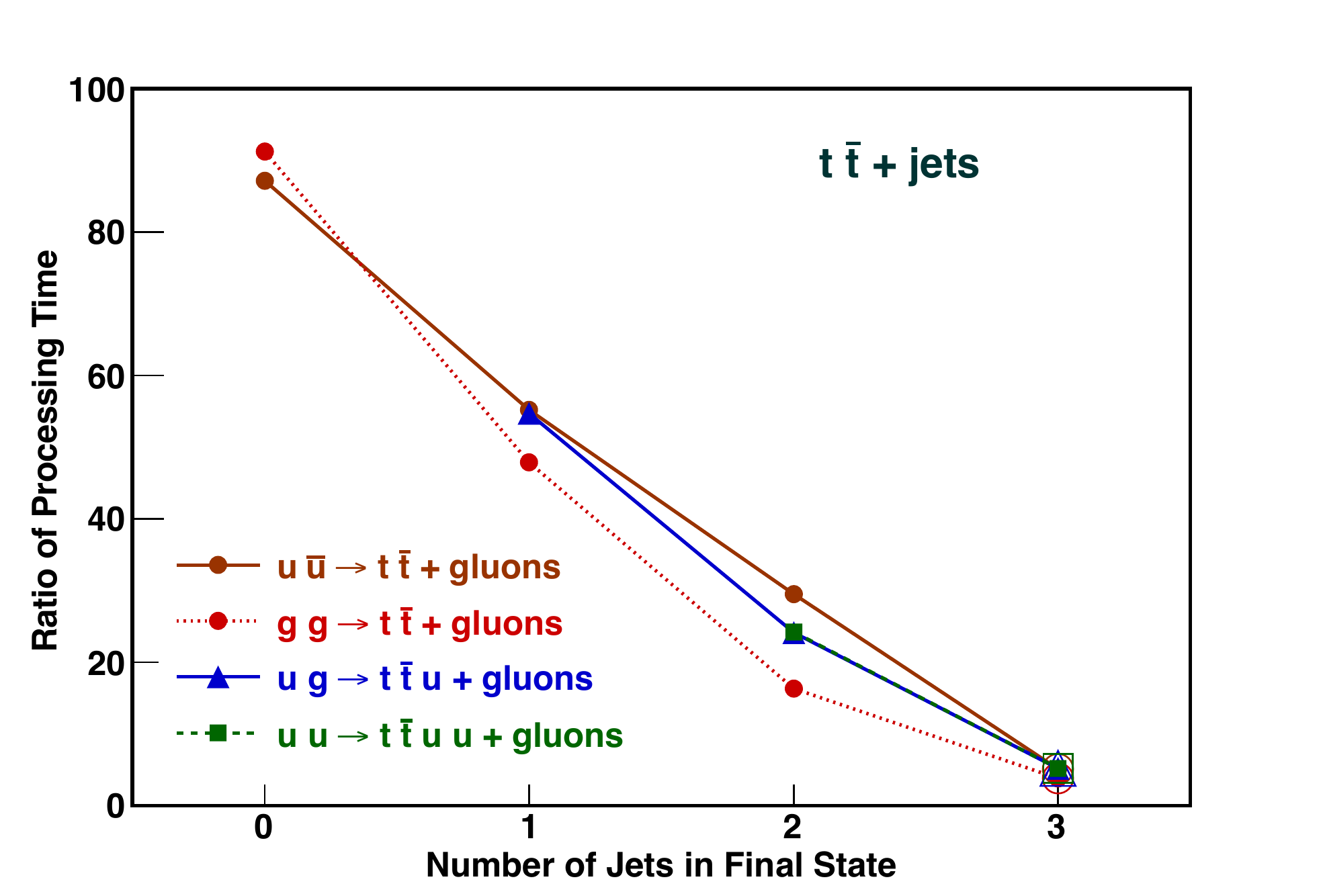}
     }
     \caption{Ratio of BASES process time (CPU/GPU) for $\ttbar + n$-jet production with $\tdecay$ and $\tbardecay\,\lemu$ for \tmass\ and \tBr\ in $pp$ collisions at \LHCenergy.
     Event selection cuts are given by Eqs.~(\ref{eq:jet-cuts-eta})-(\ref{eq:jet-cuts-ptjj}), (\ref{eq:bjet-cuts-eta})-(\ref{eq:bjet-cuts-pt}) and (\ref{eq:lepton-cuts-eta})-(\ref{eq:lepton-cuts-pt}) and the parton distributions of CTEQ6L1~\cite{cteq6} at the factorization scale of $Q\!=\!\pTjetcut\!=\!20\GeV$ is used, except for $n\!=\!0$ for which the factorization scale is chosen as $Q\!=\!\mt$.
     The strong coupling constants are set as $\alpha_{\mathrm{s}}^{2\!+\!n}\!=\!\alphas(\mt)_{\mathrm{LO}}^{2}\,\alphas(\pTjetcut)_{\mathrm{LO}}^{n}$ with $\alphas(\mt)_{\scriptscriptstyle\mathrm{LO}}\!=\!0.108$ and $\alphas(20\GeV)_{\mathrm{LO}}\!=\!0.171$.
     }
     \label{fig:time-ttbar}       
    \end{figure}
    
    \begin{table*}[htb]
     \centering
     \caption{Total cross sections and BASES process time ratios for $\ttbar + n$-jet production  with \tdecay\ and \tbardecay\ \lemu\ for \tmass\ and \tBr\ at the LHC (\LHCenergy).
     Event selection cuts are given by Eqs.~(\ref{eq:jet-cuts-eta})-(\ref{eq:jet-cuts-ptjj}), (\ref{eq:bjet-cuts-eta})-(\ref{eq:bjet-cuts-pt}) and (\ref{eq:lepton-cuts-eta})-(\ref{eq:lepton-cuts-pt}) and the parton distributions of CTEQ6L1~\cite{cteq6} at the factorization scale of $Q\!=\!\pTjetcut\!=\!20\GeV$ is used, except for $n\!=\!0$ for which the factorization scale is chosen as $Q\!=\!\mt$
     The strong coupling constants are set as $\alpha_{\mathrm{s}}^{2\!+\!n}\!=\!\alphas(\mt)_{\mathrm{LO}}^{2}\,\alphas(\pTjetcut)_{\mathrm{LO}}^{n}$ with $\alphas(\mt)_{\mathrm{LO}}\!=\!0.108$ and $\alphas(20\GeV)_{\mathrm{LO}}\!=\!0.171$.
     }
     \label{tab:result-ttbar}       
     \begin{tabular}{c|c|cccl|c} \hline
     \multirow{2}{*}{$n$} & \multirow{2}{*}{Subprocess} & \multicolumn{4}{c|}{Cross section [fb]} &
      Process time ratio \\ \cline{3-7}
      &  & gBASES & BASES & MadGraph &
      & BASES/gBASES  \\ \hline

      \multirow{2}{*}{0}
      & $u\bar{u} \to \ttbar$
      & $5.473 \pm 0.005$ & $5.469 \pm 0.005$ & $5.477 \pm 0.009$
      & $\times 10^{2}$
      & 87.1 \\

      & $gg \to \ttbar$
      & $1.085 \pm 0.001$ & $1.083 \pm 0.001$ & $1.083 \pm 0.002$
      & $\times 10^{4}$
      & 91.2 \\ \hline

      \multirow{3}{*}{1}
      & $u\bar{u} \to \ttbar + g$
      & $3.463 \pm 0.003$ & $3.454 \pm 0.003$ & $3.419 \pm 0.005$
      & $\times 10^{2}$
      & 55.2 \\

      & $gg \to \ttbar +g$
      & $2.236 \pm 0.002$ & $2.245 \pm 0.003$ & $2.217 \pm 0.003$
      & $\times 10^{4}$
      & 47.9 \\

      & $ug \to \ttbar + u$
      & $3.662 \pm 0.008$ & $3.665 \pm 0.001$ & $3.642 \pm 0.006$
      & $\times 10^{3}$
      & 54.7 \\ \hline

      \multirow{4}{*}{2}
      & $u\bar{u} \to \ttbar + gg$
      & $1.857 \pm 0.001$ & $1.855 \pm 0.001$ & $1.857 \pm 0.001$
      & $\times 10^{2}$
      & 29.5 \\

      & $gg \to \ttbar +gg$
      & $2.258 \pm 0.003$ & $2.257 \pm 0.001$ & $2.212 \pm 0.003$
      & $\times 10^{4}$
      & 16.3 \\

      & $ug \to \ttbar + ug$
      & $7.601 \pm 0.005$ & $7.584 \pm 0.005$ & $7.480 \pm 0.010$
      & $\times 10^{3}$
      & 24.2 \\

      & $uu \to \ttbar + uu$
      & $2.812 \pm 0.003$ & $2.805 \pm 0.003$ & $2.791 \pm 0.004$
      & $\times 10^{2}$
      & 24.3 \\ \hline

      \multirow{4}{*}{3}
      & $u\bar{u} \to \ttbar + ggg$
      & $9.626 \pm 0.132$ & $9.646 \pm 0.028$ & $8.908 \pm 0.043$
      & $\times 10^{1}$
      &  5.1* \\

      & $gg \to \ttbar +ggg$
      & $1.830 \pm 0.004$ & $1.847 \pm 0.004$ & $1.716 \pm 0.002$
      & $\times 10^{4}$
      &  3.9* \\

      & $ug \to \ttbar + ugg$
      & $9.267 \pm 0.003$ & $9.251 \pm 0.008$ & $8.758 \pm 0.010$
      & $\times 10^{3}$
      &  5.2* \\

      & $uu \to \ttbar + uug$
      & $6.760 \pm 0.041$ & $6.792 \pm 0.005$ & $6.462 \pm 0.009$
      & $\times 10^{2}$
      &  5.2* \\ \hline

     \end{tabular}
    \end{table*}

   Results for $\ttbar\!+\!n$-jet production processes of Eqs.~(\ref{eq1:ttbar})-(\ref{eq4:ttbar}) are presented in Fig.~\ref{fig:time-ttbar} and Table~\ref{tab:result-ttbar}, where both $t$ and \tbar\ decay semi-leptonically as $\tdecay$ and $\tbardecay\,\,\lemu$.
   Here the factorization scale is chosen as $Q\!=\!\mtt$ for $u\ubar\!\rightarrow\!\ttbar\,(n\!=\!0)$ and $Q\!=\!\mt$ for $gg\!\rightarrow\!\ttbar\,(n\!=\!0)$, while $Q\!=\!\pTjetcut\!=\!20\GeV$ for all the processes with jet productions $(n\!\geq\!1)$.
   The strong coupling constants are set as $\alpha_{\mathrm{s}}^{2\!+\!n}\!=\!\alphas(\mtt)_{\mathrm{LO}}^{2}$ $\alphas(\pTjetcut)_{\mathrm{LO}}^{n}$ for $u\ubar\!\rightarrow\!\ttbar\!+\!n$-gluon processes, while  $\alpha_{\mathrm{s}}^{2\!+\!n}\!=\!\alphas(\mt)_{\mathrm{LO}}^{2}\,\alphas(\pTjetcut)_{\mathrm{LO}}^{n}$ for the others.
The numerical values are $\alphas(2\mt)_{\mathrm{LO}}\!=\!0.108$ at the $u\ubar\!\rightarrow\!\ttbar$ threshold, $\alphas(\mt)_{\mathrm{LO}}\!=\!0.108$ and $\alphas(20\GeV)_{\mathrm{LO}}\!=\!0.171$.

   All of the cross sections in Table~\ref{tab:result-ttbar} are consistent between the GPU (HEGET) and the CPU (BASES) computations within the statistical uncertainties of the Monte Carlo integrations.

   The GPU gain over CPU on the total process time of BASES shows very similar dependence on $n$ as previous results.
   It starts from $\sim\!90$ for $n\!=\!0$ and drops to $\sim\!5$ for $n\!=\!4$.
   For all $\ttbar\!+\!3$-jet production processes, the amplitude functions have to be divided into smaller pieces in order to be processed by the CUDA compiler.
   The main cause of the long amplitudes for these processes is the proliferation of color the factor bases which has been observed for all of the QCD 5 jet production process in ref.~\cite{qcd-paper}.


   \subsection{$W$ boson associated Higgs production}

    \begin{figure}[htb]
     \centering
     \resizebox{0.48\textwidth}{!}{%
     \includegraphics{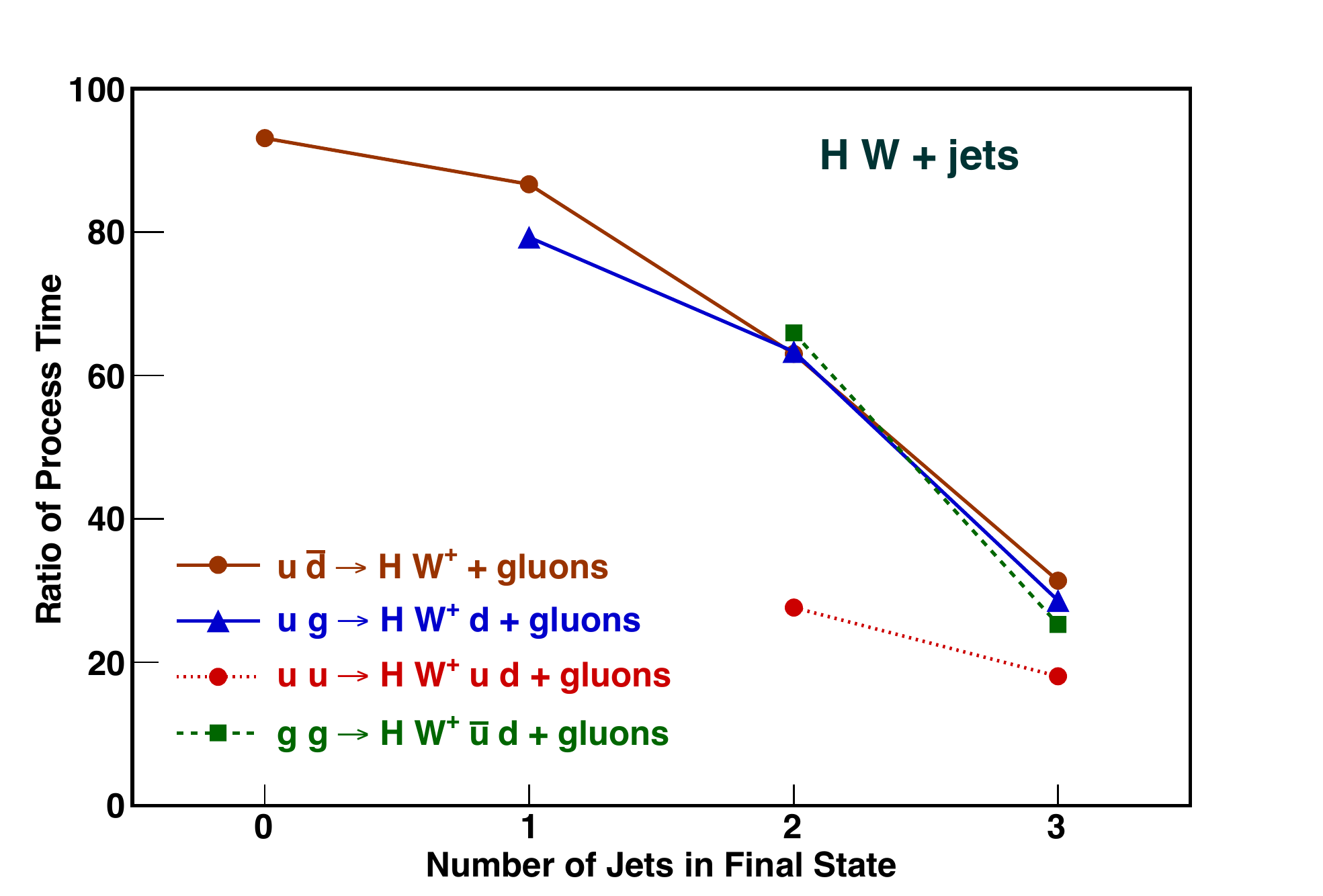}
     }
     \caption{Ratio of BASES process time (CPU/GPU) for $\HWp + n$-jet production with $\Wpdecay\,\lemu$ and $\Hdecay$ in $pp$ collisions at \LHCenergy\ for \Hmass\ and \HBr.
     Event selection cuts are given by Eqs.~(\ref{eq:jet-cuts-eta})-(\ref{eq:jet-cuts-ptjj}), (\ref{eq:bjet-cuts-eta})-(\ref{eq:bjet-cuts-pt}) and (\ref{eq:lepton-cuts-eta})-(\ref{eq:lepton-cuts-pt}) and the parton distributions of CTEQ6L1~\cite{cteq6} at the factorization scale of $Q\!=\!\pTjetcut\!=\!20\GeV$ is used, except for $n\!=\!0$ for which the factorization scale is chosen as $Q\!=\!\mHW$.
    The strong coupling is fixed at $\alphas(20\GeV)_{\mathrm{LO}}\!=\!0.171$.
     }
     \label{fig:time-HW}       
    \end{figure}

    \begin{table*}[htb]
     \centering
     \caption{Total cross sections and BASES process time ratios for $\HWp + n$-jet production with $\Wpdecay\,\lemu$ and $\Hdecay$ at the LHC (\LHCenergy), for \Hmass\ and \HBr.
     Event selection cuts are given by Eqs.~(\ref{eq:jet-cuts-eta})-(\ref{eq:jet-cuts-ptjj}), (\ref{eq:bjet-cuts-eta})-(\ref{eq:bjet-cuts-pt}) and (\ref{eq:lepton-cuts-eta})-(\ref{eq:lepton-cuts-pt}) and the parton distributions of CTEQ6L1~\cite{cteq6} at the factorization scale of $Q\!=\!\pTjetcut\!=\!20\GeV$ is used, except for $n\!=\!0$ for which the factorization scale is chosen as $Q\!=\!\mHW$.
    The strong coupling is fixed at $\alphas(20\GeV)_{\mathrm{LO}}\!=\!0.171$.
     }
     \label{tab:result-HW}       
     \begin{tabular}{c|c|cccl|c} \hline

     \multirow{2}{*}{$n$} & \multirow{2}{*}{Subprocess} & \multicolumn{4}{c|}{Cross section [fb]} &
      Process time ratio \\ \cline{3-7}
      &  & gBASES & BASES & MadGraph &
      & BASES/gBASES  \\ \hline

      0
      & $u\bar{d} \to \HWp$
      & $3.612 \pm 0.003$ & $3.614 \pm 0.003$ & $3.614 \pm 0.004$
      & $\times 10^{0}$
      & 93.0 \\ \hline

      \multirow{2}{*}{1}
      & $u\bar{d} \to \HWp + g$
      & $1.652 \pm 0.001$ & $1.653 \pm 0.001$ & $1.643 \pm 0.002$
      & $\times 10^{0}$
      & 86.7 \\ 

      & $ug \to\HWp + d$
      & $9.892 \pm 0.010$ & $9.901 \pm 0.010$ & $9.854 \pm 0.013$
      & $\times 10^{-1}$
      & 79.2 \\ \hline

      \multirow{4}{*}{2}
      & $u\bar{d} \to \HWp + gg$
      & $6.802 \pm 0.009$ & $6.804 \pm 0.008$ & $6.765 \pm 0.007$
      & $\times 10^{-1}$
      & 63.0 \\

      & $ug \to\HWp + dg$
      & $1.242 \pm 0.001$ & $1.244 \pm 0.002$ & $1.232 \pm 0.002$
      & $\times 10^{0}$
      & 63.3 \\

      & $uu \to \HWp + ud$
      & $1.824 \pm 0.001$ & $1.822 \pm 0.003$ & $1.805 \pm 0.001$
      & $\times 10^{-1}$
      & 27.7 \\

      & $gg \to\HWp + d\bar{u}$
      & $4.611 \pm 0.006$ & $4.614 \pm 0.004$ & $4.600 \pm 0.007$
      & $\times 10^{-2}$
      & 65.9 \\ \hline

      \multirow{4}{*}{3}
      & $u\bar{d} \to \HWp + ggg$
      & $2.660 \pm 0.010$ & $2.679 \pm 0.006$ & $2.604 \pm 0.003$
      & $\times 10^{-1}$
      & 31.4 \\

      & $ug \to\HWp + dgg$
      & $1.085 \pm 0.002$ & $1.084 \pm 0.001$ & $1.053 \pm 0.001$
      & $\times 10^{0}$
      & 28.6 \\

      & $uu \to \HWp + udg$
      & $1.915 \pm 0.003$ & $1.917 \pm 0.001$ & $1.835 \pm 0.002$
      & $\times 10^{-1}$
      & 18.0 \\

      & $gg \to\HWp + d\bar{u}g$
      & $5.906 \pm 0.007$ & $5.902 \pm 0.006$ & $5.818 \pm 0.007$
      & $\times 10^{-2}$
      & 25.3 \\ \hline

      \end{tabular}
    \end{table*}

    Results for $\HWp\!+n$-jet production with $\Wpdecay$ $(\ell\!=\!e,$ $\mu)$ and $\Hdecay$ are presented in Fig.~\ref{fig:time-HW} and Table \ref{tab:result-HW}.
    The SM Higgs boson of \Hmass\ and $\Br(\Hdecay)\!$ $=\!0.0405$ have been assumed, where $\taupm$ are treated the same as $\lpm$ \lemu, ignoring $\taupm$ decays.
    This allows us to estimate the cross section where the leptons and hadrons from $\tau$-decays are in the central detector region, $|\eta|\!<\!2.5$ in Eq.~(\ref{eq:lepton-cuts-eta}).

    The factorization scale of the parton distribution functions is set at $Q\!=\!\mHW$ for $n\!=\!0$, and at $Q\!=\!\pTjetcut\!=\!20\GeV$ for all the jet production processes $(n\!\geq\!1)$.
    The strong coupling is fixed at $\alphas(20\GeV)_{\mathrm{LO}}\!=\!0.171$.

    All of the cross sections in Table~\ref{tab:result-HW} are consistent between the GPU and the CPU computations within the statistical uncertainty of the Monte Carlo integrations.

     Again, the GPU gain over CPU of the total process time of BASES shows an $n$ dependence similar to those for $W\!+\!n\mbox{-jets}$ processes.
     This is because the $HW$ production in the SM can be regarded as $H$ emission from virtual $W$, in the lowest order of the electroweak couplings, and hence the number of Feynman diagrams are exactly the same between the $HW^{+}\!+\!n$-jets processes in Table~\ref{tab:result-HW}, and the corresponding $W^{+}\!+\!n$-jets processes in Table~\ref{tab:result-W}.
     This comparison confirms our observation that the GPU gain over CPU is limited mainly by the size of the amplitude function, at least in our application.
 

   \subsection{$Z$ boson associated Higgs production}

    \begin{figure}[htb]
     \centering
     \resizebox{0.48\textwidth}{!}{%
     \includegraphics{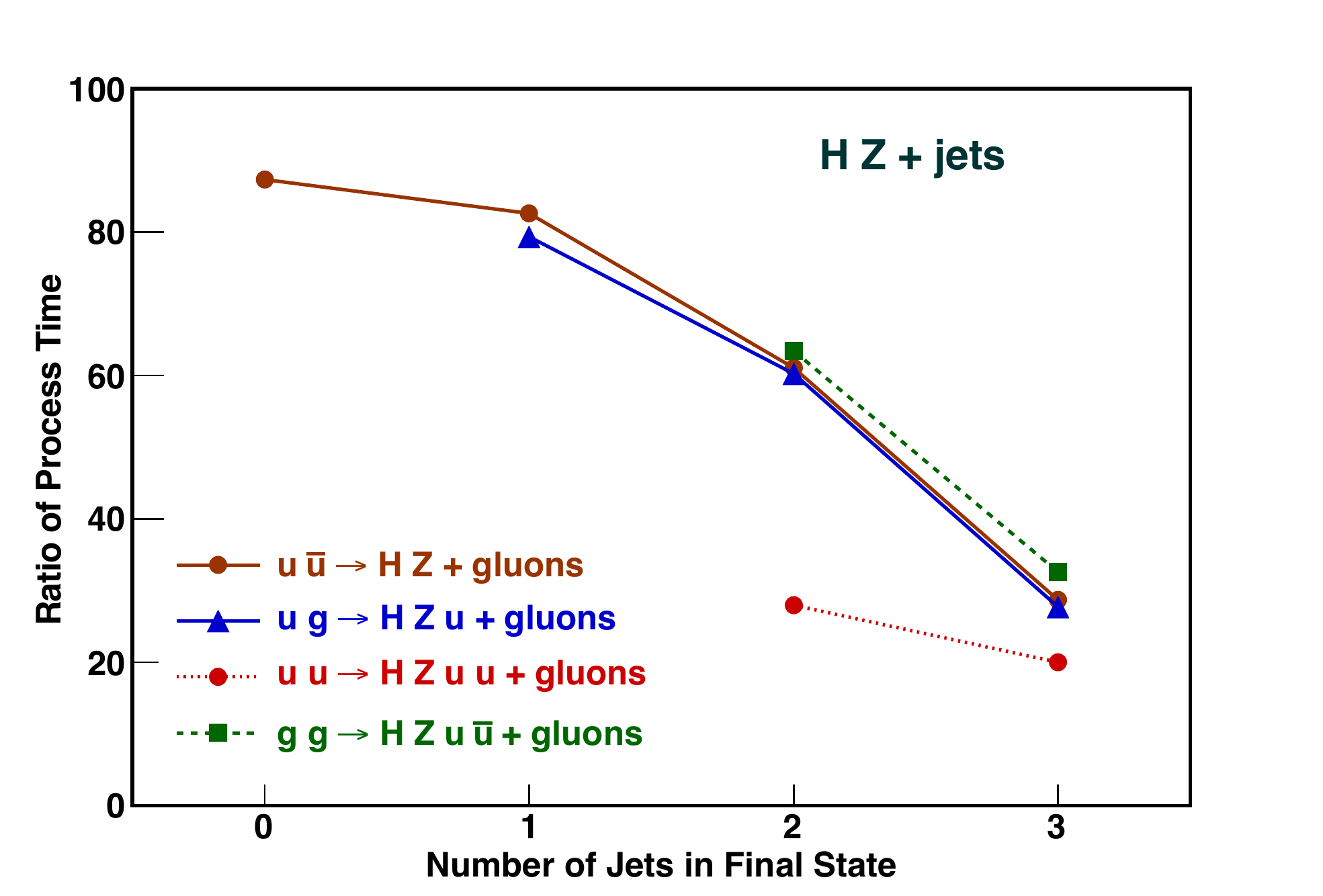}
     }
     \caption{Ratio of BASES process time (CPU/GPU) for $HZ\!+\!n$-jet production with $\Zdecay\,\lemu$ and $\Hdecay$ in $pp$ collisions at \LHCenergy\ for \Hmass\ and \HBr.
     Event selection cuts are given by Eqs.~(\ref{eq:jet-cuts-eta})-(\ref{eq:jet-cuts-ptjj}), (\ref{eq:bjet-cuts-eta})-(\ref{eq:bjet-cuts-pt}) and (\ref{eq:lepton-cuts-eta})-(\ref{eq:lepton-cuts-pt}) and the parton distributions of CTEQ6L1~\cite{cteq6} at the factorization scale of $Q\!=\!\pTjetcut\!=\!20\GeV$ is used, except for $n\!=\!0$ for which the factorization scale is chosen as $Q\!=\!\mHZ$.
    The strong coupling is fixed at $\alphas(20\GeV)_{\mathrm{LO}}\!=\!0.171$.
     }
     \label{fig:time-HZ}       
    \end{figure}

    \begin{table*}[htb]
     \centering
     \caption{Total cross sections and total process time for $HZ + n$-jet production with $\Zdecay\,\lemu$ and $\Hdecay$ at the LHC (\LHCenergy), for \Hmass\ and \HBr.
     Event selection cuts are given by Eqs.~(\ref{eq:jet-cuts-eta})-(\ref{eq:jet-cuts-ptjj}), (\ref{eq:bjet-cuts-eta})-(\ref{eq:bjet-cuts-pt}) and (\ref{eq:lepton-cuts-eta})-(\ref{eq:lepton-cuts-pt}) and the parton distributions of CTEQ6L1~\cite{cteq6} at the factorization scale of $Q\!=\!\pTjetcut\!=\!20\GeV$ is used, except for $n\!=\!0$ for which the factorization scale is chosen as $Q\!=\!\mHZ$.
    The strong coupling is fixed at $\alphas(20\GeV)_{\mathrm{LO}}\!=\!0.171$.
     }
     \label{tab:result-HZ}       
     \begin{tabular}{c|c|cccl|c} \hline
     \multirow{2}{*}{$n$} & \multirow{2}{*}{Subprocess} & \multicolumn{4}{c|}{Cross section [fb]} &
      Process time ratio \\ \cline{3-7}
      &  & gBASES & BASES & MadGraph &
      & BASES/gBASES  \\ \hline

      0
      & $u\bar{u} \to HZ$
      & $4.264 \pm 0.003$ & $4.274 \pm 0.003$ & $4.262 \pm 0.005$
      & $\times 10^{-1}$
      & 87.3 \\ \hline

      \multirow{2}{*}{1}
      & $u\bar{u} \to HZ + g$
      & $2.014 \pm 0.002$ & $2.016 \pm 0.002$ & $2.004 \pm 0.002$
      & $\times 10^{-1}$
      & 82.6 \\

      & $ug \to HZ + u$
      & $1.260 \pm 0.001$ & $1.257 \pm 0.001$ & $1.253 \pm 0.002$
      & $\times 10^{-1}$
      & 79.3 \\ \hline

      \multirow{4}{*}{2}
      & $u\bar{u} \to HZ + gg$
      & $8.390 \pm 0.006$ & $8.374 \pm 0.006$ & $8.332 \pm 0.009$
      & $\times 10^{-2}$
      & 61.1 \\

      & $ug \to HZ + ug$
      & $1.639 \pm 0.002$ & $1.640 \pm 0.002$ & $1.631 \pm 0.002$
      & $\times 10^{-1}$
      & 60.3 \\

      & $uu \to HZ + uu$
      & $1.004 \pm 0.001$ & $1.002 \pm 0.001$ & $9.908 \pm 0.012$
      & $\times 10^{-3}$
      & 28.0 \\

      & $gg \to HZ + u\bar{u}$
      & $6.417 \pm 0.004$ & $6.411 \pm 0.004$ & $6.391 \pm 0.009$
      & $\times 10^{-3}$
      & 63.4 \\ \hline

      \multirow{4}{*}{3}
      & $u\bar{u} \to HZ + ggg$
      & $3.252 \pm 0.003$ & $3.261 \pm 0.001$ & $3.154 \pm 0.003$
      & $\times 10^{-2}$
      & 28.7 \\

      & $ug \to HZ + ugg$
      & $1.472 \pm 0.001$ & $1.471 \pm 0.001$ & $1.434 \pm 0.002$
      & $\times 10^{-1}$
      & 27.7 \\ 

      & $uu \to HZ + uug$
      & $1.959 \pm 0.002$ & $1.953 \pm 0.003$ & $1.900 \pm 0.002$
      & $\times 10^{-2}$
      & 20.0 \\ 

      & $gg \to HZ + u\bar{u}g$
      & $8.339 \pm 0.003$ & $8.312 \pm 0.009$ & $8.206 \pm 0.009$
      & $\times 10^{-3}$
      & 32.6 \\ \hline

      \end{tabular}
    \end{table*}

   Results for $HZ\!+\!n$-jet production with $\Zdecay$ $(\ell\!=\!e,$ $\mu)$ and \Hdecay\ for \Hmass\ are presented in Fig.~\ref{fig:time-HZ} and Table \ref{tab:result-HZ}.
    The factorization scale of the parton distribution functions is set at $Q\!=\!\mHZ$ for $n\!=\!0$, and at $Q\!=\!\pTjetcut\!=\!20\GeV$ for all the jet production processes $(n\!\geq\!1)$.
    The strong coupling is fixed at $\alphas(20\GeV)_{\mathrm{LO}}\!=\!0.171$.
   Both the cross sections shown in Table~\ref{tab:result-HZ} and the GPU gain over CPU shown in Fig.~\ref{fig:time-HZ} are similar to those found for the $HW\!+\!n\mbox{-jet}$ processes in the previous section.
   These gains  are also consistent with those for $Z\!+\!n\mbox{-jets}$.
   As in the above case, all the Feynman diagrams for the $HZ\!+\!n$-jets processes obtained from those of the $Z\!+\!n$-jets process by allowing the external $Z$ boson to split into $Z\!+\!H$.
   The size of the amplitude functions are hence essentially the same between the two corresponding processes.
   Accordingly, the gain factors in Table~\ref{tab:result-HZ} for the $HZ\!+\!n$-jets processes are almost the same as those of the corresponding $Z\!+\!n$-jets processes in Table~\ref{tab:result-Z}.


   \subsection{Top quark associated Higgs production}

    \begin{figure}[htb]
     \centering
     \resizebox{0.48\textwidth}{!}{%
     \includegraphics{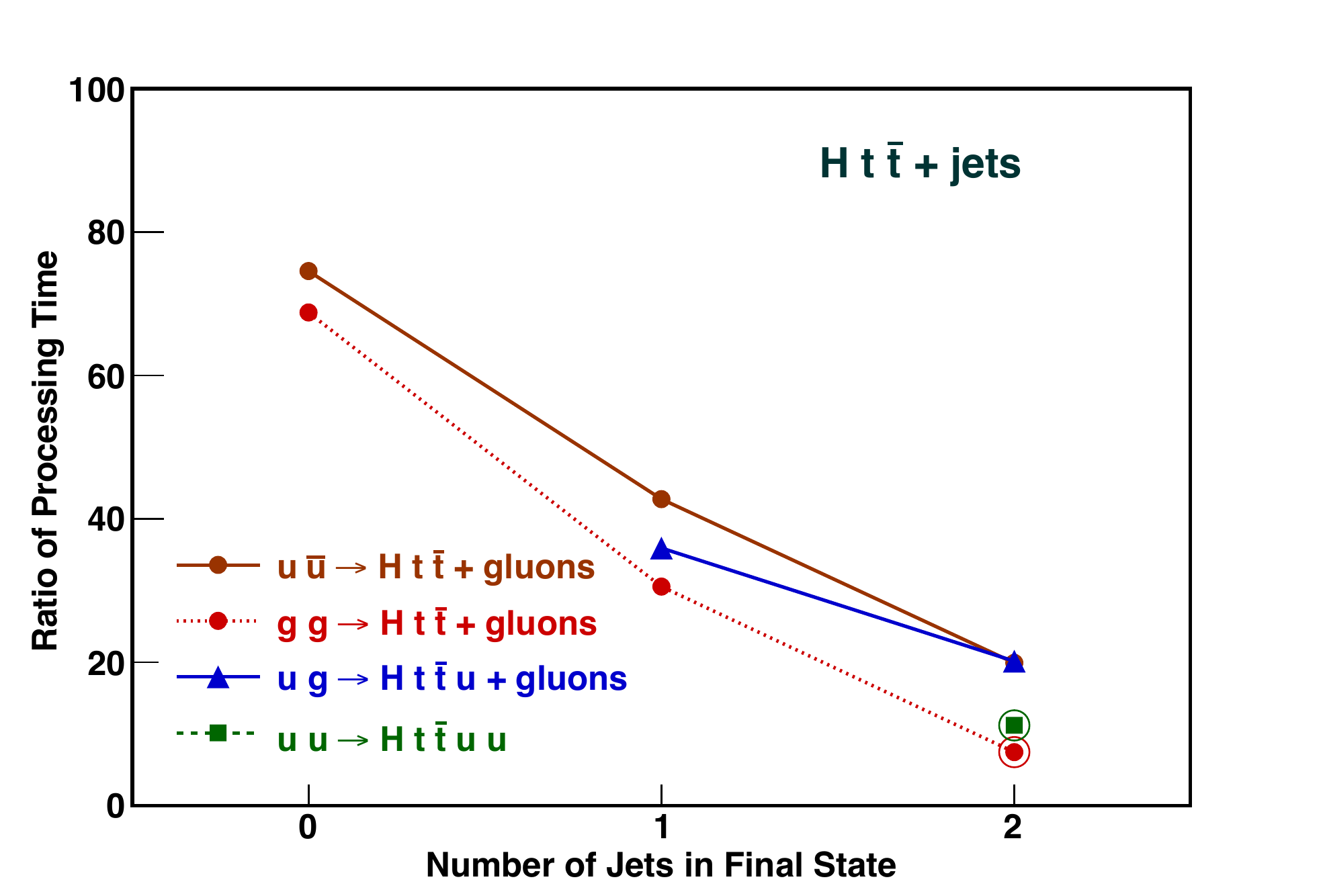}
     }
     \caption{Ratio of BASES process time (CPU/GPU) for $H\ttbar + n$-jet production with $\tdecay$ and $\tbardecay\,\,\lemu$ for \tmass\ and \tBr\ and with $\Hdecay$ in $pp$ collisions at \LHCenergy\ for \Hmass\ and \HBr.
     Event selection cuts are given by Eqs.~(\ref{eq:jet-cuts-eta})-(\ref{eq:jet-cuts-ptjj}), (\ref{eq:bjet-cuts-eta})-(\ref{eq:bjet-cuts-pt}) and (\ref{eq:lepton-cuts-eta})-(\ref{eq:lepton-cuts-pt}) and the parton distributions of CTEQ6L1~\cite{cteq6} at the factorization scale of $Q\!=\!\pTjetcut\!=\!20\GeV$ is used, except for $n\!=\!0$ for which the factorization scale is chosen as $Q\!=\!\mHtt$ for $u\ubar\!\rightarrow\!H\ttbar$ and $Q\!=\!\mt$ for $gg\!\rightarrow\!H\ttbar$. 
     The strong coupling constants are set as $\alpha_{\mathrm{s}}^{2\!+\!n}\!=\!\alphas(\mHtt)_{\mathrm{LO}}^{2}\,\alphas(\pTjetcut)_{\mathrm{LO}}^{n}$ for $u\ubar\!\rightarrow\!H\ttbar$ process, and $\alpha_{\mathrm{s}}^{2\!+\!n}\!=\!\alphas(\mt)_{\mathrm{LO}}^{2}\,\alphas(\pTjetcut)_{\mathrm{LO}}^{n}$ for all the others, with $\alphas(\mt)_{\mathrm{LO}}\!=\!0.108$ and $\alphas(20\GeV)_{\mathrm{LO}}\!=\!0.171$.
     }
     \label{fig:time-Htt}       
    \end{figure}

    \begin{table*}[htb]
     \centering
     \caption{Total cross sections and BASES process time ratios for $H\ttbar + n$-jet production with $\tdecay$ and $\tbardecay\,\,\lemu$ and $\Hdecay$ at the LHC (\LHCenergy).
     Event selection cuts are given by Eqs.~(\ref{eq:jet-cuts-eta})-(\ref{eq:jet-cuts-ptjj}), (\ref{eq:bjet-cuts-eta})-(\ref{eq:bjet-cuts-pt}) and (\ref{eq:lepton-cuts-eta})-(\ref{eq:lepton-cuts-pt}) and the parton distributions of CTEQ6L1~\cite{cteq6} at the factorization scale of $Q\!=\!\pTjetcut\!=\!20\GeV$ is used, except for $n\!=\!0$ for which the factorization scale is chosen as $Q\!=\!\mHtt$ for $u\ubar\!\rightarrow\!H\ttbar$ and $Q\!=\!\mt$ for $gg\!\rightarrow\!H\ttbar$. 
     The strong coupling constants are set as $\alpha_{\mathrm{s}}^{2\!+\!n}\!=\!\alphas(\mHtt)_{\mathrm{LO}}^{2}\,\alphas(\pTjetcut)_{\mathrm{LO}}^{n}$ for $u\ubar\!\rightarrow\!H\ttbar$ process, and $\alpha_{\mathrm{s}}^{2\!+\!n}\!=\!\alphas(\mt)_{\mathrm{LO}}^{2}\,\alphas(\pTjetcut)_{\mathrm{LO}}^{n}$ for all the others, with $\alphas(\mt)_{\mathrm{LO}}\!=\!0.108$ and $\alphas(20\GeV)_{\mathrm{LO}}\!=\!0.171$.
     }
     \label{tab:result-Htt}       
     \begin{tabular}{c|c|cccl|c} \hline
     \multirow{2}{*}{$n$} & \multirow{2}{*}{Subprocess} & \multicolumn{4}{c|}{Cross section [fb]} &
      Process time ratio \\ \cline{3-7}
      &  & gBASES & BASES & MadGraph &
      & BASES/gBASES  \\ \hline

      \multirow{2}{*}{0}
      & $u\bar{u} \to H\ttbar$
      & $5.281 \pm 0.005$ & $5.264 \pm 0.005$ & $5.272 \pm 0.009$
      & $\times 10^{-2}$
      & 74.6 \\

      & $gg \to H\ttbar$ 
      & $3.802 \pm 0.004$ & $3.798 \pm 0.004$ & $3.807 \pm 0.006$
      & $\times 10^{-1}$
      & 68.8 \\ \hline

      \multirow{3}{*}{1}
      & $u\bar{u} \to H\ttbar + g$
      & $5.424 \pm 0.002$ & $5.420 \pm 0.003$ & $5.345 \pm 0.007$
      & $\times 10^{-2}$
      & 42.7 \\

      & $gg \to H\ttbar + g$
      & $1.202 \pm 0.001$ & $1.201 \pm 0.001$ & $1.193 \pm 0.002$
      & $\times 10^{0}$
      & 30.6 \\

      & $ug \to H\ttbar + u$
      & $2.328 \pm 0.002$ & $2.323 \pm 0.001$ & $2.312 \pm 0.004$
      & $\times 10^{-1}$
      & 36.0 \\ \hline

      \multirow{4}{*}{2}
      & $u\bar{u} \to H\ttbar + gg$
      & $3.574 \pm 0.007$ & $3.574 \pm 0.001$ & $3.238 \pm 0.015$
      & $\times 10^{-2}$
      & 20.0 \\

      & $gg \to H\ttbar + gg$
      & $1.528 \pm 0.006$ & $1.525 \pm 0.006$ & $1.480 \pm 0.002$
      & $\times 10^{0}$
      &  7.5* \\

      & $ug \to H\ttbar + ug$
      & $5.815 \pm 0.004$ & $5.808 \pm 0.006$ & $5.529 \pm 0.038$
      & $\times 10^{-1}$
      & 20.2 \\ 

      & $uu \to H\ttbar + uu$
      & $2.473 \pm 0.003$ & $2.472 \pm 0.001$ & $2.314 \pm 0.027$
      & $\times 10^{-2}$
      & 11.2* \\ \hline

      \end{tabular}
    \end{table*}

   Results for $H\ttbar + n$-jet production with $\tdecay$ and $\tbar\!\rightarrow\!\bbar\,\lm\nulbar$ $\lemu$ and $\Hdecay$ are presented in Fig.~\ref{fig:time-Htt} and Table \ref{tab:result-Htt}.
   Both $t$ and \tbar\ decay semi-leptonically with $\Br(t\!\rightarrow\!b\,\lp\nul)\!=\!0.216$, and $\Hdecay$ decay with $\Br(H\rightarrow\tauptaum)\!=\!0.0405$ are assumed, and $\taupm$ are treated as $\ell=e\mbox{ or }\mu$, satisfying $|\eta_{\tau}|\!<\!2.5$ and $p_{\mathrm{T}\tau}\!>\!20\GeV$ in Eqs.~(\ref{eq:lepton-cuts-eta})-(\ref{eq:lepton-cuts-pt}), ignoring further $\tau$ decays.
   The factorization scale is chosen as $Q\!=\!\mHtt$ for $u\ubar\!\rightarrow\!H\ttbar\,(n\!=\!0)$ and $Q\!=\!\mt$ for $gg\!\rightarrow\!H\ttbar\,(n\!=\!0)$, while $Q\!=\!\pTjetcut\!=\!20\GeV$ for all the processes with jet productions $(n\!\geq\!1)$.
   The strong coupling constants are set as $\alpha_{\mathrm{s}}^{2\!+\!n}\!=\!\alphas(\mHtt)_{\mathrm{LO}}^{2}\,\alphas(\pTjetcut)_{\mathrm{LO}}^{n}$ for $u\ubar\!\rightarrow\!H\ttbar$ process, and $\alpha_{\mathrm{s}}^{2\!+\!n}\!=\!\alphas(\mt)_{\mathrm{LO}}^{2}\,\alphas(\pTjetcut)_{\mathrm{LO}}^{n}$ for all the others.
   Numerical values are $\alphas(\mt)_{\mathrm{LO}}\!=\!0.108$ and $\alphas(20\GeV)_{\mathrm{LO}}\!=\!0.171$.

   All of the cross sections in Table~\ref{tab:result-Htt} obtained by the GPU and the CPU programs as well as those by MadGraph are consistent within the statistical uncertainty of the Monte Carlo integration.
   The GPU gain over CPU for the total process time of BASES decreases from $\sim\!70$ to $\sim\!20$ as $n$ grows from $n\!=\!0$ to $n\!=\!2$ except for the $n\!=\!2$ processes $gg\rightarrow H\ttbar gg$ and $uu\rightarrow H\ttbar uu$ for which the factor is reduced to $\sim 10$ or less.
   The amplitude functions for these two processes are too long for the CUDA compiler to process, and they are divided into smaller pieces for execution.


   \subsection{Higgs boson production via weak-boson fusion}

    \begin{figure}[htb]
     \centering
     \resizebox{0.48\textwidth}{!}{%
     \includegraphics{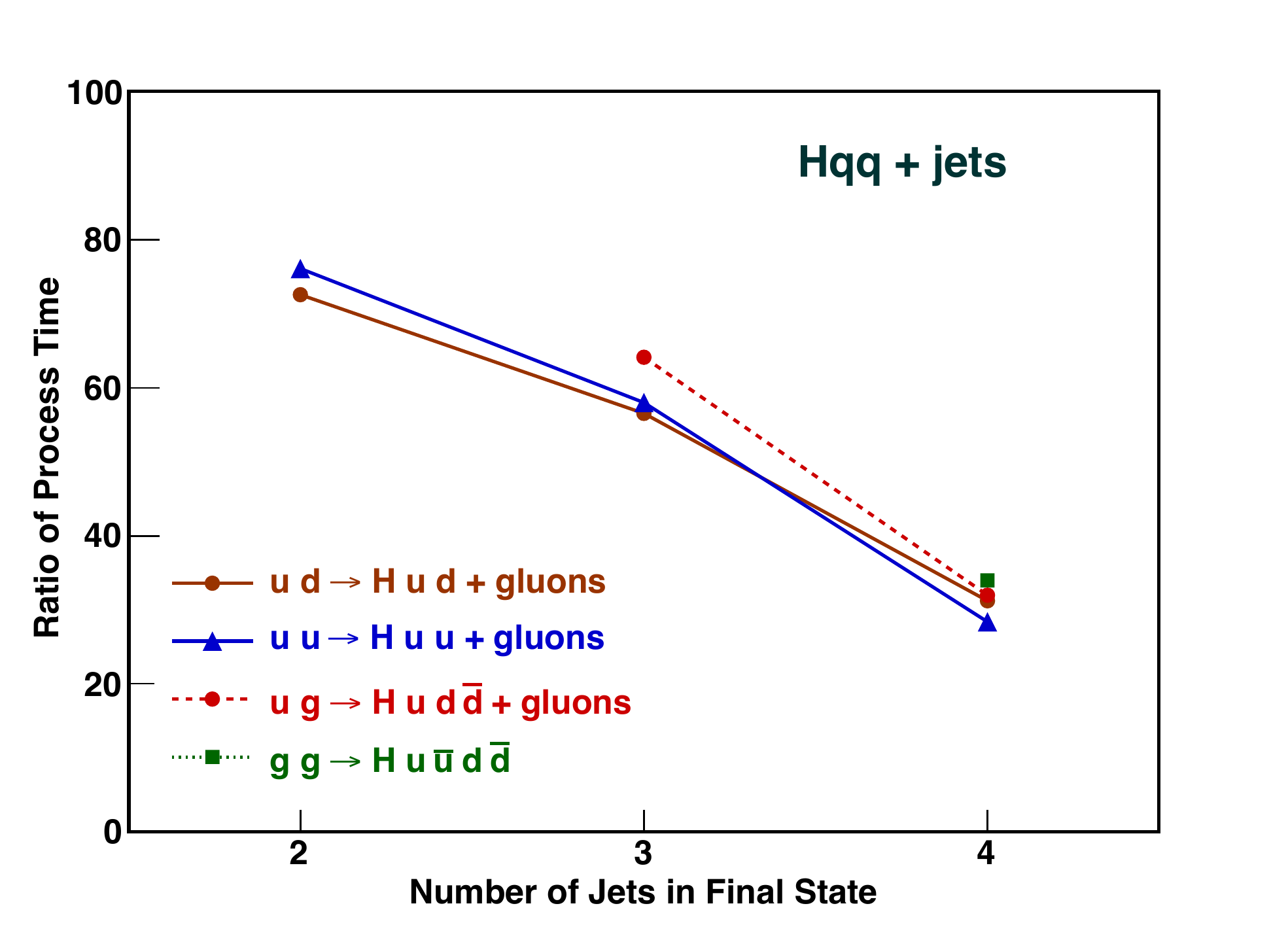}
     }
     \caption{Ratio of BASES process time (CPU/GPU) for Higgs boson plus $n$-jet production via weak-boson fusion with $\Hdecay$ in $pp$ collisions at \LHCenergy\ for \Hmass\ and \HBr.
     Event selection cuts are given by Eqs.~(\ref{eq:jet-cuts-eta})-(\ref{eq:jet-cuts-ptjj}), (\ref{eq:bjet-cuts-eta})-(\ref{eq:bjet-cuts-pt}) and (\ref{eq:lepton-cuts-eta})-(\ref{eq:lepton-cuts-pt}) and the parton distributions of CTEQ6L1~\cite{cteq6} at the factorization scale of $Q\!=\!\pTjetcut\!=\!20\GeV$ is used, except for $n\!\leq\!2$ for which the factorization scale is chosen as $Q\!=\!\mW/2$ for $ud\!\rightarrow\!Hud$, $Q\!=\!\mZ/2$ for $uu\!\rightarrow\!Huu$.
    The strong coupling is fixed at $\alphas(20\GeV)_{\mathrm{LO}}\!=\!0.171$.
     }
     \label{fig:time-Hqq}       
    \end{figure}

    \begin{table*}[htb]
     \caption{Total cross sections and BASES process time ratios for Higgs boson plus $n$-jet production via weak-boson fusion with $\Hdecay$ at the LHC (\LHCenergy).
     Event selection cuts are given by Eqs.~(\ref{eq:jet-cuts-eta})-(\ref{eq:jet-cuts-ptjj}), (\ref{eq:bjet-cuts-eta})-(\ref{eq:bjet-cuts-pt}) and (\ref{eq:lepton-cuts-eta})-(\ref{eq:lepton-cuts-pt}) and the parton distributions of CTEQ6L1~\cite{cteq6} at the factorization scale of $Q\!=\!\pTjetcut\!=\!20\GeV$ is used, except for $n\!\leq\!2$ for which the factorization scale is chosen as $Q\!=\!\mW/2$ for $ud\!\rightarrow\!Hud$, $Q\!=\!\mZ/2$ for $uu\!\rightarrow\!Huu$.
    The strong coupling is fixed at $\alphas(20\GeV)_{\mathrm{LO}}\!=\!0.171$.
     }
     \label{tab:result-Hqq}       
     \centering
     \begin{tabular}{c|c|cccl|c} \hline
     \multirow{2}{*}{$n$} & \multirow{2}{*}{Subprocess} & \multicolumn{4}{c|}{Cross section [fb]} &
      Process time ratio \\ \cline{3-7}
      &  & gBASES & BASES & MadGraph &
      & BASES/gBASES  \\ \hline

      \multirow{2}{*}{2}
      & $ud \to H+ud$
      & $4.414 \pm 0.004$ & $4.408 \pm 0.004$ & $4.406 \pm 0.007$
      & $\times 10^{1}$
      & 72.6 \\ 

      & $uu \to H+uu$
      & $4.354 \pm 0.005$ & $4.345 \pm 0.005$ & $4.338 \pm 0.007$
      & $\times 10^{0}$
      & 76.1 \\ \hline

      \multirow{3}{*}{3}
      & $ud \to H+ud + g$
      & $1.485 \pm 0.001$ & $1.486 \pm 0.002$ & $1.471 \pm 0.002$
      & $\times 10^{1}$
      & 56.6 \\ 

      & $uu \to H+uu + g$
      & $1.725 \pm 0.002$ & $1.726 \pm 0.003$ & $1.711 \pm 0.002$
      & $\times 10^{0}$
      & 58.0 \\ 

      & $ug \to H+ud + \bar{d}$ 
      & $4.573 \pm 0.004$ & $4.564 \pm 0.004$ & $4.508 \pm 0.006$
      & $\times 10^{0}$
      & 64.2 \\ \hline

      \multirow{4}{*}{4}
      & $ud \to H+ud + gg$
      & $3.808 \pm 0.011$ & $3.778 \pm 0.002$ & $3.573 \pm 0.004$
      & $\times 10^{0}$
      & 31.2 \\ 

      & $uu \to H+uu + gg$
      & $5.000 \pm 0.003$ & $5.031 \pm 0.018$ & $4.712 \pm 0.005$
      & $\times 10^{-1}$
      & 28.4 \\ 

      & $ug \to H+ud + \bar{d}g$
      & $3.631 \pm 0.008$ & $3.625 \pm 0.008$ & $3.423 \pm 0.004$
      & $\times 10^{0}$
      & 32.0 \\ 

      & $gg \to H+u\bar{u} + d\bar{d}$
      & $1.772 \pm 0.001$ & $1.773 \pm 0.002$ & $1.729 \pm 0.002$
      & $\times 10^{-1}$
      & 34.0 \\ \hline

      \end{tabular}
    \end{table*}

    Results for Higgs boson plus $n$-jet production via weak-boson fusion followed by $\Hdecay$ decay are presented in Fig.~\ref{fig:time-Hqq} and Table \ref{tab:result-Hqq}.
    The factorization scale of the parton distribution functions is set at $Q\!=\!\mW/2$ for $ud\!\rightarrow\!Hud$, $Q\!=\!\mZ/2$ for $uu\!\rightarrow\!Huu$, and $Q\!=\!\pTjetcut\!=\!20\GeV$ for all the other processes with $n\!\geq\!3$ jets.
    The above choice of factorization scales are motivated by the observation that the peak positions of the distribution of the transverse momentum of the two jets in the processes $ud\!\rightarrow\!Hud$ and $uu\!\rightarrow\!Huu$ are found to be $<\!\pT\!>\!\sim\! 40\GeV$ and $45\GeV$, respectively.
    The strong coupling is fixed at $\alphas(20\GeV)_{\mathrm{LO}}\!=\!0.171$.

   All of the cross sections in Table~\ref{tab:result-Hqq} obtained by the GPU and the CPU BASES programs as well as those by MadGraph are consistent within the statistical uncertainty of the Monte Carlo integration/event generation.
   The GPU gain over CPU for the total process time of BASES is $~\sim\!70$ for $n\!=\!2$.
   It is still around 30 even for the $n\!=\!4$ case.


  \subsection{Multiple Higgs bosons production via weak boson fusion}

    \begin{figure}[htb]
     \centering
     \resizebox{0.48\textwidth}{!}{%
     \includegraphics{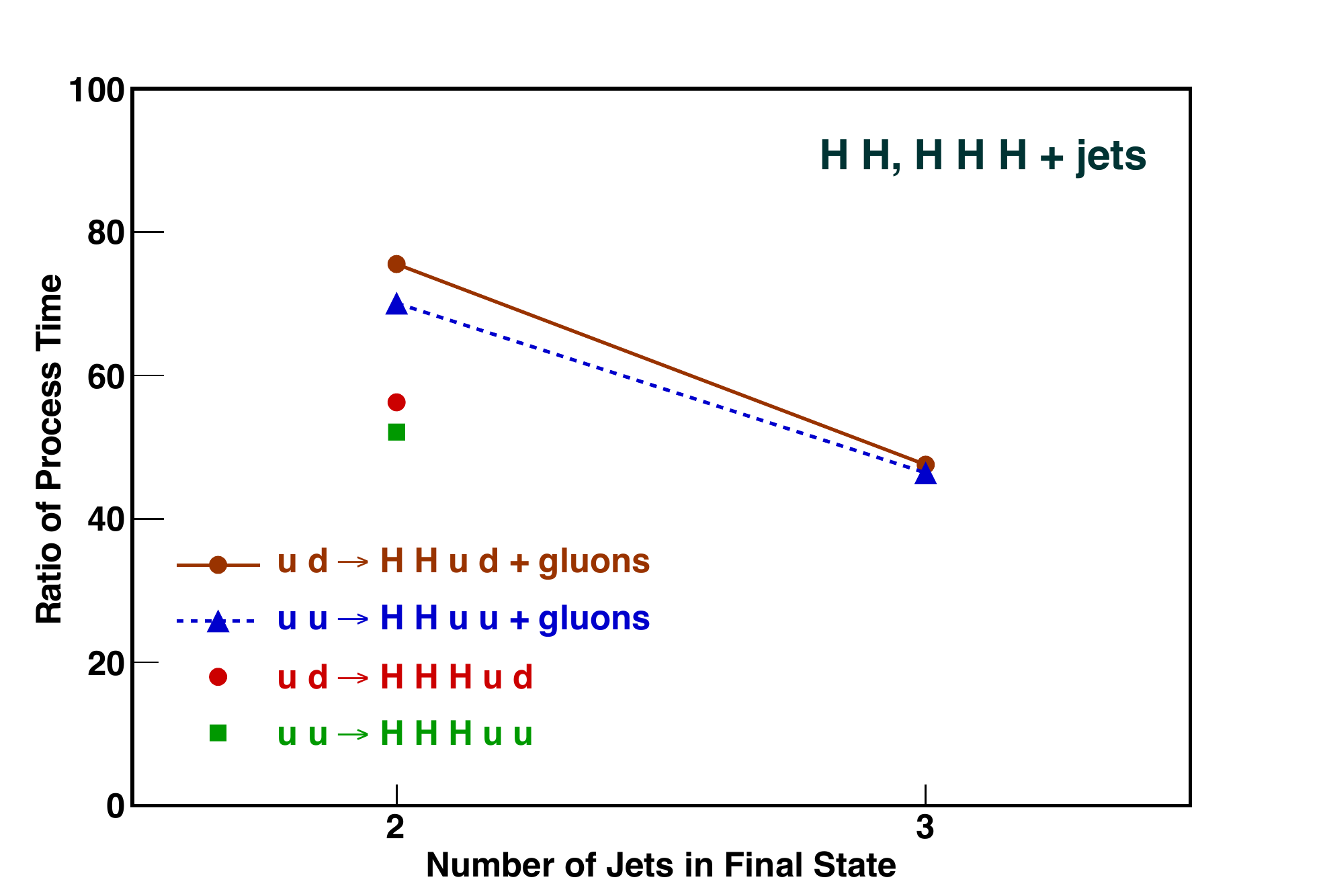}
     }
     \caption{Ratio of BASES process time (CPU/GPU) for multiple Higgs boson plus $n$-jet  production via weak boson fusion with \Hdecay\ in $pp$ collisions at \LHCenergy\ for \Hmass\ and \HBr.
     Event selection cuts are given by Eqs.~(\ref{eq:jet-cuts-eta})-(\ref{eq:jet-cuts-ptjj}), (\ref{eq:bjet-cuts-eta})-(\ref{eq:bjet-cuts-pt}) and (\ref{eq:lepton-cuts-eta})-(\ref{eq:lepton-cuts-pt}) and the parton distributions of CTEQ6L1~\cite{cteq6} at the factorization scale of $Q\!=\!\pTjetcut\!=\!20\GeV$ is used, except for $n\!\leq\!2$ for which the factorization scale is chosen as $Q\!=\mW/2$ for $ud\!\rightarrow\!HH(H)ud$, $Q\!=\mZ/2$ for $uu\!\rightarrow\!HH(H)u$.
    The strong coupling is fixed at $\alphas(20\GeV)_{\mathrm{LO}}\!=\!0.171$.
     }
     \label{fig:time-HHqq}       
    \end{figure}

    \begin{table*}[htb]
     \centering
     \caption{Total cross sections and BASES process time ratios for multiple Higgs boson plus $n$-jet production via weak boson fusion with $\Hdecay$ at the LHC (\LHCenergy).
     Event selection cuts are given by Eqs.~(\ref{eq:jet-cuts-eta})-(\ref{eq:jet-cuts-ptjj}), (\ref{eq:bjet-cuts-eta})-(\ref{eq:bjet-cuts-pt}) and (\ref{eq:lepton-cuts-eta})-(\ref{eq:lepton-cuts-pt}) and the parton distributions of CTEQ6L1~\cite{cteq6} at the factorization scale of $Q\!=\!\pTjetcut\!=\!20\GeV$ is used, except for $n\!\leq\!2$ for which the factorization scale is chosen as $Q\!=\mW/2$ for $ud\!\rightarrow\!HH(H)ud$, $Q\!=\mZ/2$ for $uu\!\rightarrow\!HH(H)u$.
    The strong coupling is fixed at $\alphas(20\GeV)_{\mathrm{LO}}\!=\!0.171$.
     }
     \label{tab:result-HHqq}       
     \begin{tabular}{c|c|cccl|c} \hline
     \multirow{2}{*}{$n$} & \multirow{2}{*}{Subprocess} & \multicolumn{4}{c|}{Cross section [fb]} &
      Process time ratio \\ \cline{3-7}
      &  & gBASES & BASES & MadGraph &
      & BASES/gBASES  \\ \hline
      
      \multirow{4}{*}{2}
      & $ud \to HH + ud$
      & $1.092 \pm 0.001$ & $1.092 \pm 0.001$ & $1.050 \pm 0.001$
      & $\times 10^{-3}$
      & 75.5 \\ 

      & $uu \to HH + uu$
      & $1.596 \pm 0.002$ & $1.593 \pm 0.002$ & $1.536 \pm 0.002$
      & $\times 10^{-4}$
      & 70.1 \\ 

      & $ud \to HHH + ud$
      & $2.718 \pm 0.001$ & $2.714 \pm 0.004$ & $2.686 \pm 0.004$
      & $\times 10^{-7}$
      & 56.3 \\ 

      & $uu \to HHH + uu$
      & $4.595 \pm 0.006$ & $4.602 \pm 0.007$ & $4.554 \pm 0.006$
      & $\times 10^{-8}$
      & 52.1 \\ \hline

      \multirow{2}{*}{3}
      & $ud \to HH + ud + g$
      & $3.964 \pm 0.003$ & $3.945 \pm 0.002$ & $3.696 \pm 0.004$
      & $\times 10^{-4}$
      & 47.6 \\

      & $uu \to HH + uu + g$
      & $6.521 \pm 0.006$ & $6.505 \pm 0.017$ & $6.169 \pm 0.006$
      & $\times 10^{-5}$
      & 46.4 \\ \hline

     \end{tabular}
    \end{table*}
    
    Results for multiple Higgs boson plus $n$-jet production via weak boson fusion followed by $\Hdecay$ decay are presented in Fig.~\ref{fig:time-HHqq} and Table \ref{tab:result-HHqq}.
    The factorization scale of the parton distribution functions is set at $Q\!=\mW/2$ for $ud\!\rightarrow\!HH(H)ud$, $Q\!=\mZ/2$ for $uu\!\rightarrow\!HH(H)uu$, and $Q\!=\!\pTjetcut\!=\!20\GeV$ for all processes.
    The strong coupling is fixed at $\alphas(20\GeV)_{\mathrm{LO}}\!=\!0.171$.

   All of the cross sections in Table~\ref{tab:result-Hqq} obtained by the GPU and the CPU BASES programs as well as those by MadGraph are consistent within the statistical uncertainty of the Monte Carlo integration/event generation.
   For double Higgs boson production processes the GPU gain over CPU for the total process time of BASES is $\sim\!70$ for $n\!=\!2$ and $\sim\!50$ for $n\!=\!3$.
   For triple Higgs boson production processes the GPU gain over CPU for the total process time of BASES is $\sim\!50$ for $n\!=\!2$.
   We study triple Higgs boson production in weak boson fusion to test the 4 scalar boson coupling function, \texttt{hsssxx.cu} (List~\ref{list:hsssxx} in \ref{sec:new-heget-codes}), even though the total cross section obtained by $\Br(H\!\rightarrow\!\tauptaum)$ is less than 0.01fb in Table~\ref{tab:result-HHqq}.
   
   It should be noted here again that we first encountered significant discrepancy of more than 100\% level for $HHH$ production processes.
   This is because subtle gauge theory cancellation among weak boson scattering amplitudes $W^{*}W^{*}$, $Z^{*}Z^{*}\!\rightarrow\!HZ^{*} (Z^{*}\!\rightarrow HH)$ can be violated significantly when the $m_{V} (V=W, Z)$ appearing in the propagators and in the $W$ vertices do not satisfy the tree level relations
   \begin{equation}
    m_{W}/m_{Z} = J_{W}/J_{Z}
   \end{equation}
   of the SM.
   The exact agreements among all the programs have been obtained after we replace all $m_{V}$'s in the couplings and and in the propagators as
   \begin{equation}
    m_{V}^{2} \rightarrow m_{V}^{2} - i\,m_{V}\Gamma_{V},
   \end{equation}
   and by imposing
   \begin{equation}
    m_{W}^{2} - i\,m_{W}\Gamma_W = \frac{g_{W}^{2}}{g_{Z}^{2}} (m_{Z}^2-i\,m_{Z}\Gamma_{Z}).
   \end{equation}
   Here $g_{Z}^{2}\!=\!c_{W}^{2}g_{W}^{2}\!=\!c_{W}^{2}s_{W}^{2}e^{2}$ at $e^{2}\!=\!4\pi\,\alpha(m_{Z})_{\MSbar}\!=\!1/128.9$, $s_{W}^2\!=\!1\!-\!c_{W}^{2}\!=\!\sin^{2}\theta_{W}(\mZ)_{\MSbar}\!=\!0.2312$, $\mZ\!=\!91.188\GeV$ and $\Gamma_{Z}\!=\!2.4952\GeV$ are used as inputs.

   \section{Summary}
   \label{sec:summary}

   We have shown the results of our attempt to extend the HEGET function package originally developed for the computations of QED and QCD processes~\cite{qed-paper,qcd-paper} on the GPU to all Standard Model (SM) processes.
   Additional HEGET functions to compute all SM amplitudes are introduced, and are listed in \ref{sec:new-heget-codes}.
   We have tested all of the functions by comparing amplitudes and cross sections of multi-jet processes associated with the production of single and double weak bosons, a top-quark pair, Higgs boson plus a weak boson or a \ttbar\ pair, as well as multiple Higgs bosons produced via weak-boson fusion, where the heavy particles $(W, Z, t, \tbar, H)$ are allowed to decay into leptons with full spin correlations.
   Based on the GPU version of the Monte Carlo integration program, gBASES~\cite{mcinteg}, we compute cross sections for all of these processes at the LHC target energy of \LHCenergy.\ \ 
   The program, MG2CUDA, has been developed to convert arbitrary MadGraph generated HELAS amplitude subroutines written in FORTRAN into HEGET codes in CUDA for the general purpose computing on GPU.
   As references for the computation of cross sections on GPU, we have performed two types of computations on CPU.
   One is based on the FORTRAN version of the BASES program whose performance is also compared with gBASES.
   Another comparison is performed with respect to the new version of MadGraph~\cite{madgraph5}.
   We find that our Monte Carlo integration program with the new HEGET functions run quite fast in the GPU environment of a TESLA C2075, as compared with the FORTRAN version of BASES programs for the same physics processes with the same integration parameters.

   Our achievements and findings may be summarized as follows.
   \begin{itemize}
    \item A set of new HEGET functions in CUDA has been developed for the general computation of amplitudes and cross sections for the SM processes.
    \item For processes with larger number of jets in the final state, we find that total cross sections obtained with MadGraph is somewhat smaller than those computed with BASES programs.
	  The difference amounts to about 5-10\% for the processes with large number of jets in the final state.
	  They might be due to the difference in the phase space generation part of the program and require further studies~\footnote{Effort to identify the source of this bias is ongoing and phase space generations for multi-parton final states are being checked by comparing generated results between different programs}.
    \item For processes with a simple final state, i.e. the number of jets in the finals state, $n$, is equal to 0, the gain of the process time of the total BASES program of GPU over those of CPU becomes nearly 100 and gradually deceases as $n$ becomes large.
    \item We use double precision version of gBASES contrary to our previous papers~\cite{qed-paper,qcd-paper}.
	  The double precision version of gBASES shows not only a good performance in process time but also good stability for various physics processes with wide range of integration parameters.
    \item Due to the limitation of the CUDA compiler, a long CUDA function cannot be compiled.
	  We have included the new mechanism to gBASES to handle successive CUDA functions calls in order to handle a long amplitude program as a set of small CUDA functions.
	  The integrated results also agree very well with those obtained by BASES in FORTRAN.
	  The performance of these divided programs is somewhat lower compared with un-divided programs.
   \end{itemize}

   \begin{acknowledgement}
    This work is supported by the Grant-in-Aid for Scientific Research from the Japan Society for the Promotion of Science (No.20340064 and No.22740155), in part by the National Natural Science Foundation of China under Grants No. 11205008 as well as a grant from the National Science Foundation (NSF PHY-1068326).
    We wish to thank Dave Rainwater for inspiring us with the idea of using GPU for fast LHC physics simulation and his initial contribution to the project.
   \end{acknowledgement}

   \appendix
   \numberwithin{equation}{section}

   \def\thesection{Appendix \Alph{section}}
   \section{Phase space parameterization}
   \def\thesection{\Alph{section}}
   \label{sec:ps}
   In this appendix, we introduce our phase space parame-trization which is useful for evaluating cross sections and distributions of hadron collider events efficiently. 
   In order to take advantage of the high parallel process capability of GPU computing, it is essential that high fraction of generated phase space points satisfy the final state cuts.
   We observe that the following final state cuts are most commonly used for quark and gluon jets in hadron collider experiments:
   \begin{align}
    &{p_T}_i>p_{T cut},\label{ptcut}\\
    &|\eta_i|<\eta_{cut}\label{ycut}.
   \end{align}
   If there are $n$ massless quark and gluon jets in the final state, our parameterization generate phase space points that satisfy all the $n$ conditions on $\eta_i$ in (\ref{ycut}), as well as the $n-1$ conditions for ${p_T}_i$ among the n transverse momenta in (\ref{ptcut}). 
   As for the jet separation cuts
   \begin{align}
   {p_T}_{ij} \equiv \min{({p_T}_i,{p_T}_j)R_{ij}>{p_T}_{cut}}, \label{durham}
   \end{align}
   our parameterization respects them only partially. 
   For $k$-gluon final states, it respects the constraints for $(k-1)$ combinations among $C_k^2$ combinations. 
   As for $1$-quark plus $k$-gluon final states, we take account of just 1 $(q, g)$ combination among the $k$ combinations. 
   Finally, as for $2$-quark plus $k$ gluon final states, we take care of 2 $(q, g)$ combinations only. 
   None of the other combinations of the relative 3-momenta are restricted, including  none of $q\bar{q}$ pairs.
   
  \subsection{$n$-body phase space: a general case}

  For all the parton collision processes at hadron colliders:
  \begin{align}
   a+b\rightarrow 1+...+n, \label{processb}
  \end{align}
  the following generalized phase space parameterization has been used including the integration over the initial parton momentum fractions:
  \begin{align}
   d\Sigma_n&\equiv dx_adx_bd\Phi_n,
   \label{ps00a}
  \end{align}
  where $d\Phi_n$ is the n-body invariant phase space elements
  \begin{align}
   &d\Phi_n=(2\pi)^4\delta^4(p_a+p_b-\sum_{i=1}^np_i)\prod_{i=1}^n\frac{d^3p_i}{(2\pi)^3 2E_i} \nonumber\\
   &\,\, =(2\pi)^4\delta^4(p_a+p_b-\sum_{i=1}^np_i)\prod_{i=1}^n\frac{dy_id^2p_{Ti}}{(4\pi^2)^2}\frac{d\phi_i}{2\pi}.
   \label{ps000}
  \end{align}
  The 4-momentum conservation $\delta$ functions are then integrated as
  \begin{align}
   d\Sigma_n=\Theta(1-x_a)\Theta(1-x_b)\frac{2\pi dy_n}{s}\prod_{i=1}^{n-1}
   \frac{dy_id^2p_{Ti}}{(4\pi^2)^2}\frac{d\phi_i}{2\pi},
   \label{ps01}
  \end{align} 
  giving our parameterization (\ref{ps02}) where 
  \begin{align}
   y_i=\frac{1}{2}\ln\frac{E_i+p_{iz}}{E_i-p_{iz}}
   \label{ps01-y}
  \end{align} 
  are the rapidity of the i-th particle, and 
  \begin{subequations}
   \begin{align}
    &\vec{p}_{Tn}=-\Sigma_{i=1}^{n-1}\vec{p}_{Ti},      \label{ps01-2a}  \\
    &x_a=\frac{1}{\sqrts}\sum_{i=1}^n(E_i+p_{iZ}),    \label{ps01-2b}  \\
    &x_b=\frac{1}{\sqrts}\sum_{i=1}^n(E_i-p_{iZ}),    \label{ps01-2c} 
   \end{align}
   \label{ps01-2}
  \end{subequations}
  are determined by the four momentum conservation.

  In the parameterization (\ref{ps02}), it is clear that all the $n$ final state cuts for $y_i$ (\ref{ycut}) and $(n-1)$ of the $p_T$ cuts (\ref{ptcut}) can be implemented directly when all final partons are massless. 
  The conditions, $x_{a,b}<1$ and
  \begin{align}
   {p_T}_n=\sqrt{p_{nx}^2+p_{ny}^2}>p_{T cut},\label{ptncut}
  \end{align}
  should be examined a posteriori, and null value should be given if any of them is violated. 
  We find that the efficiency of satisfying all the three conditions can be made high simply by parameterizing the transverse momentum as follows:
  \begin{align}
   d{p^2_T}_i=(m_i^2+{p^2_T}_i)d\ln(m_i^2+{p^2_T}_i).
  \end{align}
  Here $m_i$ can be zero for gluon and light quark, can be finite for tau and bottom, while it should be invariant mass of the decaying particles, like top, W, Z and Higgs in the SM; see Appendix~\ref{sec:psb}.

  \subsection{Final states with decaying massive particles}
  \label{sec:psb}
  For processes like $pp\rightarrow (n-m)j+\sum_{k=1}^m R_k(\to f_k)$,
  when the resonance $R_k$ decays into a final state $f_k$,
  we have the phase space parameterization as
  
  \begin{align}
   d\Sigma=d\Sigma_n\prod_{k=1}^m\frac{ds_k}{2\pi} d\Phi(R_k\to f_k \,\, {\rm at}\,\, p^2_{f_k}=s_k),
  \end{align}
  where $d\Phi(R_k\to f_k)$ is the invariant phase space for the $R_k\to f_k$ decay when the invariant mass $m_{fk}$ of the final state $f_k$ is $\sqrt{s_k}$, and $d\Sigma$ is the $n$-body generalized phase space of (\ref{ps00a}) for $(n-m)$ massless particles and $m$ massive particles of masses $\sqrt{s_k}$.

  Integration over $s_k$ is made efficient by adopting the parameterization transformation
  \begin{align}
   ds_k=\frac{(s_k-m_{k}^2)^2+m_{k}^2\Gamma^2_{k}}{m_{k}\Gamma_{k}}d\theta_k\,,
  \end{align}
  where
  \begin{align}
   \theta_k=\tan^{-1}{\bigg(\frac{s_k-m^2_{k}}{m_{k}\Gamma_{k}}\bigg)},
  \end{align}
  generates the Breit-Wigner distribution of the resonance.
  The integration region for $\theta_k$ is chosen as
  \begin{align}
   \frac{|s_k-m_{k}^2|}{m_{k}\Gamma_{k}} < \min\{ 20,\frac{m_{k}}{\Gamma_{k}} \}
  \end{align}
  as a default, where a factor of 20 gives $2\tan^{-1}20/\pi \sim 95.5 \%$ of the total rate for flat backgrounds.

  \def\thesection{Appendix \Alph{section}}
  \section{Random number generation on GPU}
  \def\thesection{\Alph{section}}

 \label{sec:RNG}

 We use the {\sf xorshift} algorithm\cite{xorshift} as the random number
 generator (RNG) on each threads of GPU.
 Because this program generate the random number on each threads of GPU
 independently,
 the time for the random number generation is faster than the total
 of the time for the random number generation on CPU and
 the translation from CPU to GPU.
 The argument of \texttt{xorshift} (List~\ref{list:xorshift}) as:
  \begin{equation}
   \begin{array}{r}
    \texttt{xorshift(unsigned int seed,int n,}\\
    \texttt{double* rndnum)}
   \end{array}
  \end{equation}
 where the inputs and the outputs are:
  \begin{equation}
   \begin{array}{ll}
    \textsc{Inputs:} & \\
    \texttt{unsigned int seed} & \textrm{seed of the random numbers}\\
    \texttt{int n}             & \textrm{number of the random numbers}\\
    \textsc{Outputs:}& \\
    \texttt{double* rndnum}    & \textrm{array of the random number}
   \end{array}
    \label{eq:xorshift}
  \end{equation}
 The \texttt{seed} is the seed of the random numbers, which
 we use the thread number and number of the repeated time.
 The number of random numbers, \texttt{n}, is same as degree of
 freedom.
 The output array \texttt{rndnum} has random number between 0 and 1.
 We set a few hundred to {\sf nLoop} in List~\ref{list:xorshift} for
 generating the better quality random numbers.
 We can get enough quality numbers, even if {\sf nLoop} is set zero.

 The periods of this \texttt{xorshift} program is $2^{128}\!-\!1$ which is
 shorter than that of the MersenneTwister (MT) 
 algorithm~\cite{MT1,MT2,MT3}  but enough for our calculation.
 We check the quality of the random number which generate with
 the \texttt{xorshift} algorithm on GPU.
 It is not found that there are any discrepancy in the equality for
 the random number distribution and the correlations among themselves.
 The \texttt{xorshift} algorithm can generate $2^{21}$ random numbers
 with $0.39$ msec on C2075 with same quality as the MT algorithm.
 This proceedings time is about 1.8 times faster than that of the MT
 program which is included in the CUDA SDK sample program.

\begin{lstlisting}[caption=xorshift.cu,label=list:xorshift]
#define nLoopRndm (32)
#define Norum (0.0000000001*2.32830643653869628906)
__device__
void xorshift(unsigned int seed,int n,double* rndnum)
{
    unsigned int x,y,z,w,t;

    seed=seed*2357U+123456789U;
    x=seed=1812433253U*(seed^(seed>>30))+1;
    y=seed=1812433253U*(seed^(seed>>30))+2;
    z=seed=1812433253U*(seed^(seed>>30))+3;
    w=seed=1812433253U*(seed^(seed>>30))+4;

    for (int i=0;i<nLoopRndm;++i) {
         t=x^x<<11;x=y;y=z;z=w;w^=(w>>19)^(t^(t>>8));
    }

    for (int i=0;i<n;++i) {
        t=x^x<<11;
        x=y;y=z;z=w;
        w^=(w>>19)^(t^(t>>8));
        rndnum[i]=(double)w*Norum;
    }
    return;
}

\end{lstlisting}

  \def\thesection{Appendix \Alph{section}}
  \section{HEGET codes}
  \def\thesection{\Alph{section}}

 \label{sec:new-heget-codes}

 In this section,
 we list all the HEGET functions that are needed to compute helicity amplitudes of all the SM processes in the tree level.
 They are listed as List~\ref{list:ixxxxx} to List~\ref{list:hsssxx}.
 All source code will be available on the web page: \\ 
 \texttt{http://madgraph.kek.jp/KEK/GPU/heget/}.
%

 \subsection{header file and constant numbers}
 \label{sec:header-file}

 We prepare the header file, \texttt{cmplx.h}, which is shown in List~\ref{list:cmplx},
 to define the complex structure for handling the complex numbers in
 HEGET functions.

\begin{lstlisting}[caption=cmplx.h,label=list:cmplx]
#ifndef __cmplx_h__
#define __cmplx_h__
typedef struct __align__(8){
  double re;double im;
} cmplx;

inline __host__ __device__ 
cmplx mkcmplx(double re, double im){
   cmplx z; z.re=re; z.im=im; 
   return z;}

inline __host__ __device__ 
cmplx mkcmplx(cmplx z){return mkcmplx(z.re,z.im);}

inline __host__ __device__ 
cmplx mkcmplx(double a){return mkcmplx(a, 0.);}

inline __host__ __device__  
double real(cmplx a){return a.re;}

inline __host__ __device__  
double imag(cmplx a){return a.im;}

inline __host__ __device__ 
cmplx conj(cmplx a){return mkcmplx(a.re,-a.im);}

inline __host__ __device__  
cmplx operator+(cmplx a, cmplx b){
   return mkcmplx(a.re + b.re, a.im + b.im);}

inline __host__ __device__  
void operator+=(cmplx &a, cmplx b){
   a.re += b.re; a.im += b.im;}

inline __host__ __device__  
cmplx operator+(cmplx a){
   return mkcmplx(+a.re, +a.im);}

inline __host__ __device__  
cmplx operator-(cmplx a, cmplx b){
   return mkcmplx(a.re - b.re, a.im - b.im);}

inline __host__ __device__  
void operator-=(cmplx &a, cmplx b){
   a.re -= b.re; a.im -= b.im;}

inline __host__ __device__  
cmplx operator-(cmplx a){
   return mkcmplx(-a.re, -a.im);}

inline __host__ __device__  
cmplx operator*(cmplx a, cmplx b){
   return mkcmplx((a.re * b.re) - (a.im * b.im),
                  (a.re * b.im) + (a.im * b.re));}

inline __host__ __device__  
cmplx operator*(cmplx a, double s){
   return mkcmplx(a.re * s, a.im * s);}

inline __host__ __device__  
cmplx operator*(double s, cmplx a){
   return mkcmplx(a.re * s, a.im * s);}

inline __host__ __device__  
void operator*=(cmplx &a, double s){
   a.re *= s; a.im *= s;}

inline __host__ __device__  
cmplx operator/(cmplx a, cmplx b){
   double tmpD=(1./(b.re*b.re+b.im*b.im));
   return mkcmplx( 
       ( (a.re * b.re) + (a.im * b.im))*tmpD,
       (-(a.re * b.im) + (a.im * b.re))*tmpD
       );}

inline __host__ __device__  
cmplx operator/(cmplx a, double s){
   double inv = 1. / s;
   return a * inv;}

inline __host__ __device__  
cmplx operator/(double s, cmplx a){
   double inv = s*(1./(a.re*a.re+a.im*a.im));
   return mkcmplx(inv*a.re,-inv*a.im);}

inline __host__ __device__  
void operator/=(cmplx &a, double s){
   double inv = 1. / s;
   a *= inv;}

inline __host__ __device__ 
double fabsc(cmplx a){
   return sqrt((a.re*a.re)+(a.im*a.im));}

inline __host__ __device__ 
double fabs2c(cmplx a){
   return (a.re*a.re)+(a.im*a.im);}
#endif
\end{lstlisting}

 \subsection{wave function}
 \label{sec:wave-function}

  \subsubsection{ixxxxx and oxxxxx}
  \label{sec:ioxxxxx}

  We have two functions to compute external fermions.
  One is for ``flowing-In'' fermions and
  the other is for ``flowing-Out'' fermions.
  The spinor wave function with a generic 3-momentum $p$ for
  ``flowing-In'' fermion is named
  \texttt{ixxxxx} (List~\ref{list:ixxxxx}),
  and for ``flowing-Out'' fermion
  is named \texttt{oxxxxx} (List~\ref{list:oxxxxx}).
  The argument of
  \texttt{ixxxxx} 
  and
  \texttt{oxxxxx} 
  as:
  \begin{equation}
   \begin{array}{r}
    \texttt{ixxxxx(double* p, double fmass,~~~~~~}\\
    \texttt{int nhel, int nsf, cmplx* fi)}\\
   \end{array}
  \end{equation}
  and
  \begin{equation}
   \begin{array}{r}
    \texttt{oxxxxx(double* p, double fmass,~~~~~~}\\
    \texttt{int nhel, int nsf, cmplx* fo)}\\
   \end{array}
  \end{equation}
  where the inputs and the outputs are:

  \begin{equation}
   \begin{array}{ll}
    \textsc{Inputs:}& \\
    \texttt{double p[4]}  & \textrm{4-momentum}\\
    \texttt{double fmass} & \textrm{fermion mass}\\
    \texttt{int nhel}    & \textrm{twice fermion helicity (-1 or 1)}\\
    \texttt{int nsf}     & \textrm{+1 for particle, -1 for anti-particle}\\
   \end{array}
  \end{equation}
  \begin{equation}
   \begin{array}{ll}
    \textsc{Outputs:}& \\
    \texttt{cmplx fi[6]} & \textrm{fermion wavefunction}~ \texttt{|fi>} \\
    &u(\texttt{p},\texttt{nhel/2})~\textrm{for}~\texttt{nsf}=+1 \\
    &v(\texttt{p},\texttt{nhel/2})~\textrm{for}~\texttt{nsf}=-1\hspace*{8ex}\\
    \label{eq:ixxxxx}
   \end{array}
  \end{equation}
  for \texttt{ixxxxx} and
  \begin{equation}
   \begin{array}{ll}
    \textsc{Outputs:}& \\
    \texttt{cmplx fo[6]} & \textrm{fermion wavefunction}~\texttt{<fo|} \\
    &\bar{u}(\texttt{p},\texttt{nhel/2})~\textrm{for}~\texttt{nsf}=+1 \\
    &\bar{v}(\texttt{p},\texttt{nhel/2})~\textrm{for}~\texttt{nsf}=-1\hspace*{8ex}\\
    \label{eq:oxxxxx}
   \end{array}
  \end{equation}
  for \texttt{oxxxxx}.

\begin{lstlisting}[caption=ixxxxx.cu,label=list:ixxxxx]
#include "cmplx.h"

__device__
void ixxxxx(double* p, double fmass, int nhel, int nsf,
            cmplx* fi)
{
   fi[4] = mkcmplx(p[0]*nsf, p[3]*nsf);
   fi[5] = mkcmplx(p[1]*nsf, p[2]*nsf);
   int nh = nhel*nsf;
   cmplx chi[2];
   if (fmass!=0.) {
      double pp = fmin(p[0], 
                       sqrt(p[1]*p[1] + p[2]*p[2] + p[3]*p[3]));
      if (pp==0.) {
         double sqm[2];
         sqm[0] = sqrt(fabs(fmass));
         sqm[1] = copysign(sqm[0], fmass);
         int ip = (1+nh)/2;
         int im = (1-nh)/2;
         fi[0] = mkcmplx((double)(ip)    *sqm[ip]);
         fi[1] = mkcmplx((double)(im*nsf)*sqm[ip]);
         fi[2] = mkcmplx((double)(ip*nsf)*sqm[im]);
         fi[3] = mkcmplx((double)(im)    *sqm[im]);
      } else {
         double sf[2],omega[2];
         sf[0] = (double)(1 + nsf + (1-nsf)*nh)*0.5;
         sf[1] = (double)(1 + nsf - (1-nsf)*nh)*0.5;
         omega[0] = sqrt(p[0]+pp);
         omega[1] = fmass*(1./omega[0]);
         double pp3 = fmax(pp+p[3],0.);
         chi[0] = mkcmplx(sqrt(pp3*0.5*(1./pp)));
         if (pp3==0.) {
            chi[1] = mkcmplx((double)(nh));
         } else {
            chi[1] = rsqrt(2.*pp*pp3) * 
               mkcmplx((double)(nh)*p[1], p[2]);
         }
         int ip = (3+nh)/2-1;
         int im = (3-nh)/2-1;
         fi[0] = sf[0]*omega[ip]*chi[im];
         fi[1] = sf[0]*omega[ip]*chi[ip];
         fi[2] = sf[1]*omega[im]*chi[im];
         fi[3] = sf[1]*omega[im]*chi[ip];
      }
   } else {
      double sqp0p3;
      if (p[1]==0. && p[2]==0. && p[3]<0.) {
         sqp0p3 = 0.;
      } else {
         sqp0p3 = sqrt(fmax(p[0]+p[3],0.))*(double)(nsf);
      }
      chi[0] = mkcmplx(sqp0p3);
      if (sqp0p3==0.) {
         chi[1] = mkcmplx((double)(nhel)*sqrt(2.*p[0]));
      } else {
         chi[1] = (1./sqp0p3) *
            mkcmplx((double)(nh)*p[1], p[2]);
      }
      cmplx czero  = mkcmplx(0.,0.);
      if (nh==1) {
         fi[0] = czero ;
         fi[1] = czero ;
         fi[2] = chi[0];
         fi[3] = chi[1];
      } else {
         fi[0] = chi[1];
         fi[1] = chi[0];
         fi[2] = czero;
         fi[3] = czero;
      }
   }
}
\end{lstlisting}

\begin{lstlisting}[caption=oxxxxx.cu,label=list:oxxxxx]
#include "cmplx.h"

__device__
void oxxxxx(double* p, double fmass, int nhel, int nsf,
            cmplx* fo)
{
   fo[4] = mkcmplx(p[0]*nsf, p[3]*nsf);
   fo[5] = mkcmplx(p[1]*nsf, p[2]*nsf);
   int nh = nhel*nsf;
   cmplx chi[2];
   if (fmass!=0.) {
      double pp = fmin(p[0],
                       sqrt(p[1]*p[1] + p[2]*p[2] + p[3]*p[3]));
      if (pp==0.) {
         double sqm[2];
         sqm[0] = sqrt(fabs(fmass));
         sqm[1] = copysign(sqm[0], fmass);
         int ip = -(1+nh)/2;
         int im =  (1-nh)/2;
         fo[0] = mkcmplx((double)(im)    *sqm[im]);
         fo[1] = mkcmplx((double)(ip*nsf)*sqm[im]);
         fo[2] = mkcmplx((double)(im*nsf)*sqm[-ip]);
         fo[3] = mkcmplx((double)(ip)    *sqm[-ip]);
      } else {
         double sf[2],omega[2];   
         sf[0] = (double)(1 + nsf + (1-nsf)*nh)*0.5;
         sf[1] = (double)(1 + nsf - (1-nsf)*nh)*0.5;
         omega[0] = sqrt(p[0]+pp);
         omega[1] = fmass*(1./omega[0]);
         double pp3 = fmax(pp+p[3],0.);
         chi[0] = mkcmplx(sqrt(pp3*0.5*(1./pp)));
         if (pp3==0.) {
            chi[1] = mkcmplx((double)(nh));
         } else {
            chi[1] = rsqrt(2.*pp*pp3) * 
               mkcmplx((double)(nh)*p[1],-p[2]);
         }
         int ip = (3+nh)/2-1;
         int im = (3-nh)/2-1;
         fo[0] = sf[1]*omega[im]*chi[im];
         fo[1] = sf[1]*omega[im]*chi[ip];
         fo[2] = sf[0]*omega[ip]*chi[im];
         fo[3] = sf[0]*omega[ip]*chi[ip];
      }
   } else {
      double sqp0p3;
      if (p[1]==0. && p[2]==0. && p[3]<0.) {
         sqp0p3 = 0.;
      } else {
         sqp0p3 = sqrt(fmax(p[0]+p[3],0.))*(double)(nsf);
      }
      chi[0] = mkcmplx(sqp0p3);
      if (sqp0p3==0.) {
         chi[1] = mkcmplx((double)(nhel)*sqrt(2.*p[0]));
      } else {
         chi[1] = (1./sqp0p3) *
            mkcmplx((double)(nh)*p[1],-p[2]);
      } 
      cmplx czero  = mkcmplx(0.,0.);
      if (nh==1) {
         fo[0] = chi[0];
         fo[1] = chi[1];
         fo[2] = czero;
         fo[3] = czero;
      } else {
         fo[0] = czero;
         fo[1] = czero;
         fo[2] = chi[1];
         fo[3] = chi[0];
      }
   }
}
\end{lstlisting}

  \subsubsection{vxxxxx}
  \label{sec:vxxxxx}

  We prepare a function named \texttt{vxxxxx} (List~\ref{list:vxxxxx})
  for the wave function of a vector boson.
  The argument of \texttt{vxxxxx} as:
  \begin{equation}
   \begin{array}{r}
    \texttt{vxxxxx(double* p, double vmass,~~~~~~~~}\\
    \texttt{int nhel, int nsv, cmplx* vc)}
   \end{array}
  \end{equation}
  where the inputs and the outputs are:
  \begin{equation}
   \begin{array}{ll}
    \textsc{Inputs:}& \\
    \texttt{double p[4]}  & \textrm{4-momentum}\\
    \texttt{double vmass} & \textrm{vector boson mass}\\
    \texttt{int nhel}    & \textrm{helicity of massive vector (-1, 0, 1)}\\
    \texttt{int nsv}     & \textrm{+1 for final, -1 for initial vector}\\
    &\\
    \textsc{Outputs:}& \\
    \texttt{cmplx vc[6]} & \textrm{vector boson wavefunction} \\
    &\epsilon^\mu(p,\texttt{nhel})^\ast~\textrm{for}~\texttt{nsv}=+1 \\
    &\epsilon^\mu(p,\texttt{nhel})~\textrm{for}~\texttt{nsv}=-1.
     \label{eq:vxxxxx}
   \end{array}
  \end{equation}

\begin{lstlisting}[caption=vxxxxx.cu,label=list:vxxxxx]
#include "cmplx.h"

__device__
void vxxxxx(double* p, double vmass, int nhel, int nsv,
            cmplx* vc)
{
   const double sqh = 0.70710678118655;
   double hel = (double)(nhel);
   double pt2 = p[1]*p[1]+p[2]*p[2];
   double pp = fmin(p[0],sqrt(pt2+p[3]*p[3]));
   double pt = fmin(pp,sqrt(pt2));
   int nsvahl = (int)(nsv*fabs(hel));
   vc[4] = mkcmplx(p[0],p[3])*nsv;
   vc[5] = mkcmplx(p[1],p[2])*nsv;
   if (vmass!=0.) {
      double hel0 = 1.-fabs(hel);
      if (pp==0.) {
         vc[0] = mkcmplx(0.,0.);
         vc[1] = mkcmplx(-hel*sqh);
         vc[2] = mkcmplx(0.,(double)nsvahl*sqh);
         vc[3] = mkcmplx(hel0);
      } else {
         double emp = p[0]*(1./(vmass*pp));
         vc[0] = mkcmplx(hel0*pp*(1./vmass));
         vc[3] = mkcmplx(hel0*p[3]*emp+hel*pt*(1./pp)*sqh);
         if (pt!=0.) {
            double pzpt = p[3]*(1./(pp*pt))*sqh*hel;
            vc[1] = mkcmplx( hel0*p[1]*emp-p[1]*pzpt,
                           (double)(-nsvahl)*p[2]*(1./pt)*sqh);
            vc[2] = mkcmplx( hel0*p[2]*emp-p[2]*pzpt,
                           (double)(nsvahl)*p[1]*(1./pt)*sqh);
         } else {
            vc[1] = mkcmplx(-hel*sqh);
            vc[2] = mkcmplx(0.,
                            (double)(nsv)*copysign(sqh,p[3]));
         }
      }
   } else {
      double pt  = sqrt(p[1]*p[1] + p[2]*p[2]);
      double rpt = 1./pt;
      double rpp = 1./pp;
      vc[0] = mkcmplx(0., 0.);
      vc[3] = mkcmplx(nhel*pt*rpp*sqh);
      if (pt!=0.) {
         double pzpt = p[3]*rpp*rpt*sqh*nhel;
         vc[1] = mkcmplx(-p[1]*pzpt,-nsv*p[2]*rpt*sqh);
         vc[2] = mkcmplx(-p[2]*pzpt, nsv*p[1]*rpt*sqh);
      } else {
         vc[1] = mkcmplx(-nhel*sqh);
         vc[2] = mkcmplx(0., nsv*copysign(sqh,p[3]));
      }
   }
}
\end{lstlisting}

  \subsubsection{sxxxxx}
  \label{sec:sxxxxx}

  The function which named \texttt{sxxxxx} (List~\ref{list:sxxxxx})
  computes a wave function of the massive scalar field.
  The argument of this function as:
  \begin{equation}
   \begin{array}{r}
    \texttt{sxxxxx(double* p, int nss, cmplx* sc)}\,,
   \end{array}
  \end{equation}
  where inputs and the outputs as
  \begin{equation}
   \begin{array}{ll}
    \textsc{Inputs:}& \\
    \texttt{double p[4]}  & \textrm{4-momentum}\\
    \texttt{double nss}   & \textrm{+1 for initial, -1 for final}\\
    &\\
    \textsc{Outputs:}& \\
    \texttt{cmplx sc[3]} & \textrm{scalar wavefunction}.
     \label{eq:sxxxxx}
   \end{array}
  \end{equation}

\begin{lstlisting}[caption=sxxxxx.cu,label=list:sxxxxx]
#include "cmplx.h"

__device__
void sxxxxx(double* p, int nss, cmplx* sc)
{
    sc[0] = mkcmplx(1.,0.);
    sc[1] = (double)(nss)*mkcmplx(p[0],p[3]);
    sc[2] = (double)(nss)*mkcmplx(p[1],p[2]);
}

\end{lstlisting}

  \subsection{FFV vertex}
  \label{sec:ffv-vertex}

  The \texttt{FFV} vertex functions are obtained from the Lagrangian
  \begin{equation}
   \mathcal{L}_{\mathrm{F}\! \mathrm{F}\! \mathrm{V}}
    \!= \overline{\psi}_{\mathrm{F}_{1}} \gamma^\mu
    \left[ \texttt{gc[0]} \frac{1\!-\!\gamma_{5}}{2}
    \!+\! \texttt{gc[1]} \frac{1\!+\!\gamma_{5}}{2} \right]
    \psi_{\mathrm{F}_{2}} V_{\mu}^{*},
  \label{HEGET-FFVb}
  \end{equation}
where the boson name is given by the flowing out quantum number.

  \subsubsection{iovxxx}
  \label{sec:iovxxx}

  This function \texttt{iovxxx} (List~\ref{list:iovxxx})
  computes an amplitude of the \texttt{FFV} vertex from a flowing-In
  fermion, a flowing-Out fermion and a vector boson wave functions,
  whether they are on-shell or off-shell.

  The argument of this function as:
 \begin{equation}
  \begin{array}{r}
  \texttt{iovxxx(cmplx* fi, cmplx* fo, cmplx* vc, } \\
  \texttt{cmplx* gc, cmplx vertex)}
   \end{array}
 \end{equation}
  where the inputs and the outputs are
 \begin{equation}
  \begin{array}{ll}
   \textsc{Inputs:} & \\
    \texttt{cmplx fi[6]} & \textrm{flowing-In fermion wavefunction} \\
    \texttt{cmplx fo[6]} & \textrm{flowing-Out fermion wavefunction} \\
    \texttt{cmplx vc[6]} & \textrm{vector wavefunction} \\
    \texttt{cmplx gc[2]} & \textrm{coupling constants of \texttt{FFV}
     vertex} \\ \\
   \textsc{Outputs:} & \\
   \texttt{cmplx vertex} & \textrm{amplitude of the \texttt{FFV}
    vertex} \\
   & \texttt{<fo|V|fi>}
    \label{eq:iovxxx}
   \end{array}
 \end{equation}

\begin{lstlisting}[caption=iovxxx.cu,label=list:iovxxx]
#include "cmplx.h"

__device__
void iovxxx(cmplx* fi, cmplx* fo, cmplx* vc, cmplx* gc,
            cmplx& vertex)
{
   vertex = 
      gc[0]*( (fo[2]*fi[0]+fo[3]*fi[1])*vc[0]
             +(fo[2]*fi[1]+fo[3]*fi[0])*vc[1]
            -((fo[2]*fi[1]-fo[3]*fi[0])*vc[2])*mkcmplx(0.,1.)
             +(fo[2]*fi[0]-fo[3]*fi[1])*vc[3] )
    + gc[1]*( (fo[0]*fi[2]+fo[1]*fi[3])*vc[0]
             -(fo[0]*fi[3]+fo[1]*fi[2])*vc[1]
            +((fo[0]*fi[3]-fo[1]*fi[2])*vc[2])*mkcmplx(0.,1.)
             -(fo[0]*fi[2]-fo[1]*fi[3])*vc[3] );
}

\end{lstlisting}

  \subsubsection{fvixxx}
  \label{sec:fvixxx}

  The function \texttt{fvixxx} (List~\ref{list:fvixxx})
  computes an off-shell fermion wave function from a ``flowing-In''
  external fermion and a vector boson.

  The argument of this function as:
  \begin{equation}
   \begin{array}{r}
    \texttt{fvixxx(cmplx* fi, cmplx* vc, cmplx* gc,}\\
    \texttt{double fmass, double fwidth,}\\
    \texttt{cmplx* fvi)}\,,
   \end{array}
  \end{equation}
  where the inputs and the outputs are
  \begin{equation}
   \begin{array}{ll}
    \textsc{Inputs:}& \\
    \texttt{cmplx fi[6]} & \textrm{flowing-In fermion wavefunction}\\
    \texttt{cmplx vc[6]} & \textrm{vector wavefunction}\\
    \texttt{cmplx gc[2]} & \textrm{coupling constants of the \texttt{FFV} vertex}
     \\
    \texttt{double fmass} & \textrm{mass of output fermion}\\
    \texttt{double fwidth} & \textrm{width of output fermion} \\
    &\\
    \textsc{Outputs:}& \\
    \texttt{cmplx fvi[6]} &\textrm{off-shell fermion wavefunction}\\
    &\texttt{|f$^\prime$: vc, fi>}
     \label{eq:fvixxx}
   \end{array}
  \end{equation}

\begin{lstlisting}[caption=fvixxx.cu,label=list:fvixxx]
#include "cmplx.h"

__device__
void fvixxx(cmplx* fi, cmplx* vc, cmplx* gc,
            double fmass, double fwidth, cmplx* fvi)
{
   fvi[4] = fi[4]-vc[4];
   fvi[5] = fi[5]-vc[5];
   double pf[4];
   pf[0] = fvi[4].re;
   pf[1] = fvi[5].re;
   pf[2] = fvi[5].im;
   pf[3] = fvi[4].im;
   double pf2 = (pf[0]+pf[3])*(pf[0]-pf[3])
              - (pf[1]*pf[1]+pf[2]*pf[2]);
   cmplx ci = mkcmplx(0.,1.);
   cmplx d = -1./mkcmplx(pf2-fmass*fmass, fmass*fwidth);
   cmplx sl1 = (vc[0] +    vc[3])*fi[0]
             + (vc[1] - ci*vc[2])*fi[1];
   cmplx sl2 = (vc[1] + ci*vc[2])*fi[0] 
             + (vc[0] -    vc[3])*fi[1];
   cmplx sr1 = (vc[0] -    vc[3])*fi[2]
             - (vc[1] - ci*vc[2])*fi[3];
   cmplx sr2 =-(vc[1] + ci*vc[2])*fi[2]
             + (vc[0] +    vc[3])*fi[3];
   fvi[0] = ( gc[0]*((pf[0]-pf[3])*sl1 -  conj(fvi[5])*sl2)
             +gc[1]*fmass*sr1)*d;
   fvi[1] = ( gc[0]*(      -fvi[5]*sl1 + (pf[0]+pf[3])*sl2)
             +gc[1]*fmass*sr2)*d;
   fvi[2] = ( gc[1]*((pf[0]+pf[3])*sr1 +  conj(fvi[5])*sr2)
             +gc[0]*fmass*sl1)*d;
   fvi[3] = ( gc[1]*(       fvi[5]*sr1 + (pf[0]-pf[3])*sr2)
             +gc[0]*fmass*sl2)*d;
}



    
\end{lstlisting}

  \subsubsection{fvoxxx}
  \label{sec:fvoxxx}

  The function \texttt{fvoxxx} (List~\ref{list:fvoxxx})
  computes an off-shell fermion wave function
  from a ``flowing-Out'' external fermion and a vector boson.

  The argument of this function as:
  \begin{equation}
   \begin{array}{r}
    \texttt{fvoxxx(cmplx* fo, cmplx* vc, cmplx* gc,}\\
    \texttt{double fmass, double fwidth,}\\
    \texttt{cmplx* fvo)}\,,
   \end{array}
  \end{equation}
  where the inputs and the outputs are
  \begin{equation}
   \begin{array}{ll}
    \textsc{Inputs:}& \\
    \texttt{cmplx fo[6]} & \textrm{flowing-Out fermion wavefunction}\\
    \texttt{cmplx vc[6]} & \textrm{vector wavefunction}\\
    \texttt{cmplx gc[2]} & \textrm{coupling constants of the \texttt{FFV} vertex}
     \\
    \texttt{double fmass} & \textrm{mass of output fermion}\\
    \texttt{double fwidth} & \textrm{width of output fermion} \\
    &\\
    \textsc{Outputs:}& \\
    \texttt{cmplx fvo[6]} & \textrm{off-shell fermion wavefunction}\\
    &\texttt{<f$^\prime$: vc, fo|}\,.
     \label{eq:fvoxxx}
   \end{array}
  \end{equation}

\begin{lstlisting}[caption=fvoxxx.cu,label=list:fvoxxx]
#include "cmplx.h"

__device__
void fvoxxx(cmplx* fo, cmplx* vc, cmplx* gc,
            double fmass, double fwidth, cmplx* fvo)
{
   fvo[4] = fo[4]+vc[4];
   fvo[5] = fo[5]+vc[5];
   double pf[4];
   pf[0] = fvo[4].re;
   pf[1] = fvo[5].re;
   pf[2] = fvo[5].im;
   pf[3] = fvo[4].im;
   double pf2 = (pf[0]+pf[3])*(pf[0]-pf[3])
              - (pf[1]*pf[1]+pf[2]*pf[2]);
   cmplx ci = mkcmplx(0.,1.);
   cmplx d = -1./mkcmplx(pf2-fmass*fmass, fmass*fwidth);
   cmplx sl1 = (vc[0] +    vc[3])*fo[2]
             + (vc[1] + ci*vc[2])*fo[3];
   cmplx sl2 = (vc[1] - ci*vc[2])*fo[2]
             + (vc[0] -    vc[3])*fo[3];
   cmplx sr1 = (vc[0] -    vc[3])*fo[0]
             - (vc[1] + ci*vc[2])*fo[1];
   cmplx sr2 =-(vc[1] - ci*vc[2])*fo[0]
             + (vc[0] +    vc[3])*fo[1];
   fvo[0] = ( gc[1]*((pf[0]+pf[3])*sr1        + fvo[5]*sr2)
             +gc[0]*fmass*sl1)*d;
   fvo[1] = ( gc[1]*( conj(fvo[5])*sr1 + (pf[0]-pf[3])*sr2)
             +gc[0]*fmass*sl2)*d;
   fvo[2] = ( gc[0]*((pf[0]-pf[3])*sl1        - fvo[5]*sl2)
             +gc[1]*fmass*sr1)*d;
   fvo[3] = ( gc[0]*(-conj(fvo[5])*sl1 + (pf[0]+pf[3])*sl2)
             +gc[1]*fmass*sr2)*d;
}
\end{lstlisting}

  \subsubsection{jioxxx}
  \label{sec:jioxxx}

  This function \texttt{jioxxx} (List~\ref{list:jioxxx})
  computes an off-shell vector current from an external
  fermion pair.
  The vector boson propagator is given in Feynman gauge
  for a massless vector and in unitary gauge for a massive vector.

  The argument of this function as:
  \begin{equation}
   \begin{array}{r}
    \texttt{jioxxx(cmplx* fi, cmplx* fo, cmplx* gc,}\\
    \texttt{double vmass, double vwidth,}\\
    \texttt{cmplx* jio)}\,,
   \end{array}
  \end{equation}
  where the inputs and the outputs are
  \begin{equation}
   \begin{array}{ll}
    \textsc{Inputs:}& \\
    \texttt{cmplx fi[6]}   & \textrm{flowing-In fermion wavefunction}\\
    \texttt{cmplx fo[6]}   & \textrm{flowing-Out fermion wavefunction}\\
    \texttt{cmplx gc[2]}  & \textrm{coupling constants of the \texttt{FFV} vertex
     }\\
    \texttt{double vmass}   & \textrm{mass of output vector boson} \\
    \texttt{double vwidth}  & \textrm{width of output vector boson} \\
    &\\
    \textsc{Outputs:}& \\
    \texttt{cmplx jio[6]} & \textrm{vector current}\\
    &j^\mu(\texttt{<fo|V|fi>})  \label{eq:jioxxx}
   \end{array}
  \end{equation}

\begin{lstlisting}[caption=jioxxx.cu,label=list:jioxxx]
#include "cmplx.h"

__device__
void jioxxx(cmplx* fi, cmplx* fo, cmplx* gc,
            double vmass, double vwidth, cmplx* jio)
{
   jio[4] = fo[4]-fi[4];
   jio[5] = fo[5]-fi[5];
   double q[4];
   q[0] = jio[4].re;
   q[1] = jio[5].re;
   q[2] = jio[5].im;
   q[3] = jio[4].im;
   double q2 = (q[0]+q[3])*(q[0]-q[3])-(q[1]*q[1]+q[2]*q[2]);
   cmplx f02 = fo[0]*fi[2];
   cmplx f03 = fo[0]*fi[3];
   cmplx f12 = fo[1]*fi[2];
   cmplx f13 = fo[1]*fi[3];
   cmplx f20 = fo[2]*fi[0];
   cmplx f21 = fo[2]*fi[1];
   cmplx f30 = fo[3]*fi[0];
   cmplx f31 = fo[3]*fi[1];
   if (vmass!=0.) {
      double vm2 = vmass*vmass;
      cmplx d = 1./mkcmplx(q2-vm2, vmass*vwidth);
      cmplx c0 = +gc[0]*(f20 + f31) +gc[1]*( f02 + f13);
      cmplx c1 = -gc[0]*(f21 + f30) +gc[1]*( f03 + f12);
      cmplx c2 =  gc[0]*(f21 - f30) +gc[1]*(-f03 + f12);
      c2 = mkcmplx(0.,1.)*c2;
      cmplx c3 = +gc[0]*(-f20 + f31) +gc[1]*( f02 - f13);
      cmplx cm2 = mkcmplx(vm2, -vmass*vwidth);
      cmplx cs = (q[0]*c0-q[1]*c1-q[2]*c2-q[3]*c3)*(1./cm2);
      jio[0] = (c0 - cs*q[0])*d;
      jio[1] = (c1 - cs*q[1])*d;
      jio[2] = (c2 - cs*q[2])*d;
      jio[3] = (c3 - cs*q[3])*d;
    } else {
      double d = 1./q2;
      jio[0] = (gc[0]*(f20+f31)+gc[1]*(f02+f13))*d;
      jio[1] = (gc[1]*(f03+f12)-gc[0]*(f21+f30))*d;
      cmplx cjio = (gc[0]*(f21-f30)-gc[1]*(f03-f12));
      jio[2] = mkcmplx(0.,1.)*cjio*d;
      jio[3] = (gc[1]*(f02-f13)-gc[0]*(f20-f31))*d;
    }
}

\end{lstlisting}

  \subsection{FFS vertex}
  \label{sec:ffs-vertex}

  \subsubsection{iosxxx}
  \label{sec:iosxxx}

  An amplitude of the \texttt{FFS} coupling is computed
  by the function \texttt{iosxxx} (List~\ref{list:iosxxx}).
  The argument of this function as:
  \begin{equation}
   \begin{array}{r}
    \texttt{iosxxx(cmplx* fi, cmplx* fo, cmplx* sc,}\\
    \texttt{cmplx* gc, cmplx\& vertex)}\,,
   \end{array}
  \end{equation}
  where the inputs and the outputs are
  \begin{equation}
   \begin{array}{ll}
    \textsc{Inputs:}& \\
    \texttt{cmplx fi[6]} & \textrm{flowing-In  fermion wavefunction}\\
    \texttt{cmplx fo[6]} & \textrm{flowing-Out fermion wavefunction}\\
    \texttt{cmplx sc[3]} & \textrm{input scalar}\\
    \texttt{cmplx gc[2]} & \textrm{coupling constant of \texttt{FFS} vertex} \\
    &\\
    \textsc{Outputs:}& \\
    \texttt{cmplx\& vertex} & \textrm{amplitude of the \texttt{FFS}}\\
    & \texttt{<fo|S|fi>}
     \label{eq:iosxxx}
   \end{array}
  \end{equation}

\begin{lstlisting}[caption=iosxxx.cu,label=list:iosxxx]
#include "cmplx.h"

__device__
void iosxxx(cmplx* fi, cmplx* fo, cmplx* sc, cmplx* g,
            cmplx& vertex)
{
   vertex = sc[0]*( g[0]*(fi[0]*fo[0]+fi[1]*fo[1])
                   +g[1]*(fi[2]*fo[2]+fi[3]*fo[3]) );
}
\end{lstlisting}

  \subsubsection{fsixxx}
  \label{sec:fsixxx}

  The function \texttt{fsixxx} (List~\ref{list:fsixxx})
  computes an off-shell fermion wave function
  from a flowing-In external fermion and a scalar boson.
  The argument of this function as:
  \begin{equation}
   \begin{array}{r}
    \texttt{fsixxx(cmplx* fi, cmplx* sc, cmplx* gc,}\\
    \texttt{double fmass, double fwidth,}\\
    \texttt{cmplx* fsi)}
   \end{array}
  \end{equation}
  where the inputs and the outputs are
  \begin{equation}
   \begin{array}{ll}
    \textsc{Inputs:}& \\
    \texttt{cmplx fi[6]}  & \textrm{flowing-In fermion wavefunction}\\
    \texttt{cmplx sc[3]}  & \textrm{input scalar wavefunction}\\
    \texttt{cmplx gc[2]}  & \textrm{coupling constant of \texttt{FFS} vertex}\\
    \texttt{double fmass}  & \textrm{mass of output fermion} \\
    \texttt{double fwidth} & \textrm{width of output fermion} \\
    &\\
    \textsc{Outputs:}& \\
    \texttt{cmplx fsi[6]} & \textrm{off-shell fermion wavefunction}\\
     &\texttt{|f$^\prime$:fi, sc>}
     \label{eq:fsixxx}
   \end{array}
  \end{equation}

\begin{lstlisting}[caption=fsixxx.cu,label=list:fsixxx]
#include "cmplx.h"

__device__
void fsixxx(cmplx* fi, cmplx* sc, cmplx* gc,
            double fmass, double fwidth, cmplx* fsi)
{
   fsi[4] = fi[4]-sc[1];
   fsi[5] = fi[5]-sc[2];
   double pf[4];
   pf[0] = fsi[4].re;
   pf[1] = fsi[5].re;
   pf[2] = fsi[5].im;
   pf[3] = fsi[4].im;
   double pf2 = (pf[0]+pf[3])*(pf[0]-pf[3])
              - (pf[1]*pf[1]+pf[2]*pf[2]);
   cmplx ds = -sc[0]*
      (1./mkcmplx(pf2 - fmass*fmass, fmass*fwidth));
   double p0p3 = pf[0]+pf[3];
   double p0m3 = pf[0]-pf[3];
   cmplx sl1 = gc[0]*(p0p3*fi[0]+conj(fsi[5])*fi[1]);
   cmplx sl2 = gc[0]*(p0m3*fi[1]     +fsi[5] *fi[0]);
   cmplx sr1 = gc[1]*(p0m3*fi[2]-conj(fsi[5])*fi[3]);
   cmplx sr2 = gc[1]*(p0p3*fi[3]     -fsi[5] *fi[2]);
   fsi[0] = ( gc[0]*fmass*fi[0] + sr1 )*ds;
   fsi[1] = ( gc[0]*fmass*fi[1] + sr2 )*ds;
   fsi[2] = ( gc[1]*fmass*fi[2] + sl1 )*ds;
   fsi[3] = ( gc[1]*fmass*fi[3] + sl2 )*ds;
}
\end{lstlisting}

  \subsubsection{fsoxxx}
  \label{sec:fsoxxx}

  The function \texttt{fsoxxx} (List~\ref{list:fsoxxx})
  computes an off-shell fermion wave function
  from a flowing-Out external fermion and a scalar boson.
  The argument of this function as:
  \begin{equation}
   \begin{array}{r}
    \texttt{fsoxxx(cmplx* fo, cmplx* sc, cmplx* gc,}\\
    \texttt{double fmass, double fwidth,}\\
    \texttt{cmplx* fso)}
   \end{array}
  \end{equation}
  where the inputs and the outputs are
  \begin{equation}
   \begin{array}{ll}
    \textsc{Inputs:}& \\
    \texttt{cmplx fo[6]}   & \textrm{flowing-Out fermion wavefunction}\\
    \texttt{cmplx sc[3]}   & \textrm{input scalar wavefunction}\\
    \texttt{cmplx gc[2]}   & \textrm{coupling constant of \texttt{FFS} vertex}\\
    \texttt{double fmass}   & \textrm{mass of output fermion} \\
    \texttt{double fwidth}  & \textrm{width of output fermion} \\
    &\\
    \textsc{Outputs:}& \\
    \texttt{cmplx fso[6]}  & \textrm{off-shell fermion wavefunction}\\
    &\texttt{<f$^\prime$:fo, sc|}
     \label{eq:fsoxxx}
   \end{array}
  \end{equation}

\begin{lstlisting}[caption=fsoxxx.cu,label=list:fsoxxx]
#include "cmplx.h"

__device__
void fsoxxx(cmplx* fo, cmplx* sc, cmplx* gc,
            double fmass, double fwidth, cmplx* fso)
{
   fso[4] = fo[4]+sc[1];
   fso[5] = fo[5]+sc[2];
   double pf[4];
   pf[0] = fso[4].re;
   pf[1] = fso[5].re;
   pf[2] = fso[5].im;
   pf[3] = fso[4].im;
   double pf2 = (pf[0]+pf[3])*(pf[0]-pf[3])
              - (pf[1]*pf[1]+pf[2]*pf[2]);
   cmplx ds = -sc[0] *
      (1./mkcmplx(pf2 - fmass*fmass, fmass*fwidth));
   double p0p3 = pf[0]+pf[3];
   double p0m3 = pf[0]-pf[3];
   cmplx sl1 = gc[1]*(p0p3*fo[2]     +fso[5] *fo[3]);
   cmplx sl2 = gc[1]*(p0m3*fo[3]+conj(fso[5])*fo[2]);
   cmplx sr1 = gc[0]*(p0m3*fo[0]     -fso[5] *fo[1]);
   cmplx sr2 = gc[0]*(p0p3*fo[1]-conj(fso[5])*fo[0]);
   fso[0] = ( gc[0]*fmass*fo[0] + sl1 )*ds;
   fso[1] = ( gc[0]*fmass*fo[1] + sl2 )*ds;
   fso[2] = ( gc[1]*fmass*fo[2] + sr1 )*ds;
   fso[3] = ( gc[1]*fmass*fo[3] + sr2 )*ds;
}



    
\end{lstlisting}

  \subsubsection{hioxxx}
  \label{sec:hioxxx}

  The scalar current from \texttt{FFS} vertex is
  computed by the function \texttt{hioxxx} (List~\ref{list:hioxxx}).
  The argument of this function as:
  \begin{equation}
   \begin{array}{r}
    \texttt{hioxxx(cmplx* fi, cmplx* fo, cmplx* gc,}\\
    \texttt{double smass, double swidth,}\\
    \texttt{cmplx* hio)}
   \end{array}
  \end{equation}
  where the inputs and the outputs are
  \begin{equation}
   \begin{array}{ll}
    \textsc{Inputs:}& \\
    \texttt{cmplx fi[6]}   & \textrm{flowing-In  fermion wavefunction}\\
    \texttt{cmplx fo[6]}   & \textrm{flowing-Out fermion wavefunction}\\
    \texttt{cmplx gc[2]}   & \textrm{coupling constant of \texttt{FFS} vertex}\\
    \texttt{double smass}   & \textrm{mass of output scalar} \\
    \texttt{double swidth}  & \textrm{width of output scalar} \\
    &\\
    \textsc{Outputs:}& \\
    \texttt{cmplx hio[3]}  & \textrm{scalar current}\\
    & j(\texttt{<fo|s|fi>})
     \label{eq:hioxxx}
   \end{array}
  \end{equation}

\begin{lstlisting}[caption=hioxxx.cu,label=list:hioxxx]
#include "cmplx.h"

__device__
void hioxxx(cmplx* fi, cmplx* fo, cmplx* g,
            double smass, double swidth, cmplx* hio)
{
   hio[1] = fo[4]-fi[4];
   hio[2] = fo[5]-fi[5];
   double q[4];
   q[0] = hio[1].re;
   q[1] = hio[2].re;
   q[2] = hio[2].im;
   q[3] = hio[1].im;
   double q2 = (q[0]+q[3])*(q[0]-q[3])-(q[1]*q[1]+q[2]*q[2]);
   cmplx dn = -mkcmplx(q2-smass*smass, smass*swidth);
   hio[0] = (1./dn)*(g[0]*(fo[0]*fi[0]+fo[1]*fi[1])
                    +g[1]*(fo[2]*fi[2]+fo[3]*fi[3]));
}
\end{lstlisting}

  \subsection{VVV vertex}
  \label{sec:vvv-vertex}
The \texttt{VVV} vertex functions are obtained from the Lagrangian 
  \begin{eqnarray}
    \mathcal{L}_{\mathrm{V}\! \mathrm{V}\! \mathrm{V}}
     \!= -i \texttt{gc}
     &&\left\{
      \left(\partial_\mu V^{\ast}_{1\nu}\right)
      \left(V^{\mu \ast}_{2}V^{\nu \ast}_{3}
      -V^{\nu \ast}_{2}V^{\mu \ast}_{3}\right) \nonumber \right.\\
   &&+
      \left(\partial_\mu V^{\ast}_{2\nu}\right)
      \left(V^{\mu \ast}_{3}V^{\nu \ast}_{1}
      -V^{\nu \ast}_{3}V^{\mu \ast}_{1}\right) \nonumber \\
   &&\left.\!
     +
      \left(\partial_\mu V^{\ast}_{3\nu}\right)
      \left(V^{\mu \ast}_{1}V^{\nu \ast}_{2}
      -V^{\nu \ast}_{1}V^{\mu \ast}_{2}\right)
     \right\}\,,
  \end{eqnarray}
where the boson names are given by the flowing out quantum numbers.
The same functions, \texttt{vvvxxx} for amplitudes and \texttt{jvvxxx} for
off-shell currents are used to compute both SU(2) vertices, $WW\gamma$ and $WWZ$,
and the QCD triple-gluon vertex. Gauge boson names and the coupling (color factors) for
each vertex are summarized in Table~\ref{tab:vvvxxx}.

  \subsubsection{vvvxxx}
  \label{sec:vvvxxx}
\begin{lstlisting}[caption=vvvxxx.cu,label=list:vvvxxx]
#include "cmplx.h"

__device__
void vvvxxx(cmplx* v1, cmplx* v2, cmplx* v3, 
            double g, cmplx& vertex)
{
   cmplx v12 = v1[0]*v2[0] - v1[1]*v2[1]
             - v1[2]*v2[2] - v1[3]*v2[3];
   cmplx v23 = v2[0]*v3[0] - v2[1]*v3[1] 
             - v2[2]*v3[2] - v2[3]*v3[3];
   cmplx v31 = v3[0]*v1[0] - v3[1]*v1[1]
             - v3[2]*v1[2] - v3[3]*v1[3];
   double pv1[4];
   double pv2[4];
   double pv3[4];
   pv1[0] = v1[4].re;
   pv1[1] = v1[5].re;
   pv1[2] = v1[5].im;
   pv1[3] = v1[4].im;
   pv2[0] = v2[4].re;
   pv2[1] = v2[5].re;
   pv2[2] = v2[5].im;
   pv2[3] = v2[4].im;
   pv3[0] = v3[4].re;
   pv3[1] = v3[5].re;
   pv3[2] = v3[5].im;
   pv3[3] = v3[4].im;
   cmplx p12 = pv1[0]*v2[0] - pv1[1]*v2[1]
             - pv1[2]*v2[2] - pv1[3]*v2[3];
   cmplx p13 = pv1[0]*v3[0] - pv1[1]*v3[1] 
             - pv1[2]*v3[2] - pv1[3]*v3[3];
   cmplx p21 = pv2[0]*v1[0] - pv2[1]*v1[1]
             - pv2[2]*v1[2] - pv2[3]*v1[3];
   cmplx p23 = pv2[0]*v3[0] - pv2[1]*v3[1] 
             - pv2[2]*v3[2] - pv2[3]*v3[3];
   cmplx p31 = pv3[0]*v1[0] - pv3[1]*v1[1] 
             - pv3[2]*v1[2] - pv3[3]*v1[3];
   cmplx p32 = pv3[0]*v2[0] - pv3[1]*v2[1] 
             - pv3[2]*v2[2] - pv3[3]*v2[3];
   vertex = -(v12*(p13-p23)
            + v23*(p21-p31)
            + v31*(p32-p12))*g;
}
\end{lstlisting}

  The function \texttt{vvvxxx} (List~\ref{list:vvvxxx})
  computes the amplitude of the \texttt{VVV} vertex from vector boson
  wave functions, whether they are on-shell or off-shell.
  The function has the arguments:
  \begin{equation} 
   \begin{array}{r}
   \texttt{vvvxxx(cmplx* v1, cmplx* v2, cmplx* v3, } \\
   \texttt{double g, cmplx \pmb{vertex})} 
    \end{array}
    \label{eq:vvv-function}
  \end{equation}
  where the inputs and the outputs are:
  \begin{equation}
   \begin{array}{ll}
    \textsc{Inputs:} & \\
    \texttt{cmplx v1[6]} & \textrm{wavefunction of vector boson} \\
    \texttt{cmplx v2[6]} & \textrm{wavefunction of vector boson} \\
    \texttt{cmplx v3[6]} & \textrm{wavefunction of vector boson} \\
    \texttt{double g} & \textrm{coupling constant of \texttt{VVV}
      vertex} \\ \\
    \textsc{Outputs:} & \\
    \texttt{cmplx vertex} & \textrm{amplitude of the \texttt{VVV} 
     vertex} 
     \label{eq:vvv}
    \end{array}
  \end{equation}
 
  \begin{table*}[htb]
   \centering
   \caption{The possible sets of the inputs for \texttt{vvvxxx} and \texttt{jvvxxx} where all the bosons are named after the flowing out quantum number}
    \begin{tabular}{ccccccc}
     \hline
     \texttt{v1} & \texttt{v2} & \texttt{v3} or \texttt{jvv} & \texttt{gc} & \texttt{vmass} & \texttt{vwidth} & \texttt{color}\\
     \hline 
     $W^-$ & $W^+$ & $Z$   & $g_{WWZ}=g_W\cos\theta_W$ & $m_Z$ & $\Gamma_Z$ &  \\
     $W^-$ & $W^+$ & $A$   & $g_{WWA}=e=g_W\sin\theta_W$ & 0 & 0 & \\
     $Z/A$ & $W^-$ & $W^+$ & $g_{WWZ}/g_{WWA}$ & $\mW$ & $\Gamma_W$ & \\
     $W^+$ & $Z/A$ & $W^-$ & $g_{WWZ}/g_{WWA}$ & $\mW$ & $\Gamma_W$ & \\
     \\
     $g^a$ & $g^b$ & $g^c$ & $g_s$ & 0 & 0 & $f^{abc}$\\
     \hline
    \end{tabular}
   \label{tab:vvvxxx}
  \end{table*}

  \subsubsection{jvvxxx}
  \label{sec:jvvxxx}

  The function \texttt{jvvxxx} (List~\ref{list:jvvxxx})
  computes an off-shell vector current from the three-point gauge boson
  coupling, \texttt{VVV}.
  The vector propagator is given in Feynman gauge for a massless
  vector and in unitary gauge for a massive vector.
  The argument of this function as:
  \begin{equation}
   \begin{array}{r}
    \texttt{jvvxxx(cmplx* v1, cmplx* v2, double g,}\\
    \texttt{double vmass, double vwidth,}\\
    \texttt{cmplx* jvv)}
   \end{array}
  \end{equation}
  where the inputs and the outputs are
  \begin{equation}
   \begin{array}{ll}
    \textsc{Inputs:}& \\
    \texttt{cmplx v1[6]} & \textrm{first vector wavefunction}\\
    \texttt{cmplx v2[6]} & \textrm{second vector wavefunction}\\
    \texttt{double g}     & \textrm{coupling constant (see
     Table~\ref{tab:vvvxxx})
     } \\
    \texttt{double vmass} & \textrm{mass of output vector v}\\
    \texttt{double vwidth}& \textrm{width of output vector v}\\
    &\\
    \textsc{Outputs:}& \\
    \texttt{cmplx jvv[6]}  & \textrm{vector current}\\
    & j^\mu(v:v1,v2)
     \label{eq:jvvxxx}
   \end{array}
  \end{equation}
  Possible sets of inputs are listed in Table~\ref{tab:vvvxxx},
  where all the bosons are defined by the flowing-Out quantum number
  and momentum direction.

\begin{lstlisting}[caption=jvvxxx.cu,label=list:jvvxxx]
#include "cmplx.h"

__device__
void jvvxxx(cmplx* v1,   cmplx* v2, double g,
            double vmass, double vwidth, cmplx* jvv)
{
   jvv[4] = v1[4] + v2[4];
   jvv[5] = v1[5] + v2[5];
   double p1[4];
   double p2[4];
   double q[4];
   p1[0] = v1[4].re;
   p1[1] = v1[5].re;
   p1[2] = v1[5].im;
   p1[3] = v1[4].im;
   p2[0] = v2[4].re;
   p2[1] = v2[5].re;
   p2[2] = v2[5].im;
   p2[3] = v2[4].im;
   q[0] = -(jvv[4].re);
   q[1] = -(jvv[5].re);
   q[2] = -(jvv[5].im);
   q[3] = -(jvv[4].im);
   double q2 = (q[0]+q[3])*(q[0]-q[3])-(q[1]*q[1]+q[2]*q[2]);
   cmplx v12 = v1[0]*v2[0] -v1[1]*v2[1]
              -v1[2]*v2[2] -v1[3]*v2[3];
   cmplx sv1 = (p2[0]-q[0])*v1[0]-(p2[1]-q[1])*v1[1]
              -(p2[2]-q[2])*v1[2]-(p2[3]-q[3])*v1[3];
   cmplx sv2 =-(p1[0]-q[0])*v2[0]+(p1[1]-q[1])*v2[1]
              +(p1[2]-q[2])*v2[2]+(p1[3]-q[3])*v2[3];
   cmplx j12[4];
   j12[0] = (p1[0]-p2[0])*v12 + sv1*v2[0] + sv2*v1[0];
   j12[1] = (p1[1]-p2[1])*v12 + sv1*v2[1] + sv2*v1[1];
   j12[2] = (p1[2]-p2[2])*v12 + sv1*v2[2] + sv2*v1[2];
   j12[3] = (p1[3]-p2[3])*v12 + sv1*v2[3] + sv2*v1[3];
   if (vmass!=0.) {
      double vm2 = vmass*vmass;
      double m1 = (p1[0])*(p1[0])-(p1[1])*(p1[1])
                 -(p1[2])*(p1[2])-(p1[3])*(p1[3]);
      double m2 = (p2[0])*(p2[0])-(p2[1])*(p2[1])
                 -(p2[2])*(p2[2])-(p2[3])*(p2[3]);
      cmplx s11 = (p1[0])*v1[0]-(p1[1])*v1[1]
                 -(p1[2])*v1[2]-(p1[3])*v1[3];
      cmplx s12 = (p1[0])*v2[0]-(p1[1])*v2[1]
                 -(p1[2])*v2[2]-(p1[3])*v2[3];
      cmplx s21 = (p2[0])*v1[0]-(p2[1])*v1[1]
                 -(p2[2])*v1[2]-(p2[3])*v1[3];
      cmplx s22 = (p2[0])*v2[0]-(p2[1])*v2[1]
                 -(p2[2])*v2[2]-(p2[3])*v2[3];
      cmplx cm2 = mkcmplx( vm2, -vmass*vwidth );
      cmplx js = (v12*(-m1+m2)+s11*s12-s21*s22)*(1./cm2);
      cmplx dg = -g*(1./mkcmplx(q2-vm2, vmass*vwidth ));
      jvv[0] = dg*(j12[0]-q[0]*js);
      jvv[1] = dg*(j12[1]-q[1]*js);
      jvv[2] = dg*(j12[2]-q[2]*js);
      jvv[3] = dg*(j12[3]-q[3]*js);
   } else {
      double gs = -g*(1./q2);
      jvv[0]=gs*j12[0];
      jvv[1]=gs*j12[1];
      jvv[2]=gs*j12[2];
      jvv[3]=gs*j12[3];
   }
}

\end{lstlisting}

  \subsection{VVVV vertex}
  \label{sec:vvvv-vertex}
  The \texttt{VVVV} vertex functions are obtained from the following Lagrangian
  \begin{eqnarray}
    \mathcal{L}_{\mathrm{V\!V\!V\!V}}
     \!= \mathtt{gg}&& \bigg[(V_1^\ast\cdot V_4^\ast)(V_2^\ast\cdot V_3^\ast)
    \nonumber \\
  &&\quad \quad -(V_1^\ast\cdot V_3^\ast)(V_2^\ast\cdot V_4^\ast)\bigg],
   \label{xab-b}
  \end{eqnarray}
  when all four vector bosons are distinct. We may express the amplitude and the $V_4$ current made by the above interaction as
  \begin{subequations}\label{appvvvv}
 \begin{align}
  & \Gamma(\texttt{gg}; V_1, V_2,V_3,V_4),     \label{appvvvv1}  \\
  & j^\mu_{V_4}(\texttt{gg}; V_1, V_2,V_3),     \label{appvvvv2} 
 \end{align}
 which are shown as List~\ref{list:ggggxx} and List~\ref{list:jgggxx}, respectively.
  \end{subequations}

  We note that the SU(2) weak boson Lagrangian
  \begin{eqnarray}
    \mathcal{L}_{\mathrm{W\!W\!W\!W}}
     &=& -\frac{\gwsq}{2} \bigg[(W^{-*}\!\cdot W^{+*})(W^{+*}\!\cdot W^{-*})
     \nonumber \\
   &&\,\,\,  -(W^{-*}\!\cdot W^{-*})(W^{+*}\!\cdot W^{+*})\bigg] \nonumber \\ 
   &&+\gw^{2} \bigg[(W^{-*}\!\cdot W^{3*})(W^{3*}\!\cdot W^{+*})
    \nonumber \\
   &&\,\,\,  -(W^{-*}\!\cdot W^{+*})(W^{3*}\!\cdot W^{3*})\bigg],
    \label{eq:xab-c}
  \end{eqnarray}
  can be expressed in terms of the Lagrangian~(\ref{xab-b}) by identifying
  \begin{subequations}\label{appv4}
  \begin{align}
   (V_1,\, V_2,\, V_3,\, V_4)  = & (W^-,\, W^+,\, W^-,\, W^+) \nonumber \\
 &\quad\quad {\rm for}\,\, W^-W^+W^-W^+    \label{appv41}  \\
   & (W^-,\, W^3,\, W^+,\, W^3) \nonumber \\
 &\quad\quad {\rm for}\,\, W^-W^3W^+W^3 ,     \label{appv42} 
  \end{align}
  \end{subequations}
  Accordingly, the amplitudes made of four weak bosons can be expressed in terms of the amplitudes of the Lagrangian~(\ref{xab-b}) 
  \begin{subequations}\label{eq:appv4b}
   \begin{align}
    & \Gamma(W^-_1, W^+_2,W^-_3,W^+_4) \nonumber\\
    & = \quad\Gamma(-g^2_w; W^-_1, W^+_2,W^-_3,W^+_4) \nonumber\\
    &\quad +\Gamma(-g^2_w; W^-_1, W^+_4,W^-_3,W^+_2),     \label{eq:appv4b1}  \\
    & \Gamma(W^-_1, W^3_2,W^+_3,W^3_4) \nonumber\\
    & = \quad\Gamma(g^2_w; W^-_1, W^3_2,W^+_3,W^3_4) \nonumber\\
    & \quad +\Gamma(g^2_w; W^-_1, W^3_4,W^+_3,W^3_2),       \label{eq:appv4b2} 
   \end{align}
  \end{subequations}
  and likewise for off-shell currents. 
  Because the functional form of the right-hand sides of Eq.~(\ref{eq:appv4b}) are identical, where only the sign of the overall coupling is opposite, we introduced only one set of HEGET functions \texttt{wwwwxx} for amplitudes and \texttt{jwwwxx} for off-shell currents. 
  These functions are listed in \ref{sec:wwwwxx} and \ref{sec:jwwwxx}, respectively, and inputs ordering and the couplings for all possible electroweak-boson vertices are summarized in Table~\ref{tab:input_jwwwxx}.

  The QCD quartic gluon vertices are obtained from the Lagrangian 
  \begin{eqnarray}
    \mathcal{L}_{\mathrm{g}\! \mathrm{g}\! \mathrm{g}\! \mathrm{g}}
     \!= -\dfrac{g_s^2}{4} f^{abe}f^{cde} (A^a\cdot A^c)(A^b\cdot A^d).
   \label{x4gb}
  \end{eqnarray}
  Since all the gluon fields are now different, we can use the functions (\ref{appvvvv}) to obtain the color amplitudes for $g_1^a$, $g_2^b$, $g_3^c$, $g_4^d$ as follows
  \begin{eqnarray}
   \lefteqn{f^{abe}f^{cde}\Gamma(g_{s}^{2};g_{1}^{a},g_{2}^{b},g_{3}^{c},g_{4}^{d})} \nonumber \\
   & + & f^{ace}f^{dbe}\Gamma(g_{s}^{2};g_{1}^{a},g_{2}^{c},g_{3}^{d},g_{4}^{b}) \nonumber \\
   & + & f^{adc}f^{bce}\Gamma(g_{s}^{2};g_{1}^{a},g_{2}^{d},g_{3}^{b},g_{4}^{c}) 
  \end{eqnarray}
  and similarly for the off-shell wave function of $g_4^d$, see Eqs.~(\ref{eq:gggg-amplitudes}) and (\ref{xxjggg}).

  \subsubsection{wwwwxx}
  \label{sec:wwwwxx}
  The function \texttt{wwwwxx} (List~\ref{list:wwwwxx})
  computes an amplitude of the four-point electroweak boson couplings
  of Eq.~(\ref{eq:xab-c}).
  All wave functions are defined by the flowing-Out quantum number.
  %
  The argument of this function is:
  \begin{equation}
   \begin{array}{r}
    \texttt{wwwwxx(cmplx* v1, cmplx* v2,}\\
    \texttt{cmplx* v3, cmplx* v4,}\\
    \texttt{double gwwa, double gwwz, cmplx\& vertex)}
   \end{array}
  \end{equation}
  where the inputs and the outputs are
  \begin{equation}
   \begin{array}{ll}
    \textsc{Inputs:}& \\
    \texttt{cmplx v1[6]} & \textrm{first flow-Out v1 vector wavefunction}\\
    \texttt{cmplx v2[6]} & \textrm{first flow-Out v2 vector wavefunction}\\
    \texttt{cmplx v3[6]} & \textrm{second flow-Out v3 vector wavefunction}\\
    \texttt{cmplx v4[6]} & \textrm{second flow-Out v4 vector wavefunction}\\
    \texttt{double gwwa} & \textrm{coupling constant of $W$ and $A$} \\
    \texttt{double gwwz} & \textrm{coupling constant of $W$ and $Z$} \\
    &\\
    \textsc{Outputs:}& \\
    \texttt{cmplx\& vertex}& \textrm{amplitude}\\
    &\Gamma(\texttt{wm1},\texttt{wp1}, \texttt{wm2},\texttt{wp2})
     \label{eq:wwwwxx}
   \end{array}
  \end{equation}

\begin{lstlisting}[caption=wwwwxx.cu,label=list:wwwwxx]
#include "cmplx.h"

__device__
void wwwwxx(cmplx* v1, cmplx* v2,
            cmplx* v3, cmplx* v4,
            double gwwa, double gwwz, cmplx& vertex)
{
   cmplx v12 = v1[0]*v2[0]-v1[1]*v2[1]
              -v1[2]*v2[2]-v1[3]*v2[3];
   cmplx v13 = v1[0]*v3[0]-v1[1]*v3[1]
              -v1[2]*v3[2]-v1[3]*v3[3];
   cmplx v14 = v1[0]*v4[0]-v1[1]*v4[1]
              -v1[2]*v4[2]-v1[3]*v4[3];
   cmplx v23 = v2[0]*v3[0]-v2[1]*v3[1]
              -v2[2]*v3[2]-v2[3]*v3[3];
   cmplx v24 = v2[0]*v4[0]-v2[1]*v4[1]
              -v2[2]*v4[2]-v2[3]*v4[3];
   cmplx v34 = v3[0]*v4[0]-v3[1]*v4[1]
              -v3[2]*v4[2]-v4[3]*v4[3];
   vertex = -g*(v12*v34 + v14*v23 -2.*v13*v24) *
      (gwwa*gwwa+gwwz*gwwz);
}

\end{lstlisting}

  \subsubsection{jwwwxx}
  \label{sec:jwwwxx}

  The function \texttt{jwwwxx} (List~\ref{list:jwwwxx})
  computes an off-shell electroweak boson current from
  the four-point gauge boson coupling, Eq.~(\ref{eq:xab-c})
  The vector propagators for the output $W^\pm$ and $Z$
  are given in the unitary gauge, while the photon current is given in the Feynman gauge.
  The arguments of this function are:
  \begin{equation}
   \begin{array}{r}
    \texttt{jwwwxx(cmplx* w1, cmplx* w2, cmplx* w3,}\\
    \texttt{double gwwa, double gwwz,}\\
    \texttt{double wmass, double wwidth,}\\
    \texttt{cmplx* jwww)}\\
   \end{array}
  \end{equation}
  where the inputs and the outputs are:
  \begin{equation}
   \begin{array}{ll}
    \textsc{Inputs:}& \\
    \texttt{cmplx w1[6]} & \textrm{first vector}\\
    \texttt{cmplx w2[6]} & \textrm{second vector}\\
    \texttt{cmplx w3[6]} & \textrm{third vector}\\
    \texttt{double gwwa} & \textrm{coupling constant of $W$ and $A$} \\
    \texttt{double gwwz} & \textrm{coupling constant of $W$ and $Z$} \\
    \texttt{double wmass} & \textrm{mass of output}~~W\\
    \texttt{double wwidth}& \textrm{width of output}~~W\\
    &\\
    \textsc{Outputs:}& \\
    \texttt{cmplx jwww[6]} & \textrm{W current}\\
    &j^\mu(\texttt{w}^\prime:
     \texttt{w1},\texttt{w2},\texttt{w3})
     \label{eq:jwww}
   \end{array}
  \end{equation}
  The possible sets of the inputs are listed in Table~\ref{tab:input_jwwwxx}, in which all the bosons are defined by the flowing-Out quantum number.
  The other arguments are
  \begin{eqnarray}
   \texttt{gwwa} = g_{WWA} &&  \texttt{gwwz} =g_{WWZ} \nonumber \\
   \texttt{zmass}= m_Z^{}  &&  \texttt{zwidth} = \Gamma_Z \nonumber \\
   \texttt{wmass}= m_W^{}  &&  \texttt{wwidth} = \Gamma_W.
    \label{eq:input_jwwwxx2}
  \end{eqnarray}

\begin{lstlisting}[caption=jwwwxx.cu,label=list:jwwwxx]
#include "cmplx.h"

__device__
void jwwwxx(cmplx* w1, cmplx* w2, cmplx* w3,
            double gwwa, double gwwz, double wmass,
            double wwidth, cmplx* jwww)
{
   jwww[4] = w1[4]+w2[4]+w3[4];
   jwww[5] = w1[5]+w2[5]+w3[5];
   double q[4];
   q[0] = -(w1[4].re + w2[4].re + w3[4].re);
   q[1] = -(w1[5].re + w2[5].re + w3[5].re);
   q[2] = -(w1[5].im + w2[5].im + w3[5].im);
   q[3] = -(w1[4].im + w2[4].im + w3[4].im);
   double q2 = (q[0]+q[3])*(q[0]-q[3])-(q[1]*q[1]+q[2]*q[2]);
   double gw2 = (gwwa*gwwa)+(gwwz*gwwz);
   double mw2 = wmass*wmass;
   cmplx  dw = -1./mkcmplx(q2-mw2,wmass*wwidth);
   cmplx w12 = w1[0]*w2[0]-w1[1]*w2[1]
              -w1[2]*w2[2]-w1[3]*w2[3];
   cmplx w32 = w3[0]*w2[0]-w3[1]*w2[1]
              -w3[2]*w2[2]-w3[3]*w2[3];
   cmplx w13 = w1[0]*w3[0]-w1[1]*w3[1]
              -w1[2]*w3[2]-w1[3]*w3[3];
   cmplx jj[4];
   jj[0] = gw2*(w1[0]*w32+w3[0]*w12-2.*w2[0]*w13);
   jj[1] = gw2*(w1[1]*w32+w3[1]*w12-2.*w2[1]*w13);
   jj[2] = gw2*(w1[2]*w32+w3[2]*w12-2.*w2[2]*w13);
   jj[3] = gw2*(w1[3]*w32+w3[3]*w12-2.*w2[3]*w13);
   cmplx cm2 = mkcmplx(mw2,-wmass*wwidth);
   cmplx jq = (jj[0]*q[0]-jj[1]*q[1]
              -jj[2]*q[2]-jj[3]*q[3])*(1./cm2);
   jwww[0] = (jj[0]-jq*q[0])*dw;
   jwww[1] = (jj[1]-jq*q[1])*dw;
   jwww[2] = (jj[2]-jq*q[2])*dw;
   jwww[3] = (jj[3]-jq*q[3])*dw;
}
\end{lstlisting}

  \subsubsection{ggggxx}
  \label{sec:ggggxx}

  The function \texttt{ggggxx} (List~\ref{list:ggggxx}) computes
  the amplitude from 4 gluon wave functions \texttt{ga}, \texttt{gb},
  \texttt{gc} and  \texttt{gd}, each with the color index $a$, $b$, $c$
  and $d$, respectively,  
  when the associated color factor is $f^{abe} f^{cde}$, whether 
  the gluons are on-shell or off-shell.  
  The function has the arguments: 
  \begin{equation} 
   \begin{array}{r}
   \texttt{ggggxx(cmplx* ga, cmplx* gb, cmplx* gc,} \\
   \texttt{cmplx* gd, double gg, cmplx vertex)} 
    \end{array}
  \end{equation}
  where the inputs and the outputs are:
  \begin{equation}
   \begin{array}{ll}
    \textsc{Inputs:} & \\
     \texttt{cmplx ga[6]} & \textrm{wavefunction of gluon with 
      color} \\
    & \textrm{index, $a$} \\ 
     \texttt{cmplx gb[6]} & \textrm{wavefunction of gluon with 
      color} \\
    & \textrm{index, $b$} \\ 
     \texttt{cmplx gc[6]} & \textrm{wavefunction of gluon with 
      color} \\
    & \textrm{index, $c$} \\ 
     \texttt{cmplx gd[6]} & \textrm{wavefunction of gluon with 
      color} \\
    & \textrm{index, $d$} \\ 
     \texttt{double gg} & \textrm{coupling constant of \texttt{VVVV}
      vertex} \\ \\
    \textsc{Outputs:} & \\
    \texttt{cmplx vertex} & \textrm{amplitude of the \texttt{VVVV} 
     vertex with the color} \\
    & \textrm{factor $f^{abe} f^{cde}$.} 
    \end{array}
  \end{equation}
  The coupling constant \texttt{gg} for the $gggg$ vertex is
  \begin{equation}
   \texttt{gg} = g_{s}^{2}\,.
  \end{equation}
  In order to obtain the complete amplitude, the function must be called
  three times (once for each color structure) with the following
  permutations:
  \begin{subequations}
   \begin{eqnarray}
    \mathtt{ggggxx(ga,gb,gc,gd,gg,\pmb{\texttt{v1}})} & \\
    \mathtt{ggggxx(ga,gc,gd,gb,gg,\pmb{\texttt{v2}})} & \\
    \mathtt{ggggxx(ga,gd,gb,gc,gg,\pmb{\texttt{v3}})} &
   \end{eqnarray}
   \label{eq:gggg-functions-calls}
  \end{subequations}
  The color amplitudes are then expressed as
  \begin{equation}
   f^{abe} f^{cde}\, (\pmb{\texttt{v1}})
    + f^{ace} f^{dbe}\, (\pmb{\texttt{v2}})
    + f^{ade} f^{bce}\, (\pmb{\texttt{v3}})\,.
   \label{eq:gggg-amplitudes}
  \end{equation}
\begin{lstlisting}[caption=ggggxx.cu,label=list:ggggxx]
#include <cmplx.h>

__device__
void ggggxx(cmplx* ga, cmplx* gb, cmplx* gc, cmplx* gd,
            double gg, cmplx& vertex)
{
   cmplx gad = ga[0]*gd[0]-ga[1]*gd[1]
              -ga[2]*gd[2]-ga[3]*gd[3];
   cmplx gbc = gb[0]*gc[0]-gb[1]*gc[1]
              -gb[2]*gc[2]-gb[3]*gc[3];
   cmplx gac = ga[0]*gc[0]-ga[1]*gc[1]
              -ga[2]*gc[2]-ga[3]*gc[3];
   cmplx gbd = gb[0]*gd[0]-gb[1]*gd[1]
              -gb[2]*gd[2]-gb[3]*gd[3];
   vertex = (gg*gg)*(gad*gbc-gac*gbd);
}
\end{lstlisting}

  \subsubsection{jgggxx}
  \label{sec:jggg}

  The function \texttt{jgggxx} (List~\ref{list:jgggxx})
  computes an off-shell gluon current from the 
  four-point gluon coupling, including the gluon propagator in the 
  Feynman gauge. 
  It has the arguments:
  \begin{equation} 
   \begin{array}{r}
   \texttt{jgggxx(cmplx* ga, cmplx* gb, cmplx* gc,} \\
   \texttt{double gg, cmplx* jggg)} 
   \end{array}
   \label{eq:jggg-func}
  \end{equation}
  where the inputs and the outputs are:
  \begin{equation}
   \begin{array}{ll}
    \textsc{Inputs:} & \\
     \texttt{cmplx ga[6]} & \textrm{wavefunction of gluon with
      color} \\ 
    & \textrm{index, $a$} \\ 
     \texttt{cmplx gb[6]} & \textrm{wavefunction of gluon with
      color} \\
    & \textrm{index, $b$} \\ 
     \texttt{cmplx gc[6]} & \textrm{wavefunction of gluon with
      color} \\
    & \textrm{index, $c$} \\ 
     \texttt{double gg} & \textrm{coupling constants of the
      \texttt{VVVV} vertex}  \\ \\
    \textsc{Outputs:} & \\
     \texttt{cmplx jggg[6]} & \textrm{vector current
      $\pmb{j}^{\mu}(d:a,b,c)$
      which has} \\ 
    & \textrm{the color index $d$ associated with the} \\
    & \textrm{color factor $f^{abe} f^{cde}$.} 
    \end{array}
  \end{equation}
   The function (\ref{eq:jggg-func}) should be called three times
  \begin{subequations}
   \begin{eqnarray}
    \texttt{jgggxx(ga,gb,gc,gg,\pmb{j1})} & \\
    \texttt{jgggxx(gc,ga,gb,gg,\pmb{j2})} & \\
    \texttt{jgggxx(gb,gc,ga,gg,\pmb{j3})} &
   \end{eqnarray}
  \end{subequations}
   as in Eq.~(\ref{eq:gggg-functions-calls}), and the off-shell gluon
   current with the color index $d$ is obtained as
  \begin{equation}
   f^{abe} f^{cde}\, (\pmb{\texttt{j1}})
    + f^{ace} f^{dbe}\, (\pmb{\texttt{j2}})
    + f^{ade} f^{bce}\, (\pmb{\texttt{j3}})\,.
    \label{xxjggg}
  \end{equation}
\begin{lstlisting}[caption=jgggxx.cu,label=list:jgggxx]
#include <cmplx.h>

__device__
void jgggxx(cmplx* ga, cmplx* gb, cmplx* gc, 
            double gg, cmplx* jggg)
{
   jggg[4] = ga[4]+gb[4]+gc[4];
   jggg[5] = ga[5]+gb[5]+gc[5];
   double q[4];
   q[0] = -jggg[4].re;
   q[1] = -jggg[5].re;
   q[2] = -jggg[5].im;
   q[3] = -jggg[4].im;
   double q2 = (q[0]+q[3])*(q[0]-q[3])-(q[1]*q[1]+q[2]*q[2]);
   double fact = gg*gg*(1./q2);
   cmplx gcb = gc[0]*gb[0]-gc[1]*gb[1]
              -gc[2]*gb[2]-gc[3]*gb[3];
   cmplx gac = ga[0]*gc[0]-ga[1]*gc[1]
              -ga[2]*gc[2]-ga[3]*gc[3];
   jggg[0] = fact*( ga[0]*gcb - gb[0]*gac );
   jggg[1] = fact*( ga[1]*gcb - gb[1]*gac );
   jggg[2] = fact*( ga[2]*gcb - gb[2]*gac );
   jggg[3] = fact*( ga[3]*gcb - gb[3]*gac );
}
\end{lstlisting}

 \begin{table}[htb]
  \centering
   \caption{Possible sets of inputs for \texttt{wwwwxx} and \texttt{jwwwxx}
   where all the boson names give the flowing-Out quantum number.}
    \begin{tabular}{|c|c|c|c|c|c|c|c|}
     \hline
     \texttt{w1}& \texttt{w2}& \texttt{w3} & \texttt{w4} or \texttt{jwww} \\
     \hline
     $W^-$& $W^+$ & $W^-$ & $W^+$ \\
     $W^+$& $W^-$ & $W^+$ & $W^-$ \\
     \hline
    \end{tabular}
    \label{tab:input_jwwwxx}
  \end{table}

  \subsection{VVS vertex}
  \label{sec:vvs-vertex}
The \texttt{VVS} vertex functions are obtained from the following
Lagrangian
  \begin{equation}
   \mathcal{L}_{\mathrm{V}\! \mathrm{V}\! \mathrm{S}}
    \!= \texttt{gc} \left(V_{1}^{\ast} \cdot V_{2}^{\ast}\right)S^{*}.
  \end{equation}

  \subsubsection{vvsxxx}
  \label{sec:vvsxxx}

  The function \texttt{vvsxxx} (List~\ref{list:vvsxxx})
  computes an amplitude of the \texttt{VVS} coupling.
  The arguments of this function as:
  \begin{equation}
   \begin{array}{r}
    \texttt{vvsxxx(cmplx* v1, cmplx* v2, cmplx* sc,}\\
    \texttt{cmplx gc, cmplx\& vertex)}\\
   \end{array}
  \end{equation}
  where the inputs and the outputs are:
  \begin{equation}
   \begin{array}{ll}
    \textsc{Inputs:}& \\
    \texttt{cmplx v1[6]} & \textrm{first vector wavefunction}\\
    \texttt{cmplx v2[6]} & \textrm{second vector wavefunction}\\
    \texttt{cmplx sc[3]} & \textrm{input scalar}\\
    \texttt{cmplx gc}    & \textrm{coupling constant of \texttt{VVS}}\\
    &\\
    \textsc{Outputs:}& \\
    \texttt{cmplx\& vertex} & \textrm{amplitude}\\
    &\Gamma(\texttt{v1},\texttt{v2},\texttt{s})
     \label{eq:vvsxxx}
   \end{array}
  \end{equation}

\begin{lstlisting}[caption=vvsxxx.cu,label=list:vvsxxx]
#include "cmplx.h"

__device__
void vvsxxx(cmplx* v1, cmplx* v2, cmplx* sc,
            cmplx g, cmplx& vertex)
{
   vertex = g*sc[0]*(v1[0]*v2[0]-v1[1]*v2[1]
                    -v1[2]*v2[2]-v1[3]*v2[3]);
}

\end{lstlisting}

  \subsubsection{jvsxxx}
  \label{sec:jvsxxx}

  The function \texttt{jvsxxx} (List~\ref{list:jvsxxx})
  computes an off-shell vector current from the \texttt{VVS} coupling.
  The vector propagator is given in Feynman gauge for a massless
  vector and in unitary gauge for a massive vector.
  The arguments of this function as:
  \begin{equation}
   \begin{array}{r}
    \texttt{jvsxxx(cmplx* vc, cmplx* sc, cmplx gc,}\\
    \texttt{double vmass, double vwidth,}
     \texttt{cmplx* jvs)}\\
   \end{array}
  \end{equation}
  where the inputs and the outputs are:
  \begin{equation}
   \begin{array}{ll}
    \textsc{Inputs:}& \\
    \texttt{cmplx vc[6]} & \textrm{input vector wavefunction}\\
    \texttt{cmplx sc[3]} & \textrm{input scalar wavefunction}\\
    \texttt{cmplx gc}    & \textrm{coupling constant of \texttt{VVS}}\\
    \texttt{double vmass} & \textrm{mass of output vector} \\
    \texttt{double vwidth}& \textrm{width of output vector} \\
    &\\
    \textsc{Outputs:}& \\
    \texttt{cmplx jvs[6]}& \textrm{vector current}\\
    &j^\mu(\texttt{v$^\prime$}: \texttt{vc}, \texttt{sc})
     \label{eq:jvsxxx}
   \end{array}
  \end{equation}

\begin{lstlisting}[caption=jvsxxx.cu,label=list:jvsxxx]
#include "cmplx.h"

__device__
void jvsxxx(cmplx* vc, cmplx* sc, cmplx  gc,
            double vmass, double vwidth, cmplx* jvs)
{
   jvs[4] = vc[4] + sc[1];
   jvs[5] = vc[5] + sc[2];
   double q[4];
   q[0] = jvs[4].re;
   q[1] = jvs[5].re;
   q[2] = jvs[5].im;
   q[3] = jvs[4].im;
   double q2 = (q[0]+q[3])*(q[0]-q[3])-(q[1]*q[1]+q[2]*q[2]);
   if (vmass==0.) { 
      cmplx dg = gc*sc[0]*(1./q2);
      jvs[0] = dg*vc[0];
      jvs[1] = dg*vc[1];
      jvs[2] = dg*vc[2];
      jvs[3] = dg*vc[3];
   } else {
      double vm2 = vmass*vmass;
      cmplx dg = gc*sc[0]
         *(1./mkcmplx(q2-vm2, vmass*vwidth));
      cmplx cm2 = (1./mkcmplx(vm2, -vmass*vwidth ));
      cmplx vk = (-q[0]*vc[0]+q[1]*vc[1]
                  +q[2]*vc[2]+q[3]*vc[3])*cm2;
      jvs[0] = dg*(q[0]*vk+vc[0]);
      jvs[1] = dg*(q[1]*vk+vc[1]);
      jvs[2] = dg*(q[2]*vk+vc[2]);
      jvs[3] = dg*(q[3]*vk+vc[3]);
   }
}

\end{lstlisting}

  \subsubsection{hvvxxx}
  \label{sec:hvvxxx}

  The function \texttt{hvvxxx} (List~\ref{list:hvvxxx}) computes
  an off-shell scalar current from the \texttt{VVS} coupling.
  The arguments of this function as:
  \begin{equation}
   \begin{array}{r}
    \texttt{hvvxxx(cmplx* v1, cmplx* v2, cmplx gc,}\\
    \texttt{double smass, double swidth,}
     \texttt{cmplx* hvv)}\\
   \end{array}
  \end{equation}
  where the inputs and the outputs are:
  \begin{equation}
   \begin{array}{ll}
    \textsc{Inputs:}& \\
    \texttt{cmplx v1[6]} & \textrm{input first vector wavefunction}\\
    \texttt{cmplx v2[6]} & \textrm{input second vector wavefunction}\\
    \texttt{cmplx gc}    & \textrm{coupling constant of \texttt{VVS}}\\
    \texttt{double smass} & \textrm{mass of output scalar} \\
    \texttt{double swidth}& \textrm{width of output scalar} \\
    &\\
    \textsc{Outputs:}& \\
    \texttt{cmplx hvv[3]}& \textrm{scalar current}\\
    &j(\texttt{s$^\prime$}: \texttt{v1}, \texttt{v2})
     \label{eq:hvvxxx}
   \end{array}
  \end{equation}

\begin{lstlisting}[caption=hvvxxx.cu,label=list:hvvxxx]
#include "cmplx.h"

__device__
void hvvxxx(cmplx* v1, cmplx* v2, cmplx  gc, 
            double smass, double swidth, cmplx* hvv)
{
   hvv[1] = v1[4] + v2[4];
   hvv[2] = v1[5] + v2[5];
   double q[4];
   q[0] = hvv[1].re;
   q[1] = hvv[2].re;
   q[2] = hvv[2].im;
   q[3] = hvv[1].im;
   double q2 = (q[0]+q[3])*(q[0]-q[3])-(q[1]*q[1]+q[2]*q[2]);
   cmplx dg = 1./mkcmplx(q2-smass*smass, smass*swidth);
   hvv[0] = -gc*dg*(v1[0]*v2[0]-v1[1]*v2[1]
                   -v1[2]*v2[2]-v1[3]*v2[3]);
}

\end{lstlisting}

  \subsection{VVSS vertex}
  \label{sec:vvss-vertex}
The \texttt{VVSS} vertex functions are obtained from the following
Lagrangian
   \begin{equation}
    \mathcal{L}_{\mathrm{V}\! \mathrm{V}\! \mathrm{S}\! \mathrm{S}}
     \!= \texttt{gc} \left(V_{1}^{\ast} \cdot V_{2}^{\ast}\right)
     S^{\ast}_{3} S^{\ast}_{4}\,.
   \end{equation}

  \subsubsection{vvssxx}
  \label{sec:vvssxx}

  The function \texttt{vvssxx} (List~\ref{list:vvssxx})
  computes an amplitude of the \texttt{VVSS} coupling.
  The arguments of this function as:
  \begin{equation}
   \begin{array}{r}
    \texttt{vvssxx(cmplx* v1, cmplx* v2,~~}\\
    \texttt{cmplx* s1, cmplx* s2,~~}\\
    \texttt{cmplx gc, cmplx\& vertex)}\\
   \end{array}
  \end{equation}
  where the inputs and the outputs are:
  \begin{equation}
   \begin{array}{ll}
    \textsc{Inputs:} & \\
    \texttt{cmplx v1[6]} & \textrm{first vector wavefunction} \\
    \texttt{cmplx v2[6]} & \textrm{second vector wavefunction} \\
    \texttt{cmplx s1[3]} & \textrm{first scalar wavefunction} \\
    \texttt{cmplx s2[3]} & \textrm{second scalar wavefunction} \\
    \texttt{cmplx gc}    & \textrm{coupling constant of \texttt{VVSS}}\\
    & \\
    \textsc{Outputs:} & \\
    \texttt{cmplx\& vertex}& \textrm{amplitude of the \texttt{VVSS}}\\
    & \Gamma(\texttt{v1},\texttt{v2},\texttt{s1},\texttt{s2})
     \label{eq:vvss}
   \end{array}
  \end{equation}

\begin{lstlisting}[caption=vvssxx.cu,label=list:vvssxx]
#include "cmplx.h"

__device__
void vvssxx(cmplx* v1, cmplx* v2, cmplx* s1, cmplx* s2,
            cmplx gc, cmplx& vertex)
{
   vertex = gc*s1[0]*s2[0]*(v1[0]*v2[0]-v1[1]*v2[1]
                           -v1[2]*v2[2]-v1[3]*v2[3]);
}
\end{lstlisting}

  \subsubsection{jvssxx}
  \label{sec:jvssxx}

  The function \texttt{jvssxx} (List~\ref{list:jvssxx})
  computes an off-shell massive vector current from the
  \texttt{VVSS} coupling.
  The vector propagator is given in unitary gauge for a massive vector.
  The arguments of this functions as:
  \begin{equation}
   \begin{array}{r}
    \texttt{jvssxx(cmplx* vc, cmplx* s1, cmplx* s2,~~~~} \\
    \texttt{cmplx gc, double vmass, double vwidth,}\\
    \texttt{cmplx* jvss)}
   \end{array}
  \end{equation}
  where the inputs and the outputs are:
  \begin{equation}
   \begin{array}{ll}
    \textsc{Inputs:} & \\
    \texttt{cmplx vc[6]} & \textrm{input vector wavefunction} \\
    \texttt{cmplx s1[3]} & \textrm{first scalar wavefunction} \\
    \texttt{cmplx s2[3]} & \textrm{second scalar wavefunction} \\
    \texttt{cmplx gc} & \textrm{coupling constants of \texttt{VVSS}
     vertex} \\
    \texttt{double vmass} & \textrm{mass of output vector}\\
    \texttt{double vwidth} & \textrm{width of output vector}\\\\
    \textsc{Outputs:} & \\
    \texttt{cmplx jvss[6]} & \textrm{vector current}\\
    &j^\mu( \texttt{v}^\prime:\texttt{vc},\texttt{s1},\texttt{s2})  \\
    \label{eq:jvss}
   \end{array}
  \end{equation}

\begin{lstlisting}[caption=jvssxx.cu,label=list:jvssxx]
#include <cmplx.h>

__device__
void jvssxx(cmplx* vc, cmplx* s1, cmplx* s2, cmplx gc,
            double vmass, double vwidth, cmplx* jvss)
{
   jvss[4] = vc[4]+s1[1]+s2[1];
   jvss[5] = vc[5]+s1[2]+s2[2];
   double q[4];
   q[0] = jvss[4].re;
   q[1] = jvss[5].re;
   q[2] = jvss[5].im;
   q[3] = jvss[4].im;
   double q2 = (q[0]+q[3])*(q[0]-q[3])-(q[1]*q[1]+q[2]*q[2]);
   if (vmass==0.){
      cmplx dg = gc*s1[0]*s2[0]*(1./q2);
      jvss[0] = dg*vc[0];
      jvss[1] = dg*vc[1];
      jvss[2] = dg*vc[2];
      jvss[3] = dg*vc[3];
   } else {
      double vm2 = vmass*vmass;
      cmplx dg = gc*s1[0]*s2[0]
         *(1./mkcmplx( q2-vm2, vmass*vwidth ));
      cmplx cm2 = mkcmplx( vm2, -vmass*vwidth );
      cmplx vk = 
         (q[0]*vc[0]-q[1]*vc[1]-q[2]*vc[2]-q[3]*vc[3])*(1./cm2);
      jvss[0] = dg*(vc[0]-vk*q[0]);
      jvss[1] = dg*(vc[1]-vk*q[1]);
      jvss[2] = dg*(vc[2]-vk*q[2]);
      jvss[3] = dg*(vc[3]-vk*q[3]);
   }
}
\end{lstlisting}

  \subsubsection{hvvsxx}
  \label{sec:hvvsxx}

  The function \texttt{hvvsxx} (List~\ref{list:hvvsxx})
  computes an off-shell scalar current of \texttt{VVSS} coupling.
  The arguments of this function as:
  \begin{equation}
   \begin{array}{r}
    \texttt{hvvsxx(cmplx* v1, cmplx* v2, cmplx* sc,~~~~} \\
    \texttt{cmplx gc, double smass, double swidth,}\\
    \texttt{cmplx* hvvs)}
   \end{array}
  \end{equation}
  where the inputs and the outputs are:
  \begin{equation}
   \begin{array}{ll}
    \textsc{Inputs:} & \\
    \texttt{cmplx v1[6]} & \textrm{first vector wavefunction} \\
    \texttt{cmplx v2[6]} & \textrm{second vector wavefunction} \\
    \texttt{cmplx sc[3]} & \textrm{input scalar wavefunction} \\
    \texttt{cmplx gc}    & \textrm{coupling constant of \texttt{VVSS}} \\
    \texttt{double smass} & \textrm{mass of output scalar}\\
    \texttt{double swidth}& \textrm{mass of output scalar} \\ \\
    \textsc{Outputs:} & \\
    \texttt{cmplx hvvs[3]} & \textrm{scalar current}\\
    &j(\texttt{s}^\prime:\texttt{v1}, \texttt{v2},\texttt{sc}) \\
    \label{eq:hvvsxx}
   \end{array}
  \end{equation}

\begin{lstlisting}[caption=hvvsxx.cu,label=list:hvvsxx]
#include "cmplx.h"

__device__
void hvvsxx(cmplx* v1, cmplx* v2, cmplx* sc, cmplx gc,
            double smass, double swidth, cmplx* hvvs)
{
   hvvs[1] = v1[4]+v2[4]+sc[1];
   hvvs[2] = v1[5]+v2[5]+sc[2];
   double q[4];
   q[0] = hvvs[1].re;
   q[1] = hvvs[2].re;
   q[2] = hvvs[2].im;
   q[3] = hvvs[1].im;
   double q2 = (q[0]+q[3])*(q[0]-q[3])-(q[1]*q[1]+q[2]*q[2]);
   cmplx dg = 1./mkcmplx(q2-smass*smass, smass*swidth);
   hvvs[0] = -gc*dg*sc[0]*(v1[0]*v2[0]-v1[1]*v2[1]
                          -v1[2]*v2[2]-v1[3]*v2[3]);
}
\end{lstlisting}

 \subsection{SSS vertex}
 \label{sec:sss-vertex}
The \texttt{SSS} vertex functions are obtained from the following
Lagrangian
  \begin{equation}
   \mathcal{L}_{\mathrm{S}\! \mathrm{S}\! \mathrm{S}}
    \!= \texttt{gc} S^{*}_{1} S^{*}_{2} S^{*}_{3}\,.
  \end{equation}

  \subsubsection{sssxxx}
  \label{sec:sssxxx}

  The function \texttt{sssxxx} (List~\ref{list:sssxxx})
  computes an amplitude of the three-scalar coupling, \texttt{SSS}.
  The arguments of this function as:
  \begin{equation}
   \begin{array}{r}
    \texttt{sssxxx(cmplx* s1, cmplx* s2, cmplx* s3,}\\
    \texttt{cmplx gc, cmplx\& vertex)}\\
   \end{array}
  \end{equation}
  where the inputs and the outputs are:
  \begin{equation}
   \begin{array}{ll}
    \textsc{Inputs:}& \\
    \texttt{cmplx s1[3]} & \textrm{first scalar wavefunction}\\
    \texttt{cmplx s2[3]} & \textrm{second scalar wavefunction}\\
    \texttt{cmplx s3[3]} & \textrm{third scalar wavefunction}\\
    \texttt{cmplx gc}    & \textrm{coupling constant of \texttt{SSS}}\\
    &\\
    \textsc{Outputs:}& \\
    \texttt{cmplx\& vertex}& \textrm{amplitude}\\
    &\Gamma(\texttt{s1}, \texttt{s2}, \texttt{s3})
     \label{eq:sssxxx}
   \end{array}
  \end{equation}

\begin{lstlisting}[caption=sssxxx.cu,label=list:sssxxx]
#include "cmplx.h"

__device__
void sssxxx(cmplx* s1, cmplx* s2, cmplx* s3,
           cmplx gc, cmplx& vertex)
{
   vertex = gc*s1[0]*s2[0]*s3[0];
}
\end{lstlisting}

  \subsubsection{hssxxx}
  \label{sec:hssxxx}

  The function \texttt{hssxxx} (List~\ref{list:hssxxx})
  computes an off-shell scalar current from the three-scalar
  coupling, \texttt{SSS}.
  The arguments of this function as:
  \begin{equation}
   \begin{array}{r}
    \texttt{hssxxx(cmplx* s1, cmplx* s2, cmplx gc,}\\
    \texttt{double smass, double swidth,}\\
    \texttt{cmplx* hss)}
   \end{array}
  \end{equation}
  where the inputs and the outputs are:
  \begin{equation}
   \begin{array}{ll}
    \textsc{Inputs:}& \\
    \texttt{cmplx s1[3]} & \textrm{first scalar wavefunction}\\
    \texttt{cmplx s2[3]} & \textrm{second scalar wavefunction}\\
    \texttt{cmplx gc}    & \textrm{coupling constant of \texttt{SSS}}\\
    \texttt{double smass} & \textrm{mass of output scalar}\\
    \texttt{double swidth}& \textrm{width of output scalar}\\
    &\\
    \textsc{Outputs:}& \\
    \texttt{cmplx hss[3]}& \textrm{scalar current}\\
     &j(s^\prime:\texttt{s1}, \texttt{s2})
     \label{eq:hssxxx}
   \end{array}
  \end{equation}

\begin{lstlisting}[caption=hssxxx.cu,label=list:hssxxx]
#include "cmplx.h"

__device__
void hssxxx(cmplx* s1, cmplx* s2, cmplx gc,
            double smass, double swidth, cmplx* hss)
{
   hss[1] = s1[1]+s2[1];
   hss[2] = s1[2]+s2[2];
   double q[4];
   q[0] = hss[1].re;
   q[1] = hss[2].re;
   q[2] = hss[2].im;
   q[3] = hss[1].im;
   double q2 = (q[0]+q[3])*(q[0]-q[3])-(q[1]*q[1]+q[2]*q[2]);
   cmplx dg = 1./mkcmplx(q2-smass*smass, smass*swidth);
   hss[0] = -gc*dg*s1[0]*s2[0];
}
\end{lstlisting}

  \subsection{SSSS vertex}
  \label{sec:ssss-vertex}
The \texttt{SSSS} vertex functions are obtained from the following
Lagrangian
  \begin{equation}
   \mathcal{L}_{\mathrm{S}\! \mathrm{S}\! \mathrm{S}\! \mathrm{S}}
    \!= \texttt{gc} S_1^{*} S_2^{*} S^{*}_{3} S^{*}_{4}\,.
  \end{equation}

  \subsubsection{ssssxx}
  \label{sec:ssssxx}

  The function \texttt{ssssxx} (List~\ref{list:ssssxx})
  computes an amplitude of the four-scalar, \texttt{SSSS}, coupling.
  The arguments of \texttt{ssssxx} as:
  \begin{equation}
   \begin{array}{r}
    \texttt{ssssxx(cmplx* s1, cmplx* s2,} \\
    \texttt{ cmplx* s3, cmplx* s4,}\\
    \texttt{ cmplx gc, cmplx\& vertex)}
   \end{array}
  \end{equation}
  where the inputs and the outputs are:
  \begin{equation}
   \begin{array}{ll}
    \textsc{Inputs:} & \\
    \texttt{cmplx s1[3]} & \textrm{first scalar wavefunction} \\
    \texttt{cmplx s2[3]} & \textrm{second scalar wavefunction} \\
    \texttt{cmplx s3[3]} & \textrm{third scalar wavefunction} \\
    \texttt{cmplx s4[3]} & \textrm{fourth scalar wavefunction} \\
    \texttt{cmplx gc}    & \textrm{coupling constants of \texttt{SSSS}
     vertex} \\ \\
    \textsc{Outputs:} & \\
    \texttt{cmplx\& vertex} & \textrm{amplitude of four scalar coupling}\\
    & \Gamma(\texttt{s1},\texttt{s2},\texttt{s3},\texttt{s4})
     \\
    \label{eq:ssssxx}
   \end{array}
  \end{equation}

\begin{lstlisting}[caption=ssssxx.cu,label=list:ssssxx]
#include "cmplx.h"

__device__
void ssssxx(cmplx* s1, cmplx* s2, cmplx* s3, cmplx* s4,
           cmplx gc, cmplx& vertex)
{
   vertex = gc*s1[0]*s2[0]*s3[0]*s4[0];
}
\end{lstlisting}

  \subsubsection{hsssxx}
  \label{sec:hsssxx}

  The function \texttt{hsssxx} (List~\ref{list:hsssxx})
  computes an off-shell scalar current from \texttt{SSSS} coupling.
  This function has the argument as:
  \begin{equation}
   \begin{array}{r}
    \texttt{hsssxx(cmplx* s1, cmplx* s2, cmplx* s3, } \\
    \texttt{cmplx gc, double smass, double swidth,}\\
    \texttt{cmplx* hsss)}\\
   \end{array}
  \end{equation}
  where the inputs and the outputs are:
  \begin{equation}
   \begin{array}{ll}
    \textsc{Inputs:} & \\
    \texttt{cmplx s1[3]} & \textrm{first scalar wavefunction} \\
    \texttt{cmplx s2[3]} & \textrm{second scalar wavefunction} \\
    \texttt{cmplx s3[3]} & \textrm{third scalar wavefunction} \\
    \texttt{cmplx gc   } & \textrm{coupling constant \texttt{SSSS}} \\
    \texttt{double smass} & \textrm{mass of output scalar}\\
    \texttt{double swidth} & \textrm{width of output scalar}\\
    &\\
    \textsc{Outputs:} & \\
    \texttt{cmplx hsss[3]} & \textrm{scalar current}\\
    &j(\texttt{s}^\prime:\texttt{s1},\texttt{s2},\texttt{s3})
     \\
    \label{eq:hsssxx}
   \end{array}
  \end{equation}

\begin{lstlisting}[caption=hsssxx.cu,label=list:hsssxx]
#include "cmplx.h"

__device__
void hsssxx(cmplx* s1, cmplx* s2, cmplx* s3, cmplx gc,
            double smass, double swidth, cmplx* hsss)
{
   hsss[1] = s1[1]+s2[1]+s3[1];
   hsss[2] = s1[2]+s2[2]+s3[2];
   double q[4];
   q[0] = hsss[1].re;
   q[1] = hsss[2].re;
   q[2] = hsss[2].im;
   q[3] = hsss[1].im;
   double q2 = (q[0]+q[3])*(q[0]-q[3])-(q[1]*q[1]+q[2]*q[2]);
   cmplx dg = 1./mkcmplx(q2-smass*smass, smass*swidth);
   hsss[0] = -gc*dg*s1[0]*s2[0]*s3[0];
}
\end{lstlisting}

  %

\end{document}